\documentclass[aps,amsmath,amssymb,prd,showpacs,floatfix,preprint,superscriptaddress,nofootinbib,12pt]{article}
\usepackage{jheppub}
\usepackage{float}
\usepackage{verbatim}
\usepackage[section]{placeins}
\usepackage{bm}
\usepackage{mathtools}
\usepackage{tcolorbox}
\usepackage{cancel}
\usepackage[utf8]{inputenc}
\usepackage{slashed,verbatim}
\pdfoutput=1
\usepackage{epigraph,lipsum}
\usepackage{fancyhdr}
\usepackage{epigraph}
\usepackage{cancel}
\usepackage{graphicx}
\makeatletter
\def\@fpheader{\relax}
\makeatother

\usepackage{lipsum}

\usepackage{wasysym}
\usepackage{amssymb}

\newcommand\blfootnote[1]{%
  \begingroup
  \renewcommand\thefootnote{}\footnote{#1}%
  \addtocounter{footnote}{-1}%
  \endgroup
}

\usepackage{bbding}
\usepackage{pifont}
\usepackage{gensymb}

\usepackage[framemethod=tikz]{mdframed}

\usepackage{amsmath,epsfig}
\usepackage{amssymb,amsfonts}
\usepackage{latexsym}
\usepackage{graphicx}

\newcommand{\cool}{\ensuremath{%
  \mathchoice{\includegraphics[height=3ex]{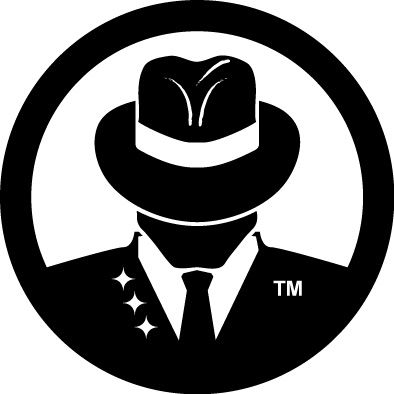}}
    {\includegraphics[height=3ex]{peace.jpg}}
    {\includegraphics[height=2.5ex]{peace.jpg}}
    {\includegraphics[height=2ex]{peace.jpg}}
}}

\usepackage{subeqnarray}
\usepackage{xcolor}

\usepackage{graphicx}
\usepackage{ulem}

\newcommand{\me}{\ensuremath{%
  \mathchoice{\includegraphics[height=2ex]{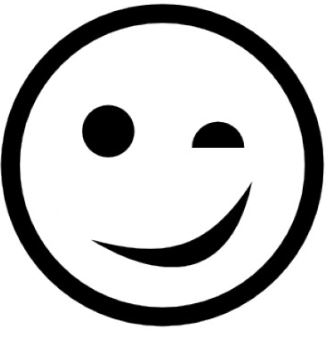}}
    {\includegraphics[height=2ex]{cat.png}}
    {\includegraphics[height=1.5ex]{cat.png}}
    {\includegraphics[height=1ex]{cat.png}}
}}

\def\be{\begin{equation}}
\def\ee{\end{equation}}
\def\bea{\begin{eqnarray}}
\def\eea{\end{eqnarray}}

\newcommand\fverb{\setbox\pippobox=\hbox\bgroup\verb}
\newcommand\fverbdo{\egroup\medskip\noindent%
                        \fbox{\unhbox\pippobox}\ }
\newcommand\fverbit{\egroup\item[\fbox{\unhbox\pippobox}]}

\newcommand{\bear}{\begin{eqnarray}}

\newcommand{\eear}{\end{eqnarray}}

\newcommand{\bsea}{\begin{subeqnarray}}
\newcommand{\esea}{\end{subeqnarray}}
\newbox\pippobox

\def\lab{\label}
\def\6{\partial}

\newcommand{\comments}[1]{}
\input Starburst.fd

\allowdisplaybreaks[3]

\preprint{IFT-UAM/CSIC-19-63}

\begin{document}



\title{%
  \Huge Zoology of Solid \& Fluid Holography \\
  \Large Goldstone Modes and Phase Relaxation}
\author[\,\cool, \,\me]{Matteo Baggioli}
\author[\,\Diamond\,,\,\star]{, Sebastian Grieninger}
\vspace{0.1cm}
\affiliation[\cool]{Instituto de Fisica Teorica UAM/CSIC,
c/ Nicolas Cabrera 13-15, Cantoblanco, 28049 Madrid, Spain}
\affiliation[\Diamond]{Theoretisch-Physikalisches Institut, Friedrich-Schiller-Universit\"at Jena,
Max-Wien-Platz 1, D-07743 Jena, Germany.}

\affiliation[\star]{Department of Physics, University of Washington, Seattle, WA 98195-1560, USA}

\emailAdd{matteo.baggioli@uam.es}
\emailAdd{sebastian.grieninger@gmail.com}

\blfootnote{\me   \,\,\,\url{https://members.ift.uam-csic.es/matteo.baggioli}}

\abstract{We provide a comprehensive classification of isotropic solid and fluid holographic models with broken translational invariance. We describe in detail the collective modes in both the transverse and longitudinal sectors. First, we discuss holographic fluid models, \textit{i.e.} systems invariant under internal volume preserving diffeomorphisms. We consider the explicit (EXB) and the spontaneous (SSB) breaking of translations and we emphasize the differences with respect to their solid counterpart. Then, we present a study of the longitudinal collective modes in simple holographic solid and fluid models exhibiting the interplay between SSB and EXB. We confirm the presence of light pseudo-phonons obeying the Gell-Mann-Oakes-Renner relation and the validity of the relation proposed in the literature between the novel phase relaxation scale, the mass of the pseudo-Golstone modes and the Goldstone diffusion. Moreover, we find very good agreement between the dispersion relation of our longitudinal sound mode and the formulae derived from the Hydro$+$ framework. Finally, our results suggest that the crystal diffusion mode does not acquire a simple damping term because of the novel relaxation scale proportional to the EXB. The dynamics is more complex and it involves the interplay of three modes: the crystal diffusion and two more arising from the splitting of the original sound mode. In this sense, the novel relaxation scale, which comes from the explicit breaking of the global internal shift symmetry of the St\"uckelberg fields, is different from the one induced by elastic defects, and depending solely on the SSB scale.}

\maketitle

\section{Introduction}

\epigraph{Symmetry is pleasing but not as sexy. Einstein is cool but Picasso knows what I'm talking about.}{\textit{Amy Poehler}}
Classifying the different phases of matter present in Nature is one of the fundamental task of Physics and Science in general. Historically, this organizational program has been initiated by looking at simple physical properties such as thermodynamical and transport features. Ice has a lower density than water; an insulator has lower electric conductivity than a metal and a superconductor has even infinite conductivity. Later on with time, physicists realized the importance of classifying the various phases of matter accordingly to the dynamics of the collective excitations therein. With this purpose in mind, the idea of building effective field theories (EFTs) in terms of these (few) low-energy degrees of freedom lead to important theoretical developments. Hydrodynamics is certainly one of them. Importantly, its formulation is based on conservation laws, already revealing the crucial importance of symmetries. In this sense, hydrodynamics can be applied not only to fluids but also to more complex systems, like solids and liquid crystals \cite{PhysRevB.22.2514,PhysRevA.6.2401}, superconductors \cite{Davison:2016hno}, charge density waves \cite{Delacretaz:2017zxd} and systems in strong magnetic fields \cite{Delacretaz:2019wzh}. In other words, hydrodynamics can still be predictive and very useful in presence of broken spacetime/internal symmetries \cite{forster1975hydrodynamic,boon1991molecular}.\\

The most important theoretical step for the classification of the various phases of matter has been the realization that symmetries are the key feature. In particular, the various phases like liquids, solids and even more exotic ones (supersolids, framids, \dots), can be simply classified according to the different spontaneous symmetry breaking patterns of Poincar\'e symmetry they correspond to \cite{Nicolis:2015sra}. The idea is that the Poincar\'e symmetry represents a fundamental and universal feature at high energy, even though it appears to be spontaneously broken at low energy\footnote{This program can be rigorously formalized using opportune Coset construction and mathematical tools \cite{Delacretaz:2014jka,Nicolis:2013lma}.}. Matter configurations select a preferred reference frame, breaking this invariance. The simplest example is that of the Ionic lattice in ordered crystals.\\

Let us explain in more detail how this classification works, focusing our attention to the most common phases of matter: fluids and solids. For simplicity, we restrict our discussion to homogeneous and isotropic systems. The description is based on a set of scalar fields:
\begin{equation}
    \Phi^I(\vec{x},t) \quad \text{with}\quad I\,=\,1,2,\dots\,,d
\end{equation}
which label the positions of the individual volume elements at fixed time; in other words, they can be thought as a set of comoving coordinates. The scalars enjoy internal shift symmetry and they define the equilibrium configuration by their expectation value:
\begin{equation}
    \langle \Phi^I \rangle \,=\,\delta^I_{\,\,i}\,x^{i} \label{vev}
\end{equation}
where $x^I$ are the spatial coordinates of the system (in our case $I=x,y$). The capital letters indicate the internal indices while the lower case denotes the spacetime ones\footnote{In the following, we will forget about this difference and denote both coordinates in the same way.}.
\begin{figure}
    \centering
  \tcbox[colframe=blue!30!black,
           colback=white!30]{ \includegraphics[width=0.6\linewidth]{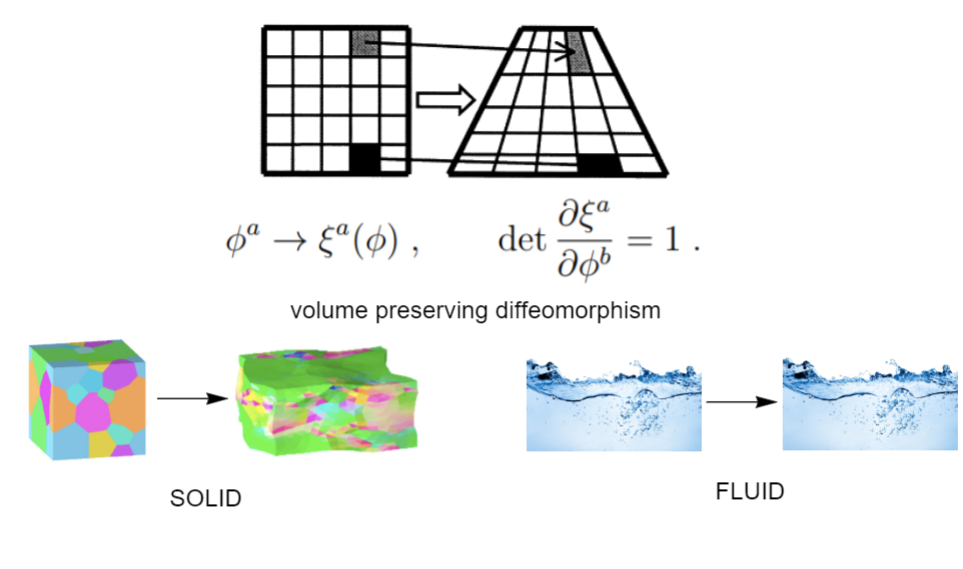}}
    \caption{The distinction between a solid and a liquid from the symmetry point of view. Both systems are invariant under $ISO(d)$ but the liquid is also invariant under a bigger symmetry group, volume preserving diffeomorphisms. That is just the mathematical formulation of the statement that ''liquids take the shape of the container in which they are placed''.}
    \label{fig0}
\end{figure}
Obviously, the equilibrium configuration, defined by eq.\eqref{vev}, breaks Lorentz invariance spontaneously. More precisely, it breaks spatial translations and rotations (and boosts). In order to retain the properties of homogeneity and isotropy, the scalar fields have to be invariant under internal translations and rotations:
\begin{align}
   \Phi^I\,\rightarrow \phi^I\,+\,a^I\,,\quad \quad \Phi^I\,\rightarrow\,SO(d)\,\cdot\,\Phi^I \label{is}
\end{align}
where $d$ indicates the number of spatial directions. The union of these two transformations is usually indicated as the group $ISO(d)$.\\
The scalars $\Phi^I$ can be identified as Goldstone modes, which appear because of the SSB induced by the equilibrium configuration \eqref{vev}. It is important to observe that despite broken spacetime translations, a diagonal group between internal translations and spacetime translations is preserved. In fact, the system is invariant under the combined transformation:
\begin{align}
    \langle \text{internal translations}\rangle \,\quad & \otimes \quad \langle \text{spacetime translations}\rangle\nonumber\\
   \Phi^I\rightarrow \Phi^I\,+\,a^I\,\quad\,\,\,\,\,\,\,\,\, &  \,\,\,\,\,\,\,\,\,\,\,\,\,\,\,\,\,\,\,\quad \, x^I\,\rightarrow x^I\,-\,a^I
\end{align}
where the index $I$ here only represents the spatial directions.\\
Once the symmetries of the system are defined, we are in the position of writing the effective action for our EFT. The fundamental building block is the kinetic matrix:
\begin{equation}
    \mathfrak{X}^{IJ}\,\equiv\,\partial_\mu \Phi^I\,\partial^\mu \Phi^J
\end{equation}
from which the two following scalar quantities can be constructed:
\begin{equation}
    \mathcal{X}\,\equiv\, Tr\left[ \mathfrak{X}^{IJ}\right]\,,\quad \mathcal{Z}\,\equiv\, det\left[ \mathfrak{X}^{IJ}\right]
\end{equation}
Both terms are invariant under an $ISO(d)$ transformation \eqref{is}. The most generic effective action can therefore be written as:
\begin{equation}
    \mathcal{S}\,=\,\int\,d^3x\,\mathcal{V}\left(\mathcal{X},\mathcal{Z}\right)\label{action1}
\end{equation}
where $\mathcal{V}$ is an arbitrary scalar potential\footnote{Strictly speaking this statement is not correct. The potential $\mathcal{V}\left(\mathcal{X},\mathcal{Z}\right)$ has to satisfy some basic conditions in order for the theory to be unitary, causal and stable. See \cite{Alberte:2015isw} for more details.}.\\
The simple action in  eq.\eqref{action1} represents the most generic effective field theory for (isotropic and homogeneous) solids and liquids. All the thermodynamic properties and the relevant low energy collective excitations can be obtained introducing perturbations on top of it. The full theory of elasticity and (non dissipative) hydrodynamics can be recovered as well using this framework \cite{Alberte:2018doe}.\\

Until this point, we have not made any distinction between solids and fluids; both phases of matter enjoy invariance under the $ISO(d)$ group defined in \eqref{is}. The difference between the two phases are clear when we think about our everyday experience. Fluids take the shape of the container in which they are placed; solids do not. From a more mathematical perspective, this means that fluids enjoy a larger symmetry, which takes the name of volume preserving internal diffeomorphism (VPdiffs). In particular, fluids are invariant under the internal transformation:
\begin{equation}
    \Phi^I\,\rightarrow \xi^I(\Phi)\,,\quad \quad det \frac{\partial \xi^I}{\partial \Phi^J}\,=\,1\label{diff}
\end{equation}
with unitary Jacobian matrix. Invariance under the transformations of eq.\eqref{diff} strongly constraints the original action \eqref{action1} and in particular it does not permit any possible dependence on the scalar object $X$. Finally, the effective action for fluids take the form:
\begin{equation}
    \mathcal{S}_{\text{fluids}}\,=\,\int\,d^3x\,\mathcal{V}(\mathcal{Z})
\end{equation}
At the level of the physical observables, the fluid choice above forces the shear elastic modulus and the propagation speed of transverse phonons to be zero:
\begin{equation}
    G\,=\,0\,,\quad \,c_T^2\,=\,0
\end{equation}
since both these quantities are proportional to the $X$-derivative of the potential appearing in eq.\eqref{action1}. In summary, using this framework, we can recover the well-known properties of fluids simply using symmetry arguments.\\

The EFT framework just presented is elegant and powerful but it inherently has two important flaws. First, it is not easily generalizable to finite temperature. It is extremely hard, and still an open question, to include dissipation within this framework. Including dissipation within this framework is not only extremely difficult but still an open question \cite{Endlich:2012vt}. The same problem has been recently discussed in the context of the Lagrangian formulation for hydrodynamics \cite{Glorioso:2017fpd,Haehl:2018lcu,Jensen:2018hse,Crossley:2015evo}. Second, the EFT picture is not valid in a scale invariant system like a quantum critical material. Scale invariance implies the absence of a mass gap and the presence of a continuum of states with no separation of scales, making invalid the basic assumptions of the effective field theory methods. Therefore, it is far from obvious how to apply this construction to critical materials, especially if they are strongly coupled.\\

Given the outlined shortcomings, it is valuable and important to extend the EFT methods to dissipative and quantum critical systems. The Holographic methods \cite{Hartnoll:2016apf} are a concrete possibility to overcome these problems. In order to consider the previous picture in the context of the gauge-gravity duality, all the spacetime symmetries have to be gauged\footnote{The question regarding the internal symmetries of the St\"uckelberg fields is more subtle. We will not gauge those symmetries in the bulk. See \cite{Esposito:2017qpj} for more discussions about this point.} and the following action has to be considered:
\begin{equation}
    S\,=\,\int d^4X\,\sqrt{-g}\,\left[R\,-\,2\,\Lambda\,-\,m^2\,V(X,Z)\right]\label{ee}
\end{equation}
where $g$ is the metric tensor of the curved spacetime background, $R$ the corresponding Ricci scalar and $\Lambda$ a possible cosmological constant\footnote{A negative cosmological constant is actually necessary to formally apply the AdS-CFT correspondence.}.\\
The holographic action in eq.\eqref{ee} is built using a set of bulk scalar fields $\phi^I(\vec{x},t,u)$ from which:
\begin{equation}
I^{IJ}\,=\,\partial_\mu \phi^I \partial^\mu \phi^J\,,\quad X=Tr[I^{IJ}]\,,\quad Z=Det[I^{IJ}],
\end{equation}
in analogy with the previous field theory discussion.
Before proceeding, it is important to distinguish the bulk fields $\phi^I(\vec{x},t,u)$ from the EFT scalar fields $\Phi^I(\vec{x},t)$. The fields $\phi^I$, living in our four dimensional bulk geometry, are the duals, in the holographic sense, of the EFT fields $\Phi^I$. At the same time, the relation between the bulk potential $V(X,Z)$ and the EFT potential $\mathcal{V}(\mathcal{X,\mathcal{Z})}$ is not direct and transparent. In other words, a bulk theory defined by a potential $V(X,Z)$ does not correspond to a dual field theory defined by the same potential $V(\mathcal{X},\mathcal{Z})$.\\

That said, it is well known in the literature \cite{Dubovsky:2004sg,Rubakov:2008nh,Hinterbichler:2011tt} that the theory defined above in eq.\eqref{ee} defines a Lorentz violating massive gravity theory, where the mass of the graviton $m$ can be thought as the consequence of the fixed reference frame induced by the matter content. The set of scalars $\phi^I$ are simply the St\"uckelberg fields for the massive gravity theory or in other words the Goldstone bosons for the broken symmetries. The role of these theories in the context of holography has been discussed in several previous works \cite{Baggioli:2014roa,Baggioli:2015gsa,Alberte:2015isw,Alberte:2016xja,Alberte:2017cch,Alberte:2017oqx,Andrade:2019zey,Ammon:2019wci}, for a simple review see \cite{Baggioli:2016rdj}.\\

Before proceeding, let us make an important remark. From the bulk spacetime perspective, the equilibrium scalars profile $\phi^I=x^I$ breaks the spacetime symmetries (translations) spontaneously. The scalars are just the bulk phonons. From the point of view of the dual field theory, the situation is different. The bulk scalars $\phi^I$ correspond to a set of operators in the dual field theory $\mathcal{O}^I$ breaking translations. Nevertheless, the nature of the symmetry breaking, whether explicit (EXB) or spontaneous (SSB), depends on the asymptotic behaviour of the bulk scalars $\phi^I$ close to the UV AdS boundary. In particular, the generic asymptotic behavior for the scalars reads:
\begin{equation}
    \phi\,=\,\phi_0\,+\,\phi_1\,u^{p}\,+\,\dots
\end{equation}
where the UV boundary is set at $u=0$. Here, following the AdS-CFT dictionary \cite{Aharony:1999ti}, two possibilities arise\footnote{We do not consider here the possibility of taking the ''alternative quantization'' scheme nor that of doing a double trace deformation with mixed boundary conditions.}:
\begin{enumerate}
    \item $p>0$: $\phi_0$ is the source for the dual operator $\mathcal{O}$ and $\phi_1$ its VEV $\phi_1\equiv \langle \mathcal{O}\rangle$. In this case, the background solution $\phi^I=x^I$ represents a source for the dual operator and therefore the breaking of translations is explicit.
    \item $p<0$: $\phi_0$, is the VEV this time, and $\phi_1$ is the source. In this scenario the breaking is spontaneous, indeed totally dynamical\footnote{For subtleties regarding the minimization of the Free energy and thermodynamic stability see \cite{Alberte:2017oqx}.}.
\end{enumerate}
As explained in detail in \cite{Alberte:2017oqx}, in order to have SSB, we have to choose a potential with:
\begin{align}
    &V(X,Z)\,=\,X^N\,,\quad \quad N>5/2\,\quad \text{for solids}\\
    &V(X,Z)\,=\,Z^N\,,\quad \quad N>5/4\,\quad \text{for fluids}
\end{align}
while for smaller powers the breaking is always explicit (see for example \cite{Andrade:2013gsa}).
Moreover, with the same class of models, it is possible to implement the pseudo-spontaneous breaking of translations, where the breaking is mostly, but not totally, spontaneous. This was done in \cite{Alberte:2017cch,Ammon:2019wci} and will be explained further in the following.\\
In this manuscript we analyze in detail the collective modes in  holographic models breaking translations. We consider both solids but most importantly fluids, which have received less attention in the existing literature so far. We discuss the explicit (EXB), spontaneous (SSB), and pseudo-spontaneous breaking patterns. We aim to complete and collect in a comprehensive picture the study of solid and fluid holographic models with broken translational invariance. We provide a systematic description of the transport properties and the low energy collective modes and we show the agreement between the numerical results and the expected behaviours obtained via hydrodynamic methods.
\begin{figure}[h]
    \centering
    \includegraphics[width=0.6\linewidth]{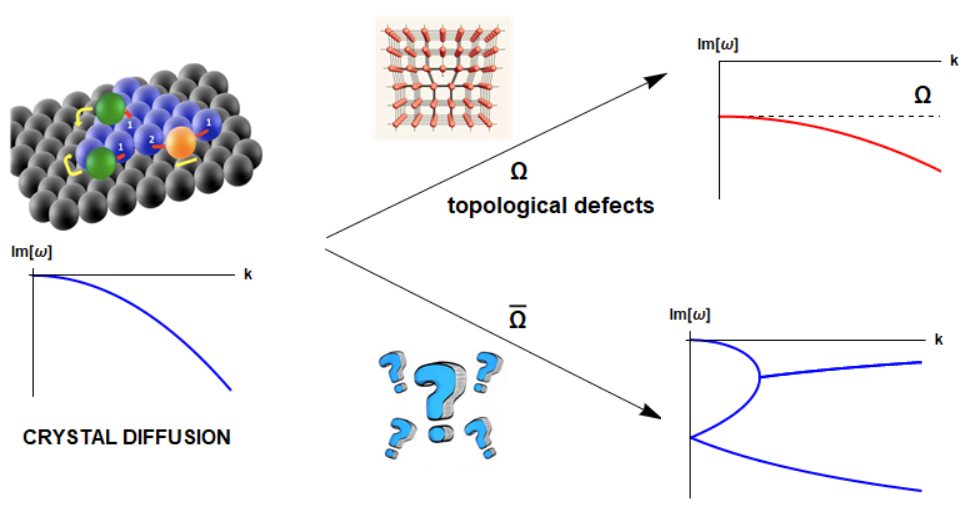}
    \caption{The dynamics of crystal diffusion in presence of phase relaxation. Phase relaxation mechanisms, $\Omega$, induced by the proliferation of elastic defects, as dislocations, produce a damping term for  the crystal diffusion mode $\omega\,=\,-\,i\,\Omega\,-\,D_\phi k^2$. In the holographic models, the relaxation mechanism $\bar{\Omega}$ is proportional to the explicit breaking of translations and does induce a complicated dynamics between three different modes. In summary, the two mechanisms can not be simply thought as different contributions to the total phase relaxation $\Omega_{rel}$.}
    \label{figd1}
\end{figure}\\
Finally, we address a very specific question, which has been discussed in recent literature \cite{Alberte:2017cch,Donos:2019txg,Ammon:2019wci,Andrade:2018gqk,Amoretti:2018tzw}. Can the interplay of explicit and spontaneous translations breaking induce a novel contribution to the phase relaxation of the phonons? Phase relaxation, $\Omega$, is usually attributed to the presence and the proliferation of topological defects, such as dislocations or disclinations \cite{kosevich2006crystal}. In that sense, from the microscopic point of view, it is related to the SSB of translational invariance and it should be insensitive to the explicit breaking mechanisms. Its presence has important consequences on the hydrodynamic modes and transport coefficients \cite{Delacretaz:2017zxd}. At the moment, there are no holographic models able to describe the physics of those elastic defects. Probably, the closest model is that introduced in \cite{Grozdanov:2018ewh}. Nevertheless, recent works \cite{Alberte:2017cch,Donos:2019txg,Ammon:2019wci,Andrade:2018gqk,Amoretti:2018tzw} noticed the presence of a similar contribution, denoted by $\bar{\Omega}$, which crucially depends on the EXB scale. This relaxation mechanism, independently of its microscopic nature, is significantly different from the one we just mentioned. Apparently, it has already been discussed in the condensed matter community \cite{fogler2000dynamical,doi:10.1143/JPSJ.45.1474}, but its complete hydrodynamic formulation is missing. Curiously, it enters in most of the observables\footnote{It does not produce anyway a pole in the frequency dependent viscosity $\eta(\omega)$, as shown in \cite{Ammon:2019wci}. Whether the reason for this is simply the range of parameters chosen or some more fundamental reason is not clear yet.} as the ''proper'' phase relaxation $\Omega$. Here, we perform an additional test of this hypothesis by checking directly the dispersion relation of the crystal diffusion mode in the pseudo-spontaneous regime. More concretely, we question the presence of a mode with dispersion relation
\begin{equation}
    \omega\,=\,-\,i\,\overset{?}{\Omega_{rel}}\,-\,i\,D_\phi\,k^2\,+\,\dots \quad \quad \Omega_{rel}\,=\,\Omega\,+\,\bar{\Omega}\label{sisi}
\end{equation}
where $\Omega_{rel}$ is the total phase relaxation rate, namely the sum of all the possible contributions, whether dependent on the SSB or the EXB scale.\\
Given that in our model $\Omega=0$, the only possibility to have the mode in eq.\eqref{sisi} damped is given by the coexistence of EXB and SSB, encoded in the novel parameter $\bar{\Omega}$. This analysis does not aim to shed light on the microscopic mechanism behind the novel relaxation scale $\bar{\Omega}$, but it can certainly give more indications whether it acts exactly like the phase relaxation induced by topological defects or not.\\

The numerical results suggest that this novel relaxation scale $\bar{\Omega}$ does not induce a simple dynamics as that explained above in eq.\eqref{sisi}. In particular, $\bar{\Omega}$ does not immediately give any finite damping contribution to the crystal diffusion mode present in the longitudinal spectrum. On the contrary, it produces a much more complicated phenomenology, which shows non-trivial interactions between three different soft modes. One of them is the crystal diffusion mode and the other two come from the splitting of the original longitudinal sound mode. For a simplified representation see fig.\ref{figd1}. In summary, the interplay of SSB and EXB produces a novel relaxation time scale whose effects on the hydrodynamic mode are different from those of the phase relaxation discussed in \cite{Delacretaz:2017zxd}. Unfortunately, this means that not only a microscopic understanding of such novel phase relaxation mechanism is absent. An effective and hydrodynamic treatment in accordance with numerical results is to the best of our knowledge, still lacking.\\
On the positive side, our numerical results concerning the holographic fluid models confirm the universal relation proposed in \cite{Amoretti:2018tzw,Andrade:2018gqk} between the phase relaxation rate $\bar{\Omega}$, the pinning frequency $\omega_0$ and the Goldstone diffusion $\xi$ :
 \begin{equation}
        \bar{\Omega}\,\sim\,\mathrm{M}^2\,\xi\,\sim\,\frac{\omega_0^2\,\chi_{PP}}{G}\,\xi
 \end{equation}
 This might represent the first step to understand the fundamental nature of this new phase relaxation mechanism dependent on the explicit breaking of translations.
    \\[0.5cm]
\section{The class of holographic models}\label{sec:model}
We consider the most generic Lorentz violating holographic massive gravity theory introduced in \cite{Alberte:2015isw} and defined by the following action: 
\begin{equation}\label{action}
S\,=\, M_P^2\int d^4x \sqrt{-g}
\left[\frac{R}2+\frac{3}{\ell^2}- \, m^2 V(X,Z)\right]\, ,
\end{equation}
in $d+1=4$ bulk dimensions. We define the kinetic matrix $\mathfrak{X}^{IJ} \equiv \frac12 \, g^{\mu\nu} \,\partial_\mu \phi^I \partial_\nu \phi^J$ and the corresponding scalar invariants $X \equiv Tr[\mathfrak{X}^{IJ}]$ and $Z \equiv det[\mathfrak{X}^{IJ}]$. The $\phi^I$ are the St\"uckelberg fields for the massive gravity theory \cite{Dubovsky:2004sg,Alberte:2015isw}. They display a simple bulk profile:
\begin{equation}
    \phi^I\,=\,x^I
\end{equation}
where we indicate the internal and spatial coordinates with the same latin index. The scalar fields are the Goldstone bosons for spacetime bulk translations. From the dual field theory perspective, they break spacetime translations and internal shifts down to the diagonal group, as explained in the introduction.\\
We study 4D asymptotically AdS black hole configurations using Eddington-Finkelstein (EF) coordinates:
\begin{equation}
\label{backg}
ds^2=\frac{1}{u^2} \left[-f(u)\,dt^2-2\,dt\,du + dx^2+dy^2\right]\, ,
\end{equation}
with $u\in [0,u_h]$ the radial holographic direction spanning from the boundary $u=0$ to the horizon, defined through $f(u_h)=0$. We consider backgrounds with zero charge density; the blackening factor reads:
\begin{equation}\label{backf}
f(u)= u^3 \int_u^{u_h} d\xi\;\left[ \frac{3}{\xi^4} -\frac{m^2}{\xi^4}\, 
V(\xi^2,\xi^4) \right] \, .
\end{equation}
The corresponding temperature of the dual QFT takes the following form:
\begin{equation}
T=-\frac{f'(u_h)}{4\pi}=\frac{6 -  2 m^2 V\left(u_h^2,u_h^4 \right)}{8 \pi u_h}\, ,\label{eq:temperature}
\end{equation}
while the entropy density is simply $s=2\pi/u_h^2$. Without loss of generality, we will set the radius of the BH horizon to $u_h=1$ throughout the manuscript. Finally, we can write down the energy density of the system as:
\begin{equation}
    \epsilon\,=\,1\,+\,m^2\,\int_0^1\,\frac{V(\zeta^2,\zeta^4)}{\zeta^4}\,d\zeta
\end{equation}
and the momentum susceptibility as $\chi_{PP}=3/2\, \epsilon$.\\

In the following we will consider both the transverse and longitudinal sectors of the fluctuations and in particular we will analyze the corresponding collective modes in detail. We will assume different choices for the potential $V(X,Z)$; in other words, we will focus on different symmetry breaking patterns. For details about the technical computations and the numerical techniques we refer the reader to the appendices \ref{app1} and  \ref{app2}.
\section{An holographic fluid with broken translations}\label{fluidEXB}
The first model we consider is a fluid model with explicitly broken translational invariance. It is defined by the potential:
\begin{equation}
    V(X,Z)\,=\,Z \label{aa}
\end{equation}
and it can be thought as the fluid counterpart of the widely used linear axions model \cite{Andrade:2013gsa}. From a technical point of view, it differs from the solid model due to the presence of an additional internal symmetry. In specific, the model defined in \eqref{aa} is invariant under internal volume preserving diffeomorphisms (VPDiffs):
\begin{equation}
    \phi^I\,\longrightarrow \xi^I(\phi)\,,\quad \quad det \,\frac{\partial \xi^I}{\partial \phi^J}\,=\,1\label{sym}
\end{equation}
which are the fundamental difference between solids and fluids. Moreover, the invariance under VPDiffs forces the mass of the helicity-2 component of the graviton to be zero. As a consequence, the equation of motion for the transverse and traceless component of the graviton $h^{TT}$ simply reads:
\begin{equation}
    \Box \,h^{TT}\,=\,0 \label{bb1}
\end{equation}
as in translational invariant systems, \textit{e.g.} Schwarzschild background.\\
From equation \eqref{bb1}, using the membrane paradigm \cite{Iqbal:2008by}, we immediately obtain that in this fluid model:
\begin{equation}
    \frac{\eta}{s}\,=\,\frac{1}{4\,\pi}
\end{equation}
namely the Kovtun-Son-Starinets (KSS) bound \cite{Policastro:2001yc} is saturated.
\begin{figure}[h!]
    \centering
    \includegraphics[width=0.35\linewidth]{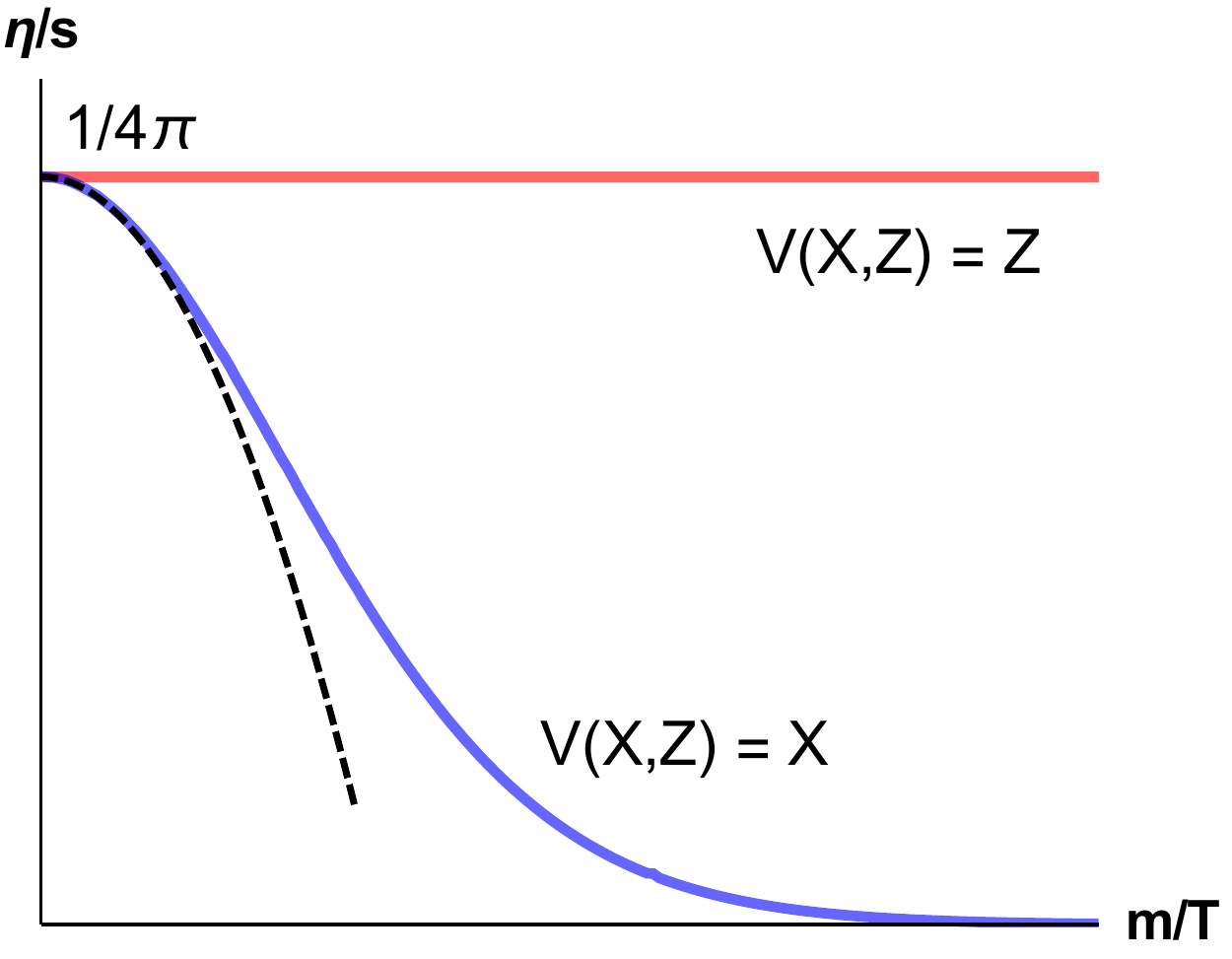}
    \quad
    \includegraphics[width=0.39\linewidth]{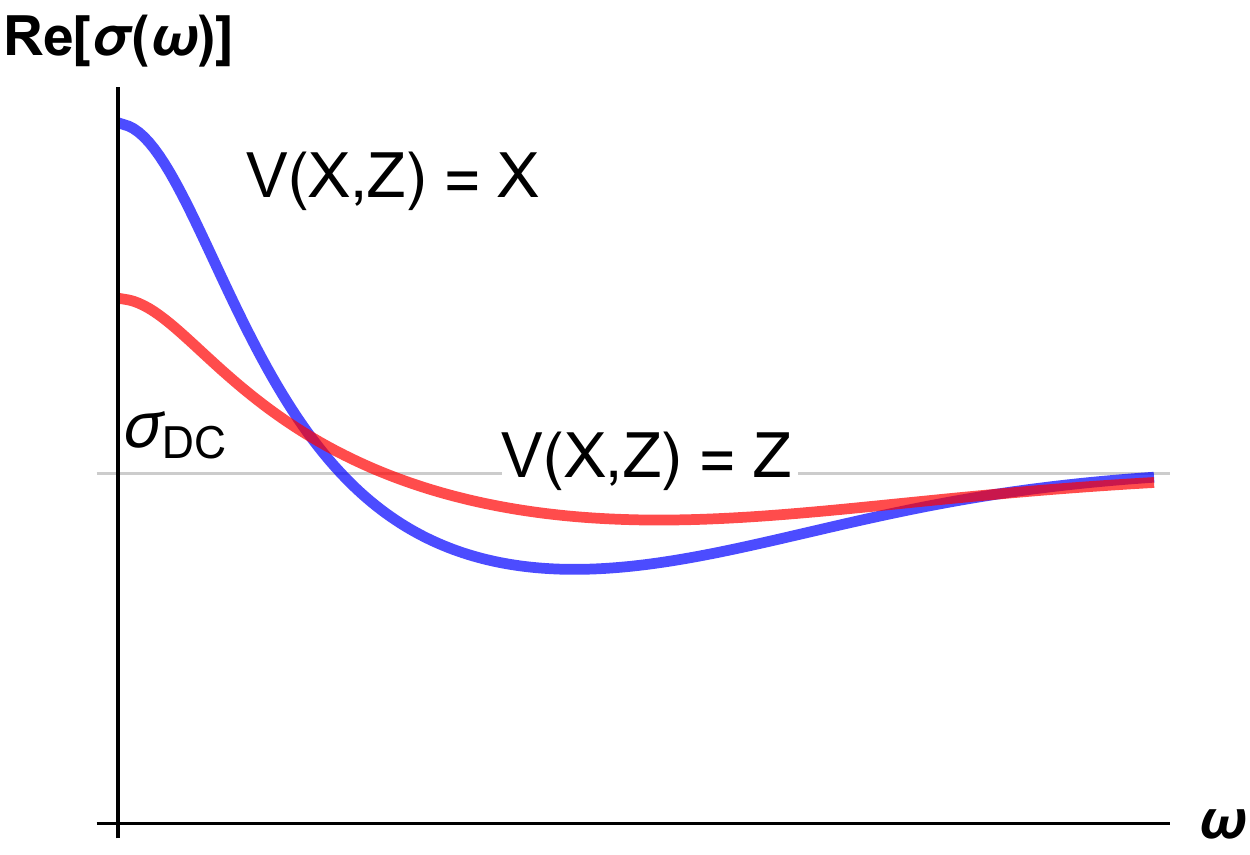}
    \caption{The difference between the linear axion model $V(X,Z)=X$ \cite{Andrade:2013gsa} and the fluid model $V(X,Z)=Z$. Both models break translations explicitly and have a finite DC conductivity. Nevertheless the first violates the Kovtun-Starinets-Son (KSS) bound while the second does not. The violation of the KSS bound in the first case can be derived analytically (dashed line) \cite{Alberte:2016xja}.}
    \label{fig1}
\end{figure}\\
This represents a striking difference with the solid linear axion models \cite{Andrade:2013gsa}, which is known to violate the viscosity-entropy ratio bound \cite{Hartnoll:2016tri,Alberte:2016xja,Burikham:2016roo,Ling:2016ien}.
This is an important point given the confusion in the literature regarding the explicit breaking of translations and the violation of the KSS bound. From the results in this model, it is evident that momentum dissipation does not necessarily imply such a violation. The key technical point is the vanishing of the helicity-2 graviton mass, which is the real reason behind the failure of the KSS bound, as explained in general terms in \cite{Hartnoll:2016tri,Alberte:2016xja}.\\
Nonetheless, both the solid and fluid models display a finite mass for the helicity-1 component of the graviton \cite{Alberte:2015isw}. That accounts for the non-conservation of the momentum operator of the dual field theory and more practically for a finite DC conductivity \cite{Alberte:2015isw}. The comparison between the solid and fluid models is shown in fig.\ref{fig1}.\\
Let us now move to discuss in detail the collective modes of the fluid model \eqref{aa}.
\begin{figure}[h]
    \centering \includegraphics[width=5cm]{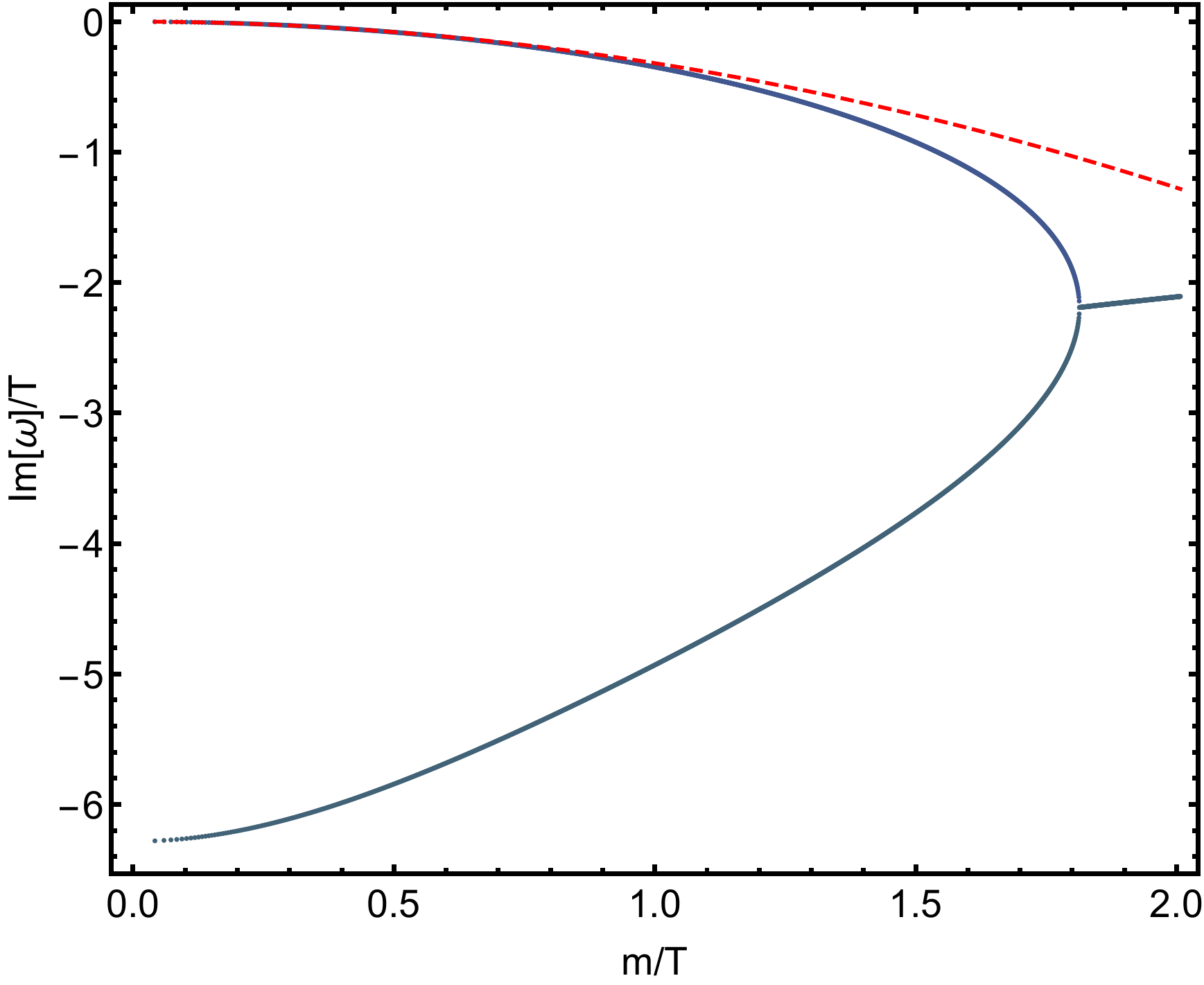}
    \includegraphics[width=5.3cm]{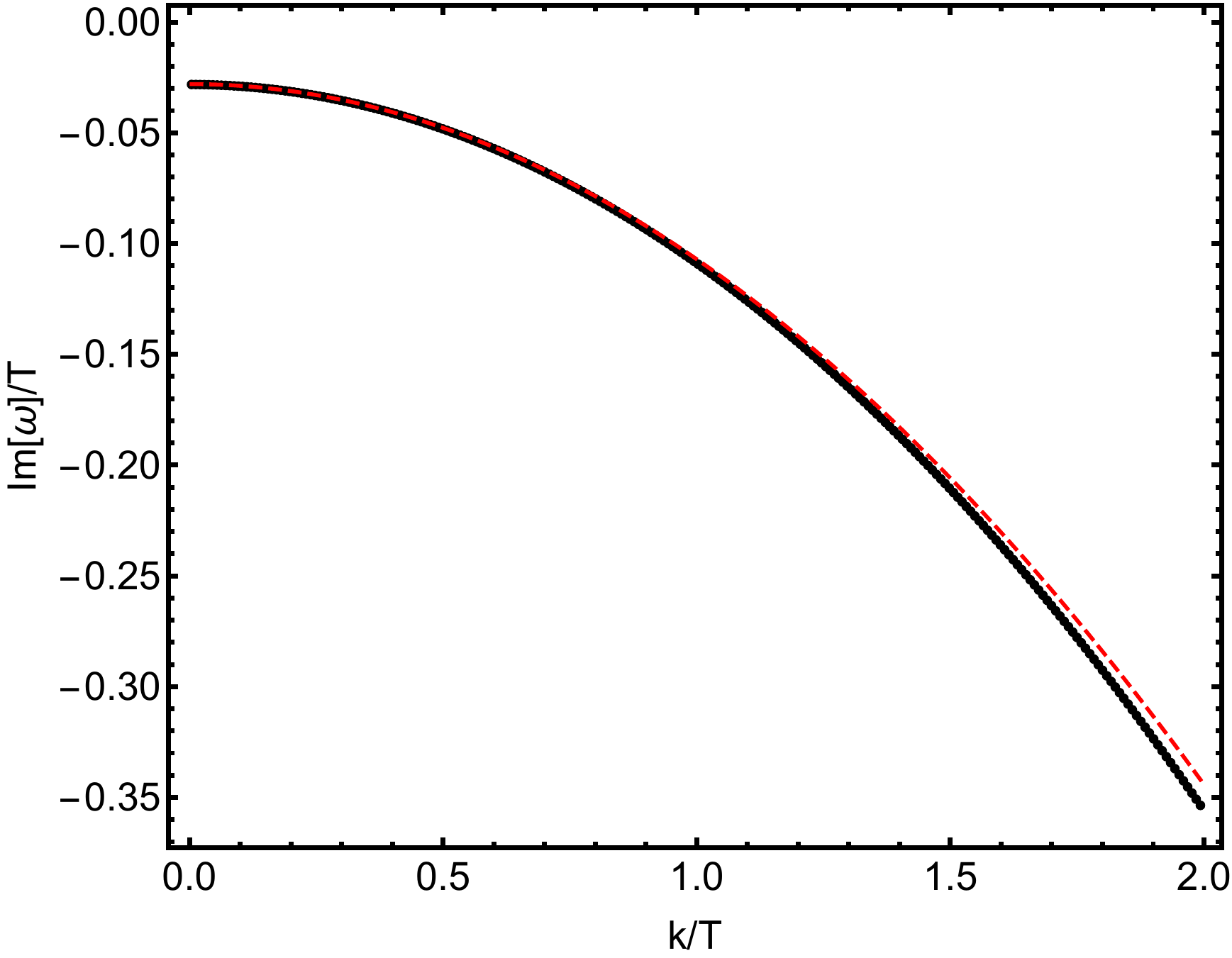}
        \caption{\textbf{Left: } The imaginary part of the lowest mode of the transverse spectrum for the fluid model \eqref{aa} in function of the dimensionless parameter $m/T$. The red dashed line is formula \eqref{d1}. Around $m/T \sim 1.8$ the collision indicating the coherent-incoherent transition happens. \textbf{Right. } The dispersion relation of the lowest mode for $m/T=0.2966$. The dashed line is the expression obtained formula \eqref{d1}. The agreement at $k/T \ll 1$ is good.}
        \label{figuno}
\end{figure}
\subsection*{The transverse sector}\label{tran1}
We start considering the transverse sector of the fluctuations and the corresponding QNMs.
For this concrete model, the masses of the helicity-2 and helicity-1 components of the graviton read:
\begin{equation}
    m_T^2\,=\,0,\quad m_V^2\,=\,2\,u^2\,m^2\,V_Z\,,
\end{equation}
where the abbreviation $V_Z$ stands for $\partial_Z V(X,Z)$, \textit{e.g.} $V_Z=1$ for the specific choice in eq.\eqref{aa}. Moreover, the DC conductivity was computed in \cite{Alberte:2015isw} and it is given by:
\begin{equation}
    \sigma_{DC}\,=\,1\,+\,\frac{\mu^2}{2\,m^2\,V_Z}
\end{equation}
where $\mu$ is the chemical potential, and it is indeed finite for $m \neq 0$. From the point of view of the electric transport properties, this model is completely analogous to the linear axion model \cite{Davison:2014lua,Kim:2014bza}.\\
For slow momentum dissipation $m/T \ll 1$, the least damped QNM is a pseudo-diffusive mode defined by the following dispersion relation:
\begin{equation}
    \omega\,=\,-\,i\,\Gamma\,-\,i\,D\,k^2\,+\,\mathcal{O}(k^4)\,, \quad \quad  \Gamma\,=\,\frac{m^2\,V_Z}{\pi\,T}\,+\,\mathcal{O}(m^4)\,,\quad \quad D\,=\,\frac{\eta}{\chi_{PP}}+\dots \label{d1}
\end{equation}
where $\Gamma$ is the momentum relaxation rate \cite{Davison:2013jba} and $D$ the diffusion constant.
We show the imaginary part of the lowest QNM in expression \ref{d1}, at zero momentum $k=0$, in fig.\ref{figuno}. The results indicate that for slow momentum dissipation, $m/T \ll 1$, formula \eqref{d1} is in agreement with the numerical data. Increasing further the strength of the explicit breaking, a coherent-incoherent transition appears as the first QNM collides with a second one producing a pair of off-axes modes (see fig.\ref{figuno}). This phenomenon is totally analogous to what is observed in the linear axions model in \cite{Davison:2014lua,Kim:2014bza}.
\begin{figure}[h]
    \centering
   \includegraphics[width=5cm]{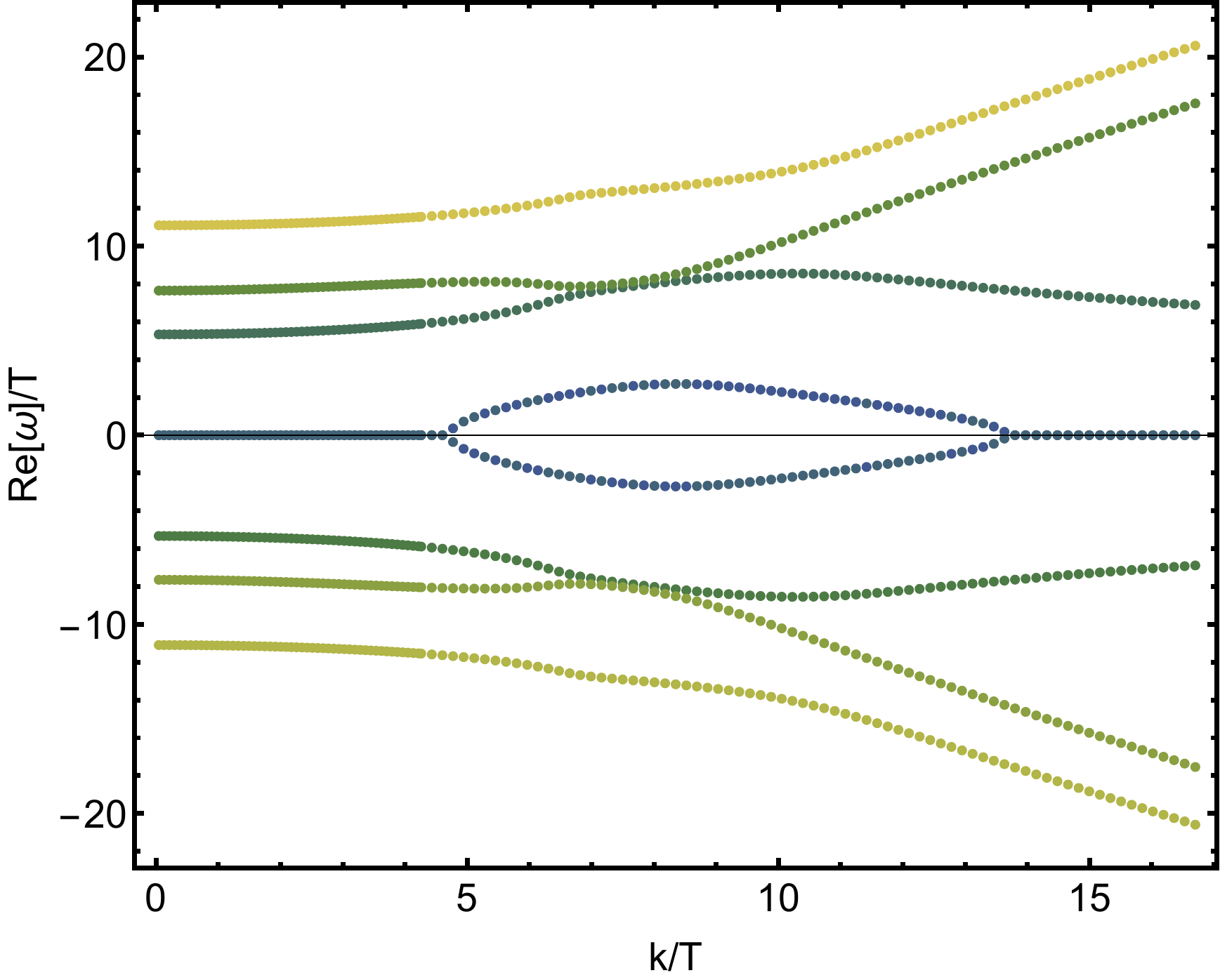}\quad  \includegraphics[width=5cm]{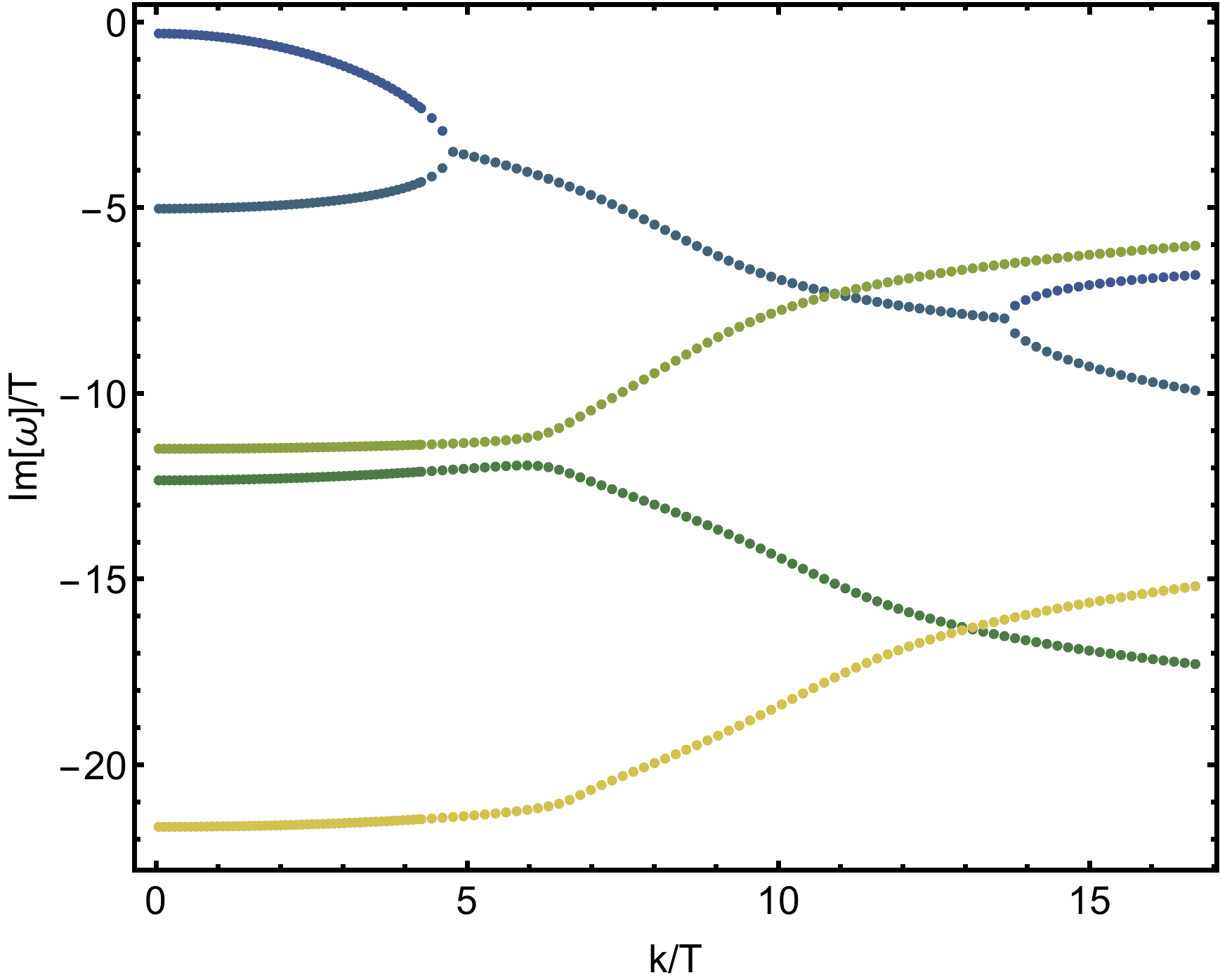}
        \caption{A snapshot of the first four modes in the transverse spectrum of the fluid model \eqref{aa}. We fixed $m/T=0.9525$.}
        \label{figdue}
\end{figure}\\
In fig.\ref{figuno} we exhibit the dispersion relation of the lowest QNMs \eqref{d1}. As expected, the hydrodynamic prediction \eqref{d1} provides a good approximation of the diffusion constant. Clearly, at $k/T \gg 1$, higher order corrections enter and the dispersion relation is no longer purely diffusive\footnote{The same corrections are present in pure relativistic hydrodynamics and can be found for example in \cite{Grozdanov:2015kqa}.}. This is the reason of the discrepancies in that regime.\\
Finally, we are interested in analyzing the behaviour of the QNMs beyond the hydrodynamic limit $\omega/T,k/T \ll 1$. This is motivated by the interesting features observed in the solid version of the model \cite{Baggioli:2018vfc,Baggioli:2018nnp}. In particular it is plausible that this model also exhibits the so-called $k-$gap phenomenon \cite{Baggioli:2019jcm}, namely the appearance of a propagating shear wave beyond a certain momentum $k>k_g$.
\begin{figure}[h]
    \centering
   \includegraphics[width=5cm]{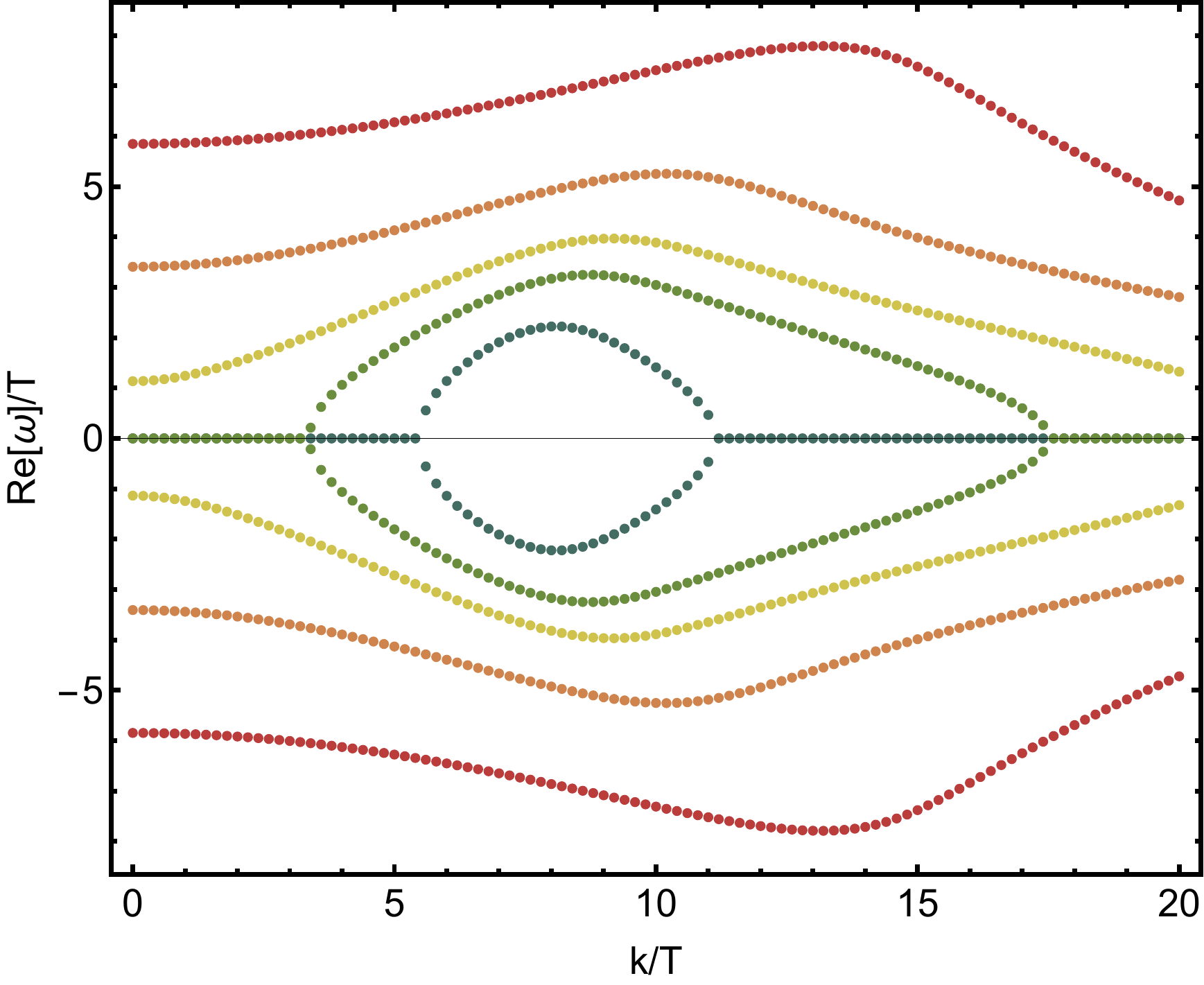}\quad  \includegraphics[width=5.1cm]{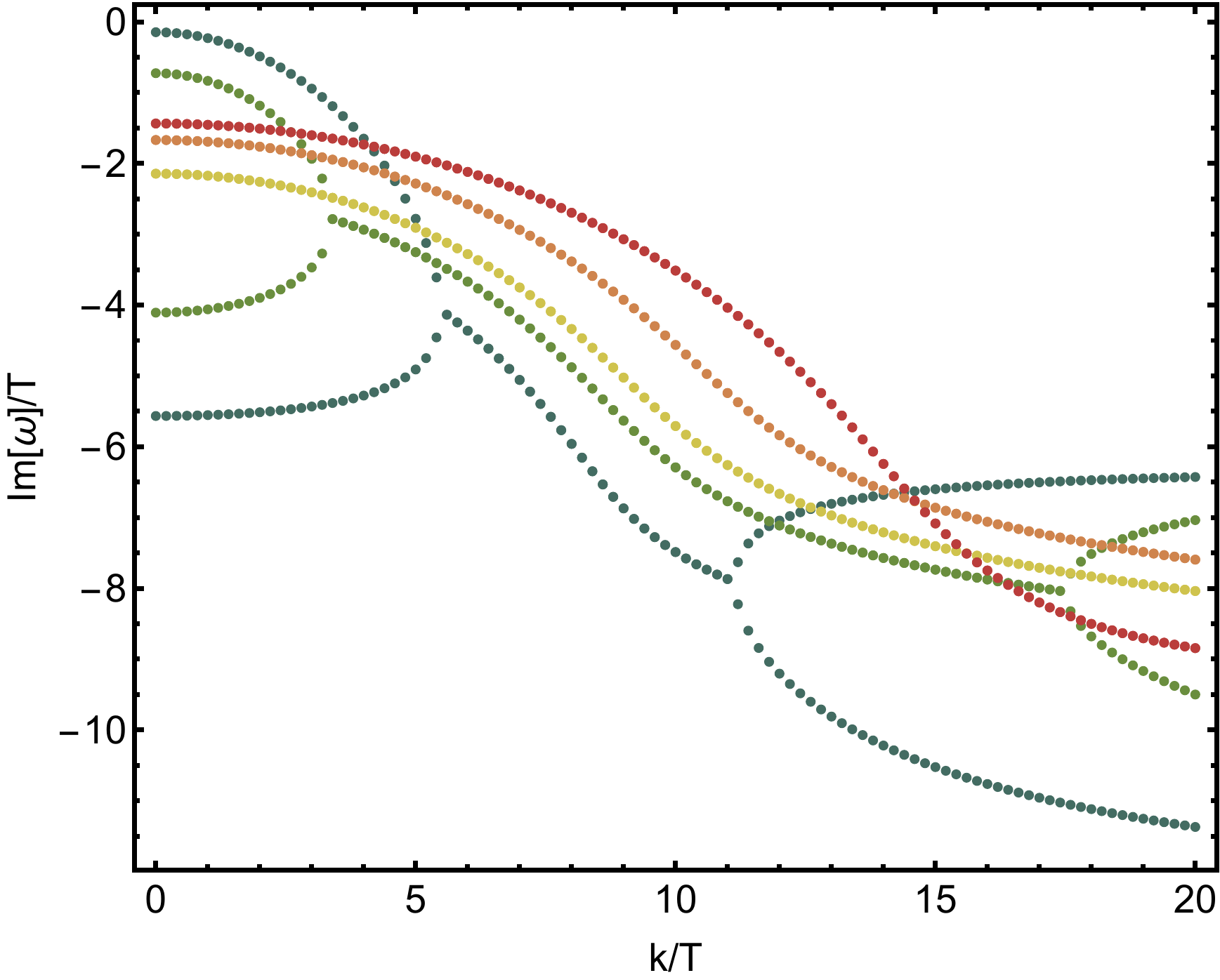}
        \caption{The two lowest modes in the transverse spectrum of the fluid model \eqref{aa} for  $m/T \in [0.67,6.28]$ (from blue to red).}
        \label{figdueb}
\end{figure}\\
First, we give a snapshot of the first four modes in the spectrum at a specific value of $m/T$ in fig.\ref{figdue}. The dynamics beyond the hydrodynamic limit, is quite complicated and qualitative similar to what is found in \cite{Alberte:2017cch}. In particular, the first two modes display a curious interplay which seems more complex than a single $k$-gap phenomenon.\\ In order to study this better, we isolate the first two modes and we follow their dispersion relation until large momenta $k/T \gg 1$, varying the value of the dimensionless parameter $m/T$.
We show our results in fig.\ref{figdueb}. We observe that, for small values of the dimensionless parameter $m/T$, the two modes collide at an initial value of momentum $k_1$, creating a finite real part in a $k-$gap fashion \cite{Baggioli:2019jcm}. Nevertheless, at a second value of the momentum $k_2 > k_1$ the modes split up and the real part becomes zero. In summary, the real part of the dispersion relation is non-zero only in a finite interval of momenta $k_1<k<k_2$. This is similar to what is already observed in \cite{Alberte:2017cch}. Increasing the value of $m/T$, the $[k_1, k_2]$ interval gets larger and larger until the first momentum $k_1$ reaches the origin, $k_1=0$. At that point, the modes become gapped and they display, at low momenta, a dispersion relation of a massive particle type $\omega^2\,=\,M^2\,+\,v^2\,k^2$. This tendency is very similar to what happens in the linear axions model \cite{Baggioli:2018vfc}, but in that case no closing up at larger momentum $k_2$ happens. This difference can be qualitative motivated by the fact that the $V(X,Z)=Z$ model contains higher derivatives with respect to its solid counterpart $V(X,Z)=X$, and therefore more complex dynamics at high momenta.
\subsection*{The longitudinal sector}\label{cc}
We now consider the longitudinal part of the spectrum. We follow closely the analysis of \cite{Davison:2014lua}. First, we consider the regime in which momentum is slowly dissipated; this happens whenever $\Gamma/T \ll 1$ or in terms of our parameters when $m/T \ll 1$. Using hydrodynamic arguments, which can be found in \cite{Davison:2014lua}, in this regime, the generic dispersion relation of the lowest mode, reads:
\begin{equation}
        \omega\,=\,\pm\,k\,\sqrt{\frac{\partial p}{\partial \epsilon}\,-\,\frac{1}{4}\,\left(\Gamma\,k^{-1}\,+\,\frac{\eta}{\chi_{PP}}\,k\right)^2}\,-\,\frac{i}{2}\,\left(\Gamma\,+\,\frac{\eta}{\chi_{PP}}\,k^2\right)\,+\,\dots\label{disp2}
    \end{equation}
which comes from solving the equation:
\begin{equation}
    i\,\omega\,\left(-\,i\,\omega\,\Gamma\,+\,\frac{\eta}{\epsilon\,+\,p}\,k^2\right)\,=\,k^2\,\frac{\partial p}{\partial \epsilon}
\end{equation}
and expanding the result at small momentum. In absence of momentum dissipation, $\Gamma=0$, we recover the sound mode of relativistic hydrodynamics:
\begin{equation}
    \omega\,=\,v_s\,k\,-\,\frac{1}{2}\,i\,\frac{\eta}{\epsilon\,+\,p}\,k^2\,+\,\dots
\end{equation}
where $v_s^2=1/2$ and $\eta/(\epsilon+p)=1/(8\pi T)$ \cite{Policastro:2002tn}.\\

\begin{figure}[h]
    \centering
   \includegraphics[width=5cm]{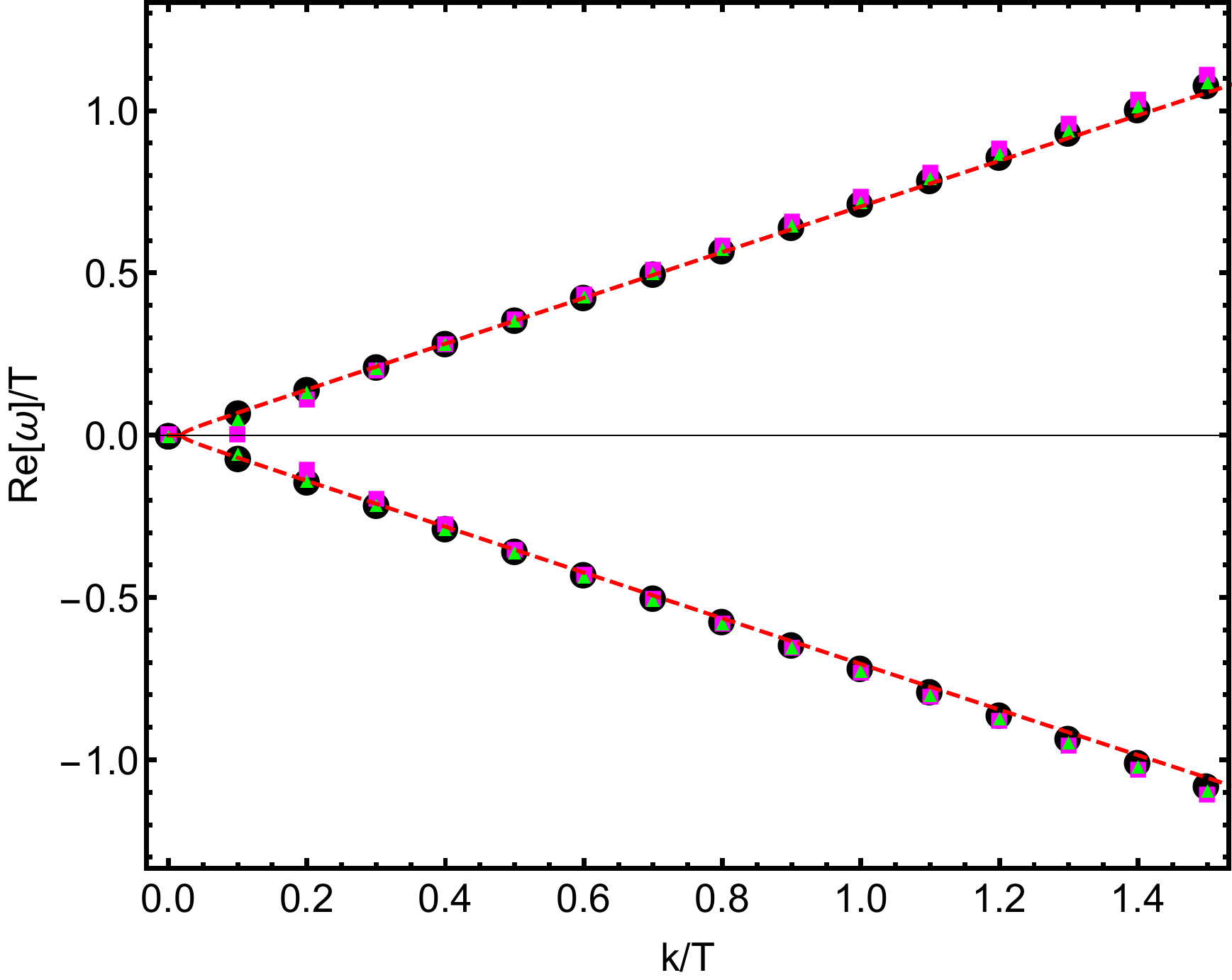}\quad  \includegraphics[width=5.1cm]{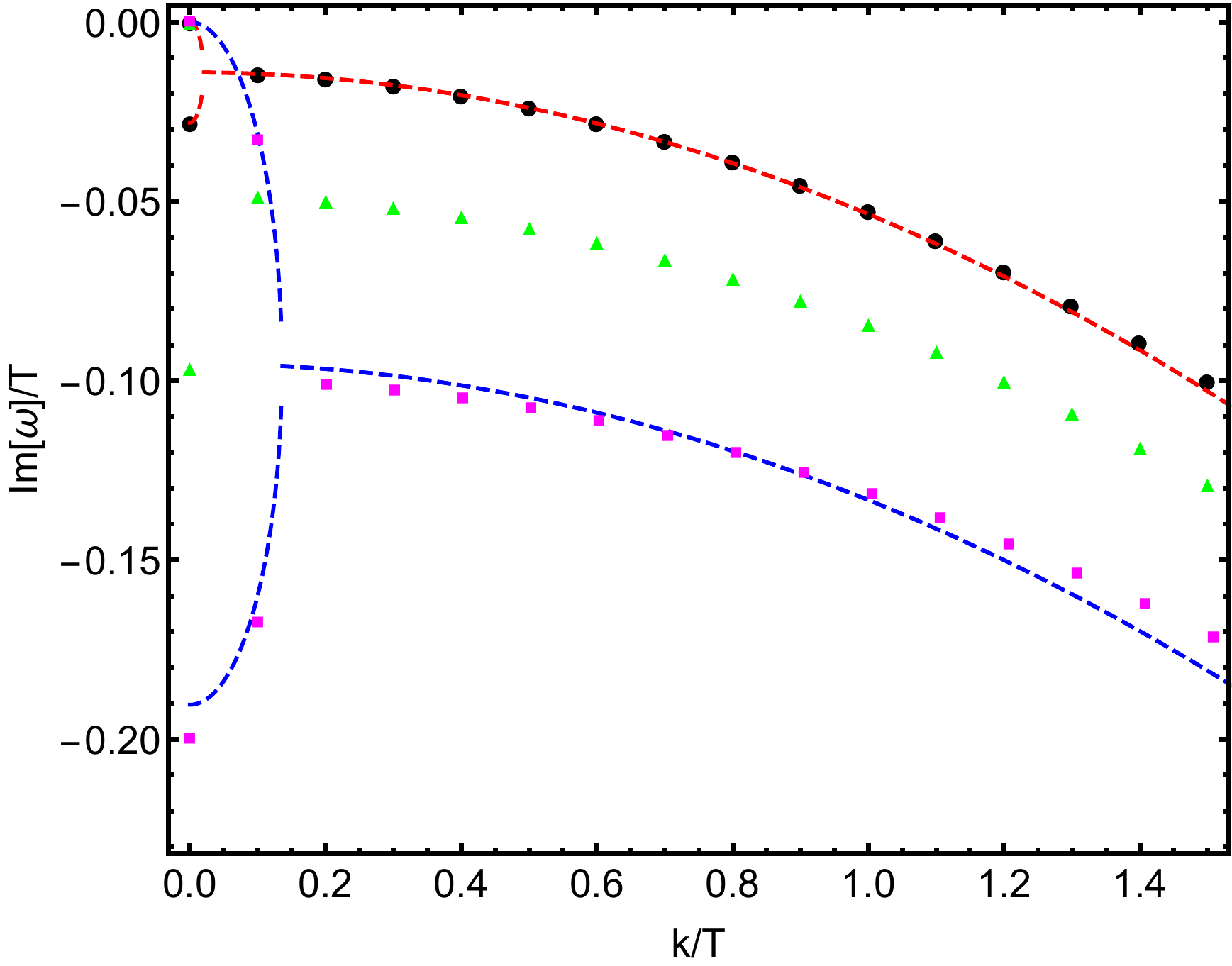}
        \caption{The dispersion relation of the lowest modes for $m^2=\{0.297, 0.773, 0.544\}$ (from black to magenta). The dashed lines are the hydrodynamic formula \eqref{disp2} which works well for $m/T \ll 1$.}
        \label{figsmall}
\end{figure}
At small momenta, $k/\Gamma \ll 1$, sound is destroyed by momentum dissipation and the relevant hydrodynamic poles are\footnote{The ellipsis in all the following expressions represent higher order corrections in $\omega/T,k/T$ which go beyond the low energy hydrodynamic approximation.}:
\begin{equation}
    \omega\,=\,-\,i\,\frac{\partial p}{\partial \epsilon}\,\Gamma^{-1}\,k^2\,+\,\dots\,,\quad \quad \omega\,=\,-\,i\,\Gamma\,+\,i\,k^2\,\left(\frac{\partial p}{\partial \epsilon}\,\Gamma^{-1}\,-\,\frac{\eta}{\epsilon\,+\,p}\right)\,+\,\dots
\end{equation}
Increasing the momentum, these two modes collide with each other and they form a propagating sound mode:
\begin{equation}
    \omega\,=\,\frac{\partial p}{\partial \epsilon}\,k\,-\,i\,\left(\Gamma\,+\,\frac{\eta}{\epsilon\,+\,p}\right)\,+\,\dots
\end{equation}
\begin{figure}[h]
    \centering
  \includegraphics[width=5cm]{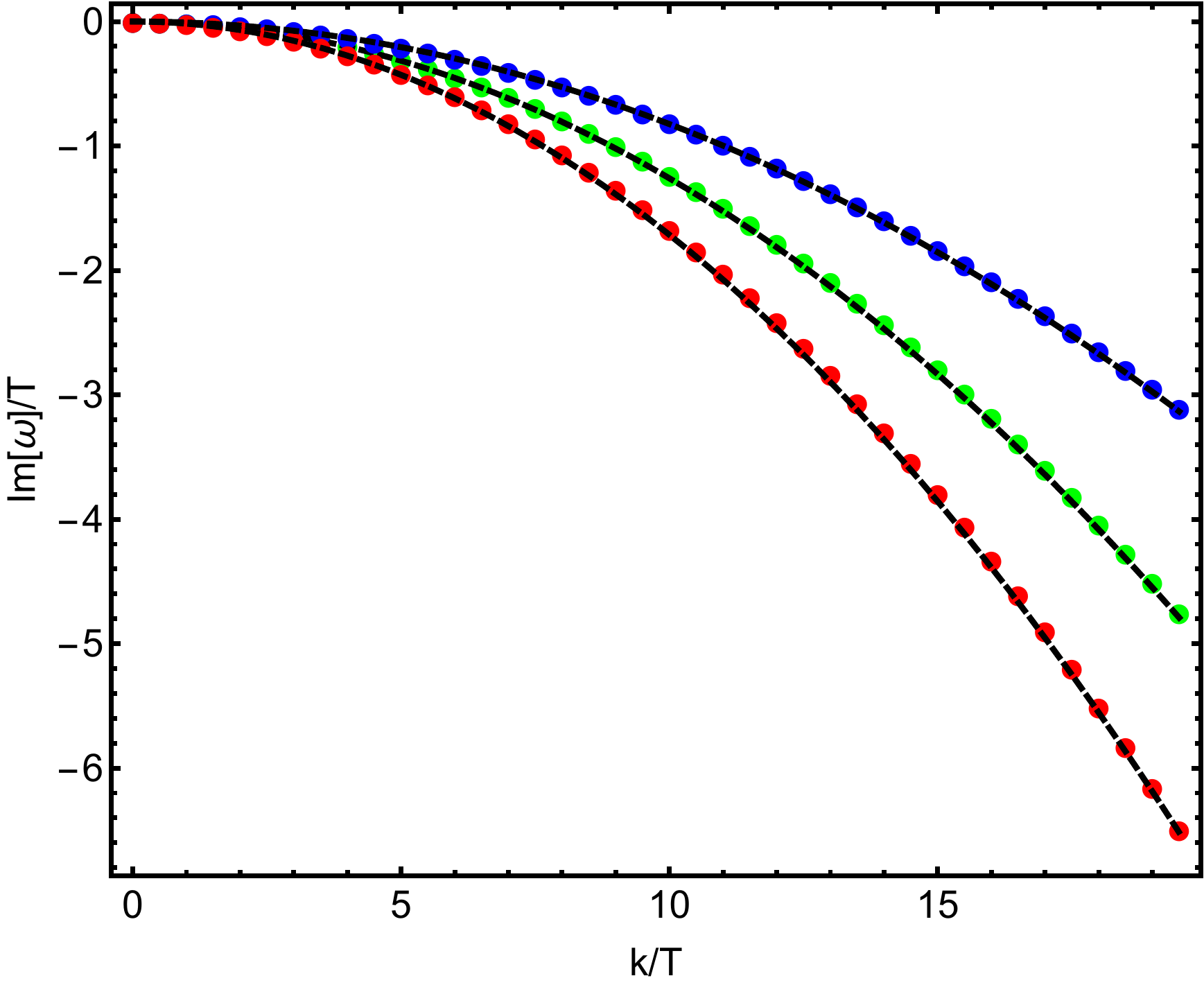}\quad  \includegraphics[width=5.1cm]{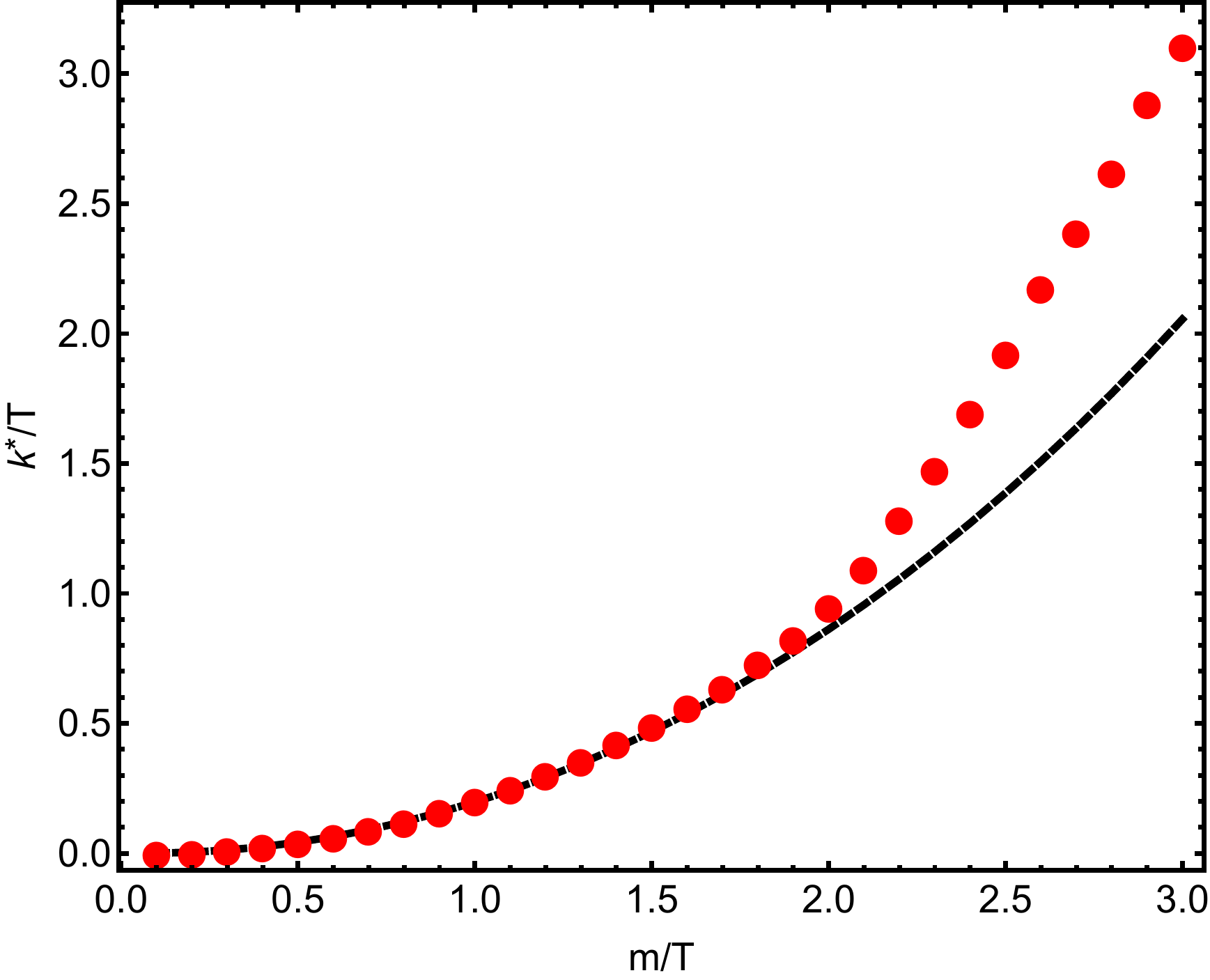}
        \caption{\textbf{Left: }The lowest mode in the longitudinal sector for very large graviton mass $m/T \gg 1$ ($m/T\in\{50.66,105.13\}$ red to blue), \textit{i.e.} in the incoherent regime.  The dashed black line is eq. \eqref{eqlargemt}.  \textbf{Right: }The momentum of the poles collision $k^\star$ extracted from the numerical data (red bullets). The dashed line is the hydrodynamic approximation of eq.\eqref{kstareq} which show good agreement for $m/T \ll 1$. At very large $m/T$, the  data increase linearly in the dimensionless EXB strength $m/T$.}
        \label{boh1}
\end{figure}
This diffusion to sound crossover, appearing from the usual pole collision, happens at a specific momentum $k^*$. We can see this behaviour explicitly  in fig.\ref{figsmall} for different values of the graviton mass, in the limit of weak momentum dissipation. At small $m/T$, using hydrodynamics, we can define the momentum of the poles collision via:
\begin{equation}
    \frac{\Gamma}{k^*}\,+\,\frac{\eta}{\epsilon\,+\,p}\,k^*\,=\,2\,\sqrt{\frac{\partial p}{\partial \epsilon}}\label{kstareq}
\end{equation}
which comes directly from eq.\ref{disp2}. The agreement between the numerical data and this formula is shown in the right panel of fig.\ref{boh1}. At very large explicit breaking our data suggests that the momentum of the collision grows linearly with the EXB strength $m/T$.
On the other side, when momentum is strongly dissipate we have $\Gamma/T \gg 1$, which relates to the regime $m/T \gg 1$, the physics is totally diffusive and there is a single hydrodynamic pole:
\begin{equation}
    \omega\,=\,-\,i\,D_{\parallel}\,k^2\,+\,\dots\,,\quad \quad D_{\parallel}\,=\,\frac{\kappa}{c_v}\,,\label{eqlargemt}
\end{equation}
where the diffusion constant is given by the thermal conductivity $\kappa$ and the specific heat $c_v=T ds/dT$. In other words, the left diffusive mode simply corresponds to the conservation of energy. The appearance of this mode is shown in fig.\ref{boh1}. Its diffusion constant is in good agreement with formula \eqref{eqlargemt}, where we have used:
\begin{equation}
    c_v\,=\,\frac{16\,\pi^2\,T}{3\,+\,m^2}\,,\quad   \kappa\,=\,\frac{(2\pi)^2\,T}{m^2}
\end{equation}
and fixed $u_h=1$, as usual.
\begin{figure}[h]
    \centering
   \includegraphics[width=5.2cm]{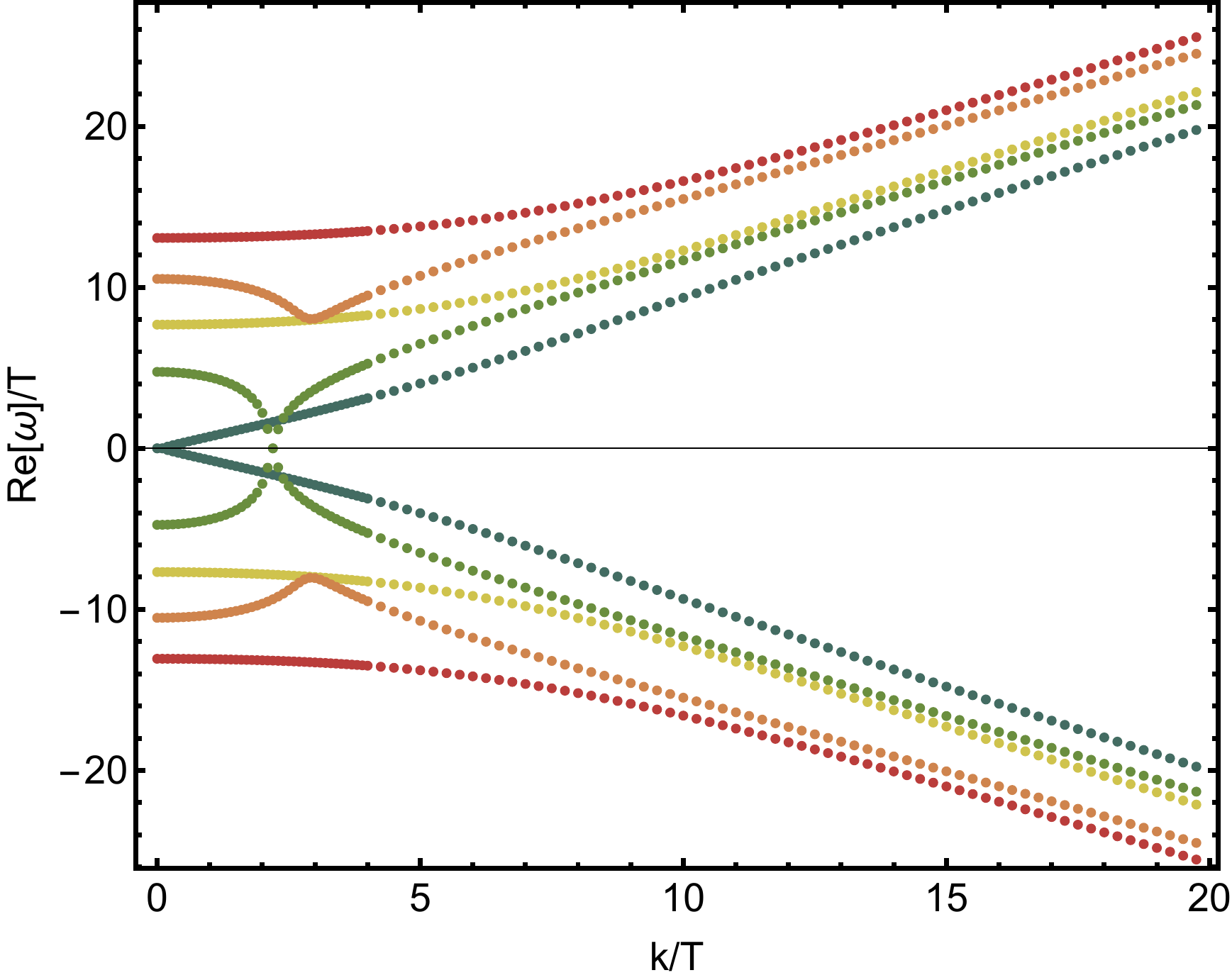}\quad  \includegraphics[width=5.2cm]{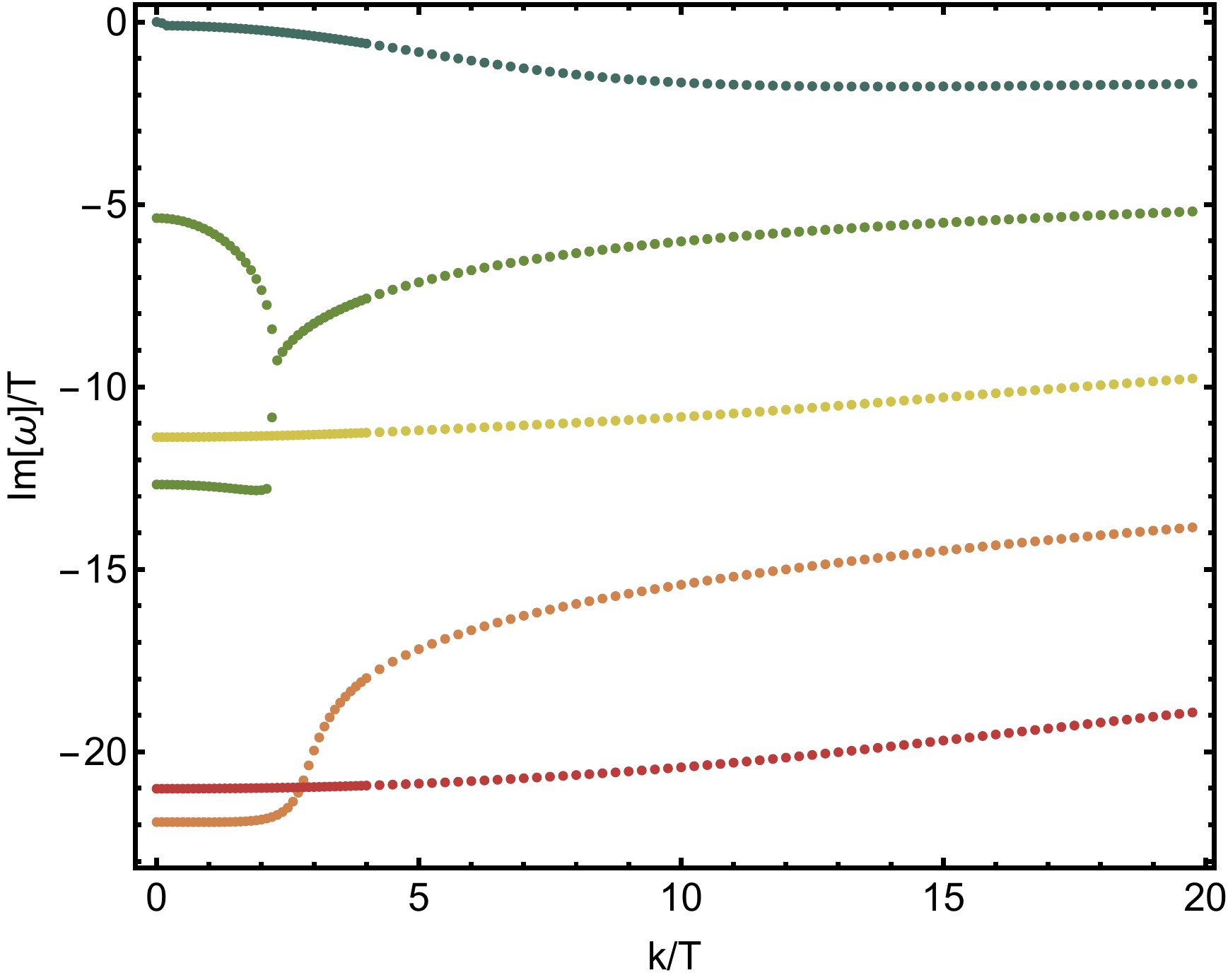}
        \caption{The longitudinal spectrum of the fluid model $V(Z)=Z$  for $m/T=0.773$.}
        \label{boh2}
\end{figure}\\
Given the agreement between the lowest modes and the hydrodynamic predictions, we extend our analysis to the higher modes in the spectrum. In fig.\ref{boh2} we show the large frequency and momentum dynamics for a specific value of the EXB. Several gapped and strongly damped modes appear in the spectrum producing a very non trivial interplay of collisions and crossovers. Let us notice that at large $k/T \gg 1$ all the modes obey $Re(\omega)=k$ which is simply dictated by the relativistic symmetry of the UV fixed point.
\section{Phonons in an holographic fluid with SSB}\label{fluidSSB}
In this section we consider a second fluid model defined by the potential:
\begin{equation}
    V(X,Z)\,=\,Z^N\,,\quad\quad N>5/4 \label{bb}
\end{equation}
and for concreteness we just set $N=2$. We have checked explicitly that the results are qualitative the same for higher $N$. This choice is still invariant under VPdiffs \eqref{sym} but it displays a different symmetry breaking pattern of the translational invariance. In particular, in this case, translations are broken spontaneously. The main difference between this fluid model and its solid counterpart described in \cite{Alberte:2017oqx} is again the vanishing of the helicity-2 mass. As a consequence, the elastic shear modulus $G$ is zero, as in realistic fluids, and the KSS bound is saturated.\\
Understanding what is the field theory description dual to the bulk action \eqref{bb} and to which extent it differs from the relativistic hydrodynamics, encoded in the simple Schwarzschild background, are the aims of this section. Both models, simple Einstein gravity and \eqref{bb}, represent fluid configurations at finite temperature. In the Schwarzschild case the solution is purely conformal, and therefore the value of the temperature itself is not meaningful since it can not be compared with any other scale. The potential \eqref{bb} is different from this point of view. More than that, is there any difference in the hydrodynamic modes? How does the $m/T$ parameter in \eqref{bb} modify the transport coefficients and the dynamics?
\subsection*{The transverse sector}\label{tran2}
In a system with spontaneously broken translations, we generically expect the appearance of propagating transverse modes satisfying:
\begin{equation}
    \omega\,=\,v_T\,k\,-\,i\,D\,k^2\,+\,\dots
\end{equation}
which take the name of transverse phonons.
\begin{figure}[h]
    \centering
   \includegraphics[width=5.2cm]{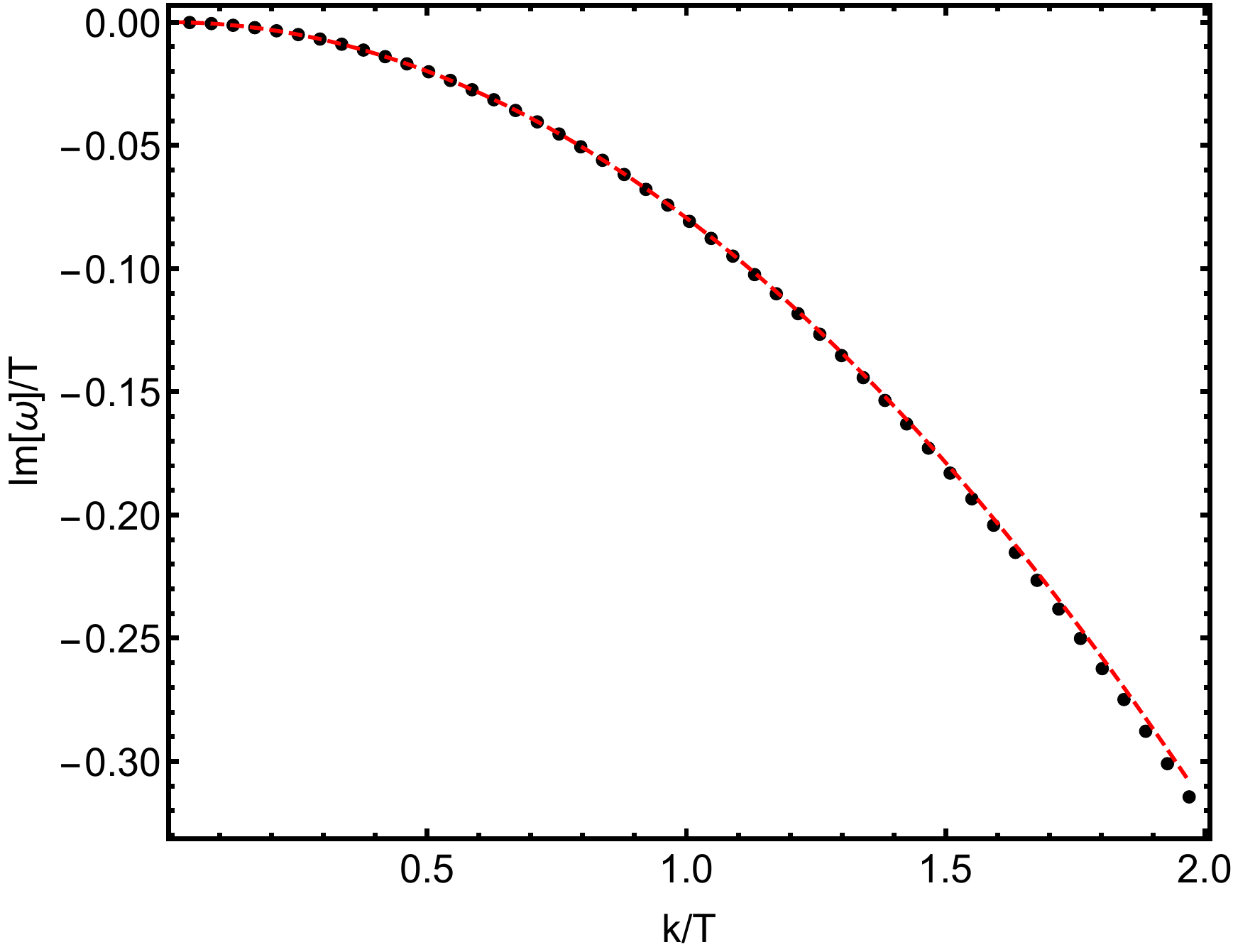}\quad  \includegraphics[width=5.1cm]{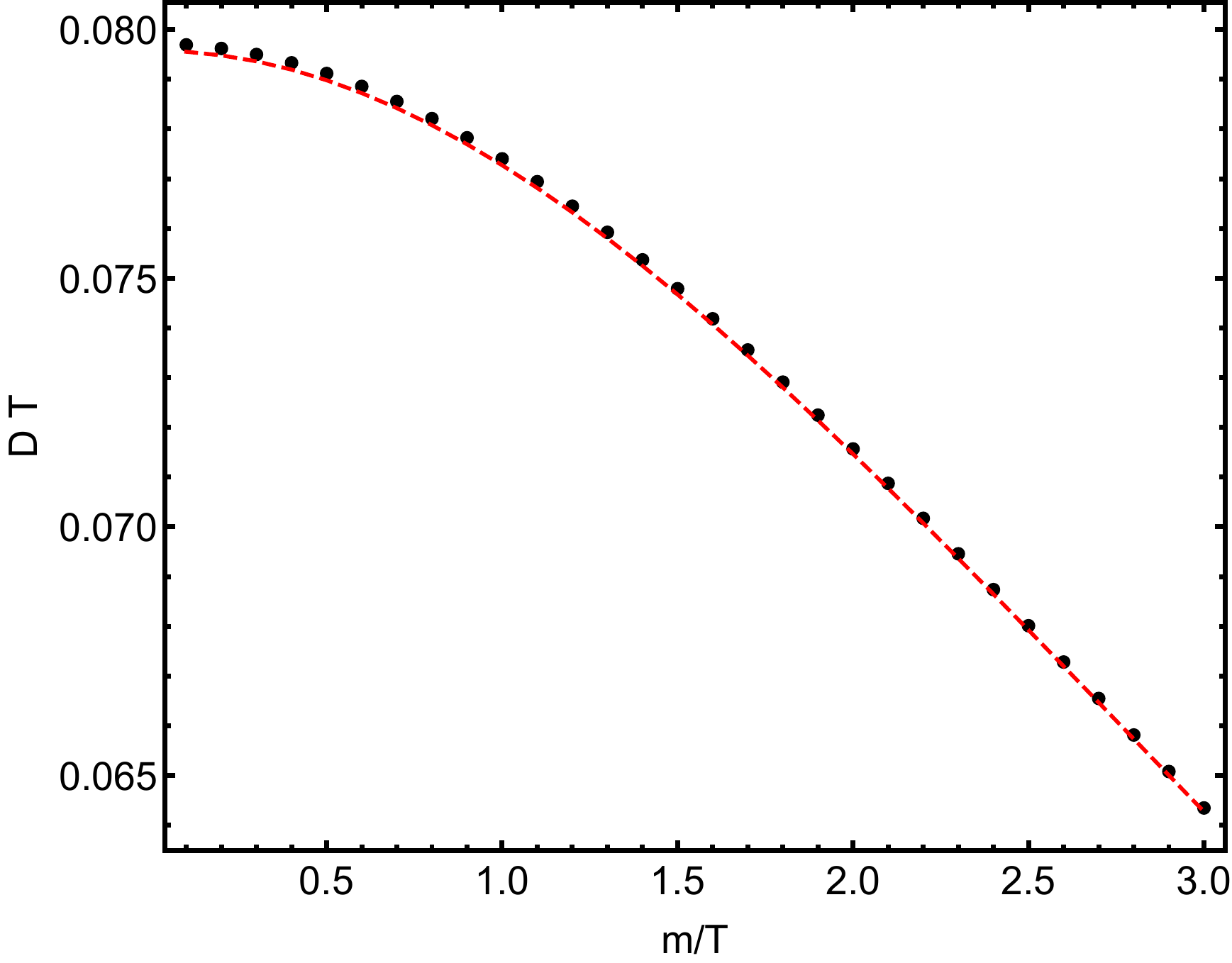}
        \caption{\textbf{Left: }The shear diffusion mode in the transverse spectrum of the fluid model \eqref{bb}. The red dashed line is the hydrodynamic formula \eqref{anz2}. We fix the dimensionless SSB parameter to $m/T=0.148$. \textbf{Right: } The shear diffusion constant in function of the dimensionless parameter $m/T$. The dashed line indicates the approximated formula \eqref{spsp}.}\label{figtre}
\end{figure}\\
 Importantly, their propagation speed is fixed by the elastic properties of the material to be:
\begin{equation}
    v_T^2\,=\,\frac{G}{\chi_{PP}}
\end{equation}
where $G$ is the shear elastic modulus and $\chi_{PP}$ the momentum susceptibility. As a consequence of the fluid symmetry \eqref{sym} of this model, the shear modulus vanishes and therefore there are no propagating transverse phonons. This is a very well-known fact for fluids and it is usually taken as their main difference with respect to solids. All in all, we expect instead of the propagating transverse sound, a simple hydrodynamic diffusive mode of the type:
\begin{equation}
    \omega\,=\,-\,i\,D\,k^2\,+\,\dots\,\,,\quad D\,=\,\frac{\eta}{\chi_{PP}}\,+\,\dots\label{anz2}
\end{equation}
where the diffusion constant is determined by the viscosity of the fluid.\\
More specifically, fixing $u_h=1$, we can obtain the expression:
\begin{equation}
    D\,T\,=\,\frac{5\,\left(3\,-\,m^2\right)}{12\,\pi\,\left(m^2\,+\,5\right)}\label{spsp}
\end{equation}
The presence of this mode is confirmed numerically in fig.\ref{figtre}. The diffusive mode appears independently of the value of the SSB strength, \textit{i.e.} $m/T$, and the diffusion constant is indeed given by the hydrodynamic formula \eqref{anz2}.\\
Let us spend some words about this mode. It is well known that transverse phonons have zero speed of propagation in liquids. Nevertheless, it is still interesting to think about them as Goldstone bosons for translational symmetry. This suggests that, in dissipative liquids\footnote{In absence of dissipation and viscosity (e.g. a perfect fluid) the dispersion relation of the transverse phonons will be just $\omega=0$.}, the transverse Goldstone bosons are diffusive and not propagating. Apparently, in dissipative systems, the presence of Goldstone modes which are diffusive is expected from field theory arguments \cite{Minami:2018oxl}. It would be valuable to understand them better in terms of the type A Goldstone bosons \cite{Amado:2013xya,Nitta:2014jta,Watanabe:2014zza,Watanabe:2012hr,Watanabe:2011dk}\footnote{We thank Amadeo Jimenez for discussions regarding this point.}. We will see later on how these diffusive Goldstone bosons are affected by the additional EXB of translations.\\

\begin{figure}[h]
    \centering
   \includegraphics[width=5cm]{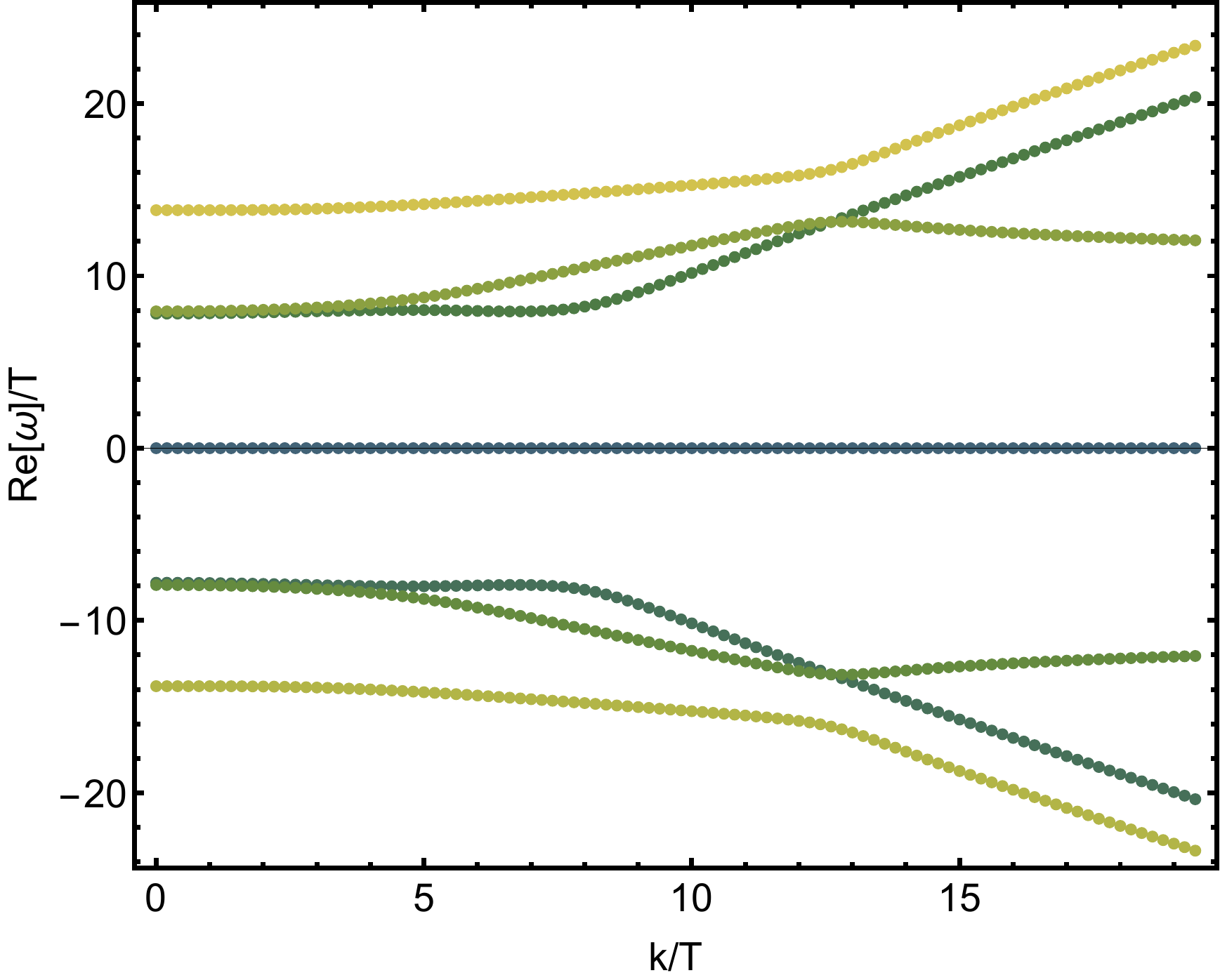}\quad  \includegraphics[width=5.2cm]{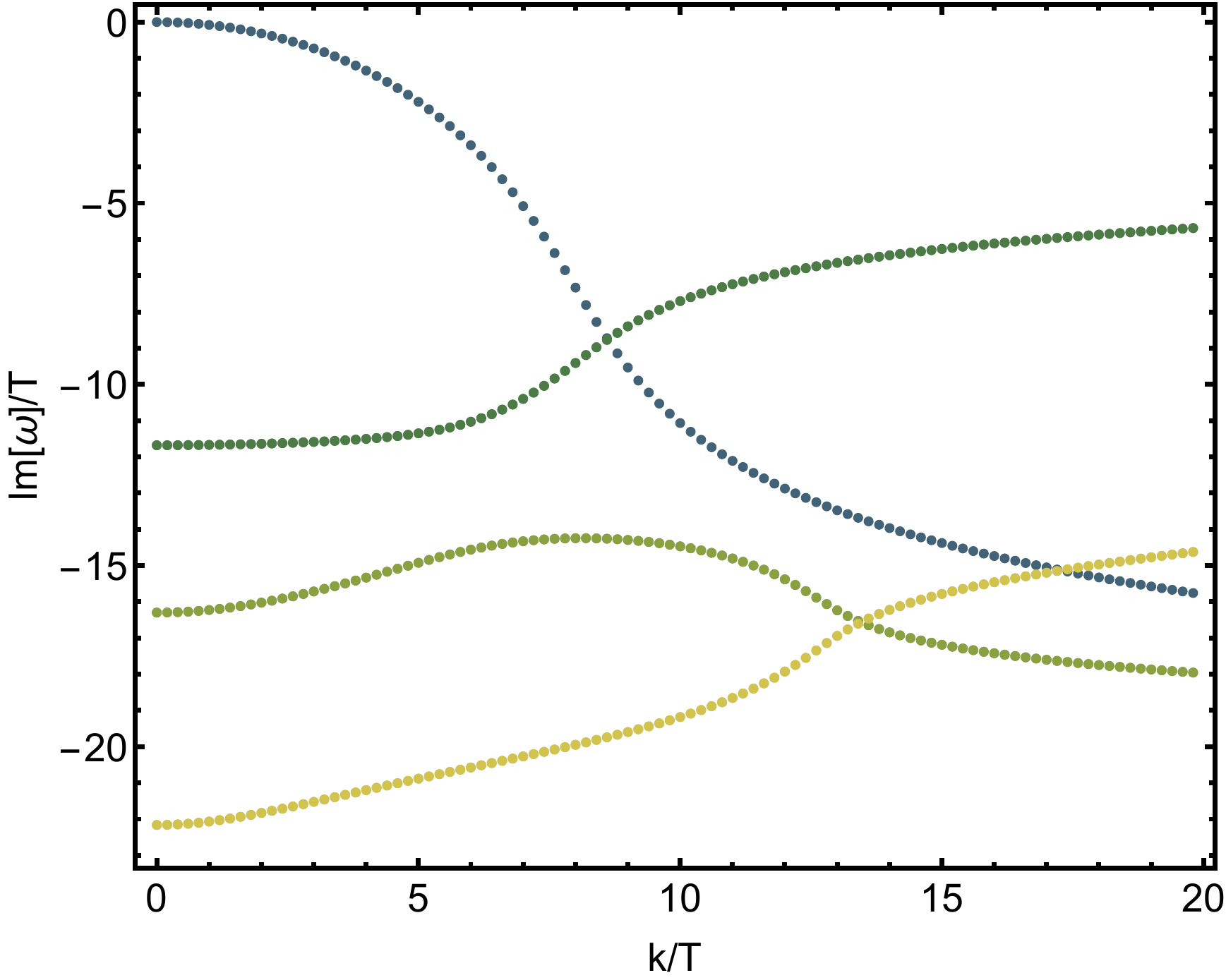}
        \caption{The higher modes of the transverse spectrum of the model \eqref{bb} for $m/T=0.953$.}\label{figsnap}
\end{figure}

After confirming the nature of the hydrodynamic modes, we proceed by considering higher and more damped excitations in the system. We show a snapshot of the results in fig.\ref{figsnap}. We observe a set of gapped and highly damped modes. Moreover, as in the simplest relativistic hydrodynamics case, we observe the crossing (but not the collision) of the hydrodynamic diffusive mode and a second non-hydro mode. This crossover is usually taken as the definition for the breakdown of the hydrodynamic approximation \cite{Grozdanov:2019kge}. 

\subsection*{The longitudinal sector}
Fluids and solids share the same qualitative features in the longitudinal spectrum. In both cases, we have a propagating longitudinal sound mode with dispersion relation:
\begin{equation}
    \omega\,=\,v_L\,k\,-\,i\,D_s\,k^2\,+\,\dots\label{long1}
\end{equation}
where $D_s$ is the related sound attenuation constant. The speed of longitudinal sound can be obtained from the elastic moduli as:
\begin{equation}
    v_L^2\,=\,\frac{G\,+\,K}{\chi_{PP}}
\end{equation}
where $G$ and $K$ are the elastic shear and bulk moduli. In this case, due to the fluid symmetry \eqref{aa}, the shear modulus is zero. The bulk modulus can be derived using thermodynamics as:
\begin{equation}
    K\,=\,-\,V\,\frac{d p}{dV}\,=\,\frac{3}{4}\,\epsilon
\end{equation}
where $p$ and $V$ are the pressure and the volume of our system.
\begin{figure}[h]
    \centering
   \includegraphics[width=5.2cm]{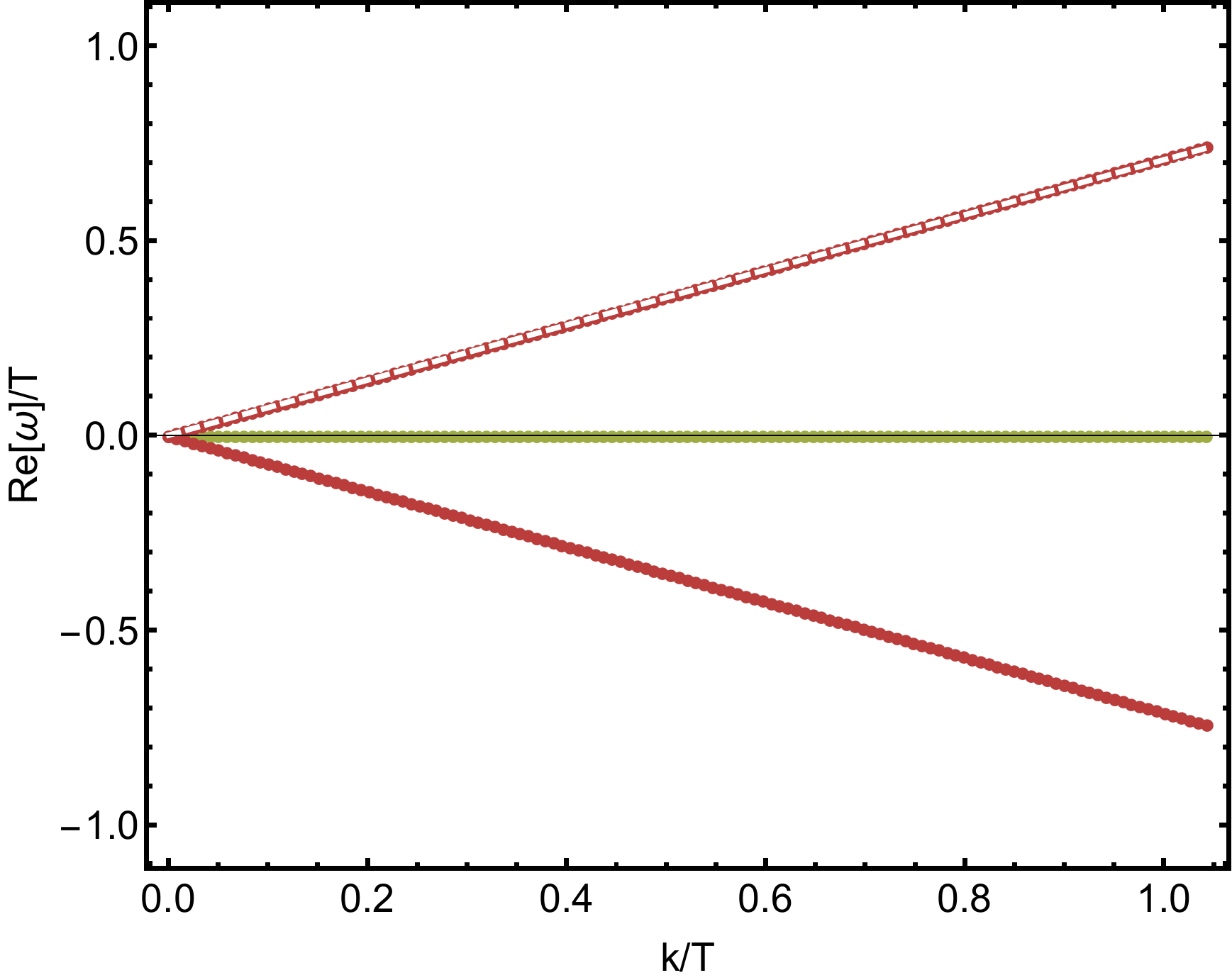}\quad  \includegraphics[width=5.2cm]{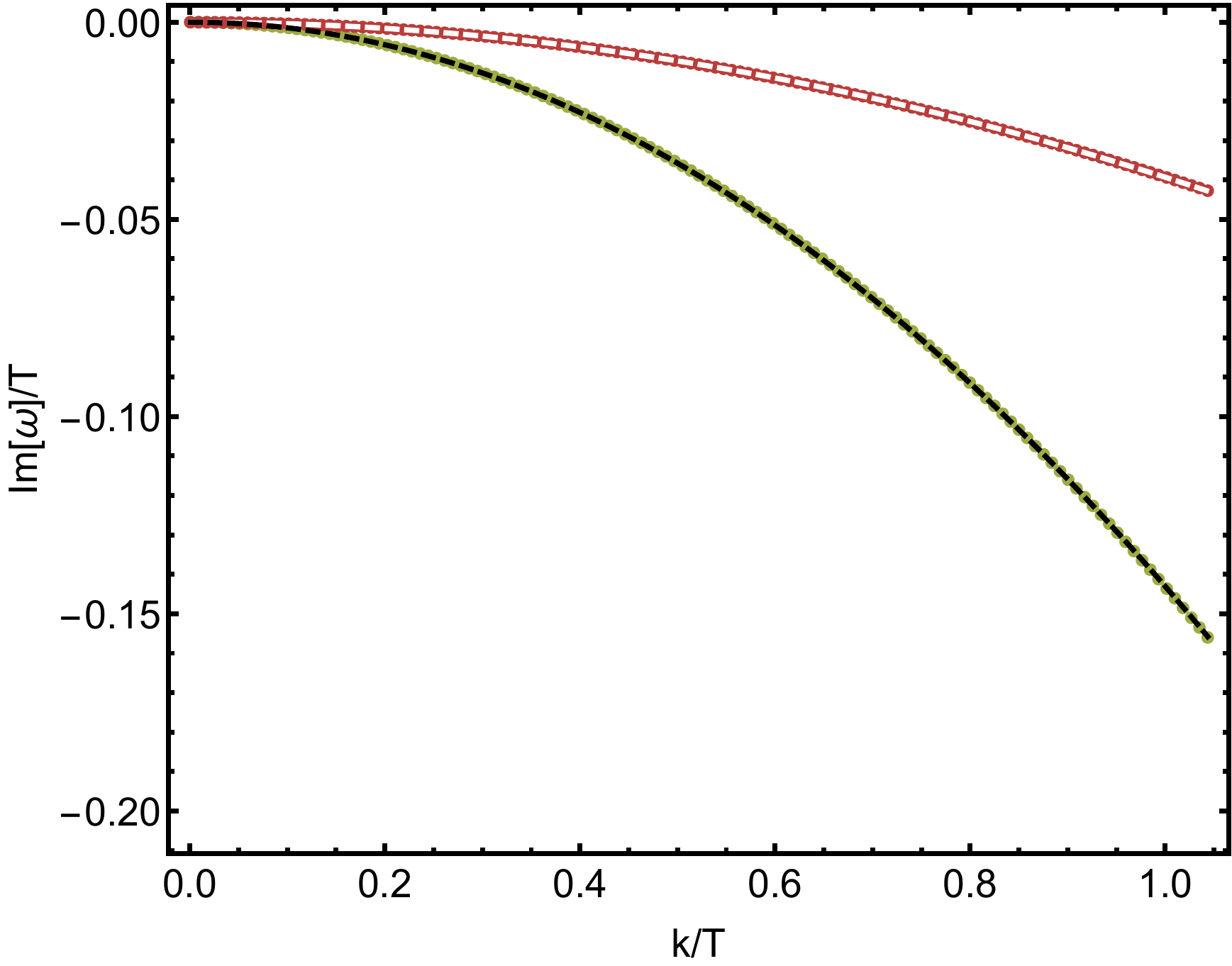}
        \caption{The sound (red) and crystal diffusion (green) modes present in the longitudinal spectrum of the fluid model \eqref{bb}for $m/T=0.470$. \textbf{Left:} The real part of the modes. The white dashed line is the hydrodynamic formula \eqref{long1} with speed \eqref{speed1}. \textbf{Right:} The white dashed line is the hydrodynamic formula \eqref{long1} with diffusion constant \eqref{att1}. The black dashed line is the fit for the crystal diffusion mode.}
        \label{f1}
\end{figure}\\
Using the definition of the momentum susceptibility, $\chi_{PP}=3/2 \,\epsilon$, we obtain the final value for the longitudinal speed:
\begin{equation}
    v_s^2\,=\,\frac{1}{2}\label{speed1}
\end{equation}
which is surprisingly independent of the value of the SSB strength $m/T$\footnote{This is true only at zero chemical potential. At finite chemical potential, or in presence of other deformations of the CFT, $K\neq \frac{3}{4}\epsilon$ and this result is modified.}. We have confirmed this result numerically for several values of the SSB strength. A specific example is shown in fig.\ref{f1}. Both the speed and the attenuation constant of the sound mode are in agreement with our formulas \eqref{speed1},\eqref{att1}. It is interesting to notice that, despite the introduction of the SSB of translations, the speed of longitudinal sound is exactly that of an un-deformed CFT \cite{Policastro:2002tn}. Likewise, the sound attenuation constant, at leading order in the SSB breaking parameter, is given by:
\begin{equation} D_s\,=\,\frac{1}{2}\,\frac{\eta}{\chi_{PP}}\,+\,\dots\lab{att1}
\end{equation}
where the ellipsis indicates higher corrections in $m/T$.\\

\begin{figure}[h]
    \centering
   \includegraphics[width=5.1cm]{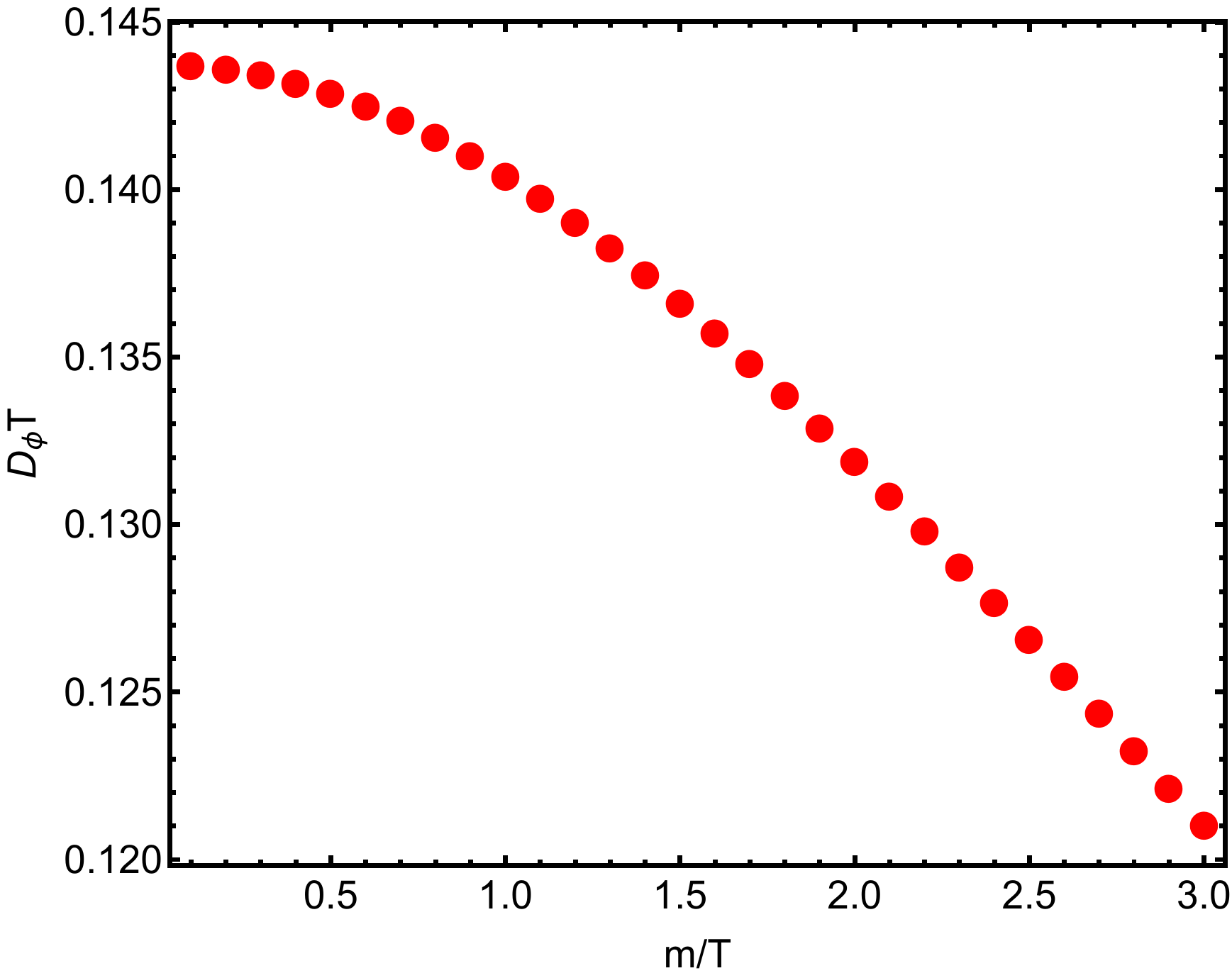}
   \quad  \includegraphics[width=5.1cm]{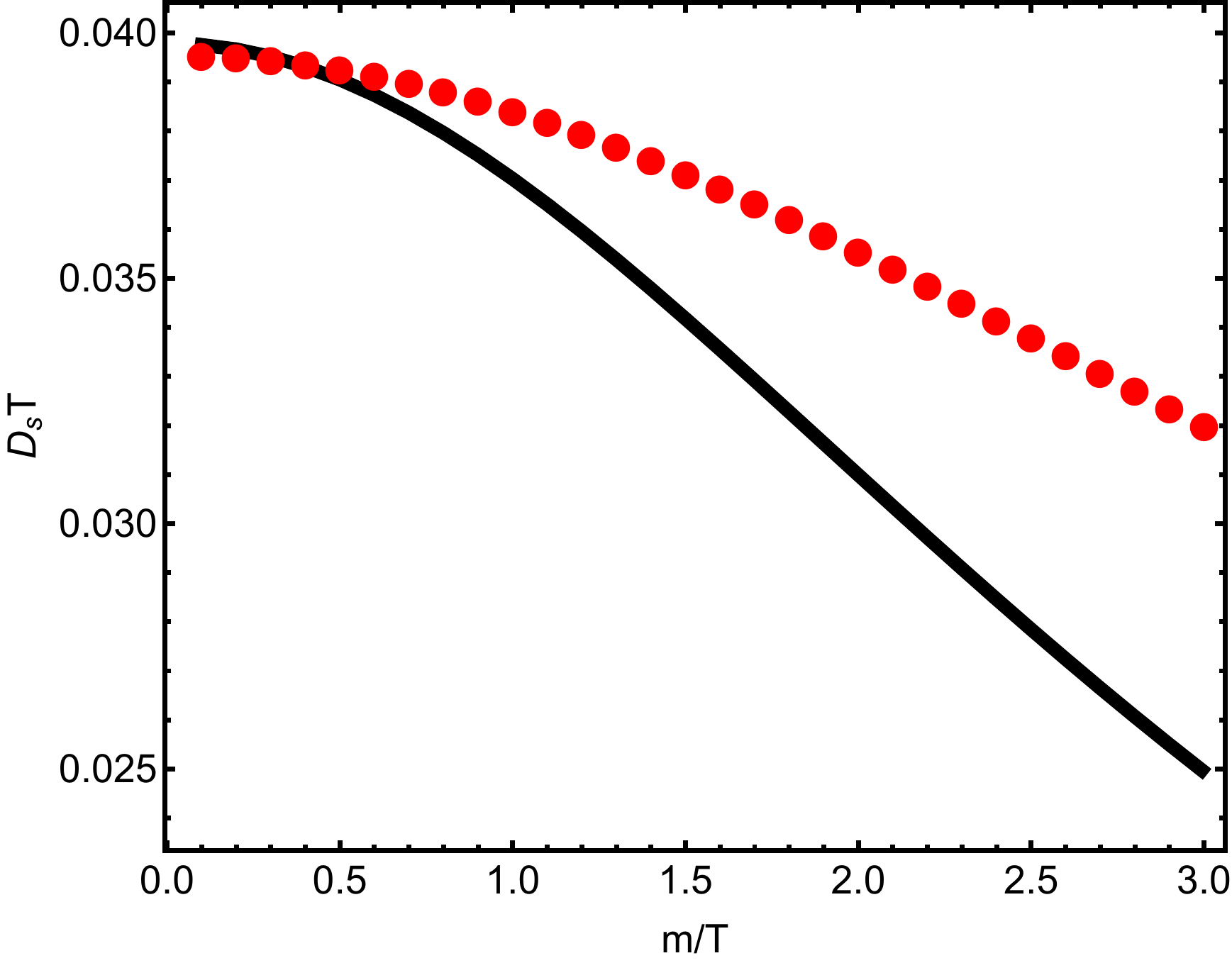}
        \caption{\textbf{Left: }The crystal diffusion constant of the mode \eqref{difdisp} in function of the dimensionless SSB parameter $m/T$. \textbf{Right: } The sound attenuation constant $D_s$ in function of $m/T$. The black line is the hydrodynamic formula \eqref{att1}. Within the precision of the numerical data, the agreement is good for $m/T \ll 1$.}
        \label{dfi}
\end{figure}
Moreover, because of the introduction of the new dynamical Goldstone degrees of freedom, we have an additional hydrodynamic mode in the longitudinal sector. In particular, there is an additional diffusive mode, sometimes referred to as ''crystal diffusion'', whose dispersion relation reads:
\begin{equation}
    \omega\,=\,-\,i\,D_\phi\,k^2\,+\,\dots\label{difdisp}
\end{equation}
where the ellipsis indicates higher order corrections in the dimensionless parameter $m/T$. This second hydrodynamic modes appears in the spectrum obtained numerically and it was already observed in \cite{Andrade:2017cnc,Baggioli:2019aqf}. In fig.\ref{f1} we show the results for a specific value of the SSB parameter $m/T$. By fitting the numerical dispersion relation, we are able to extract the behaviour of the crystal diffusion constant $D_\phi$ as function of the dimensionless parameter $m/T$. The results are shown in fig.\ref{dfi}. The crystal diffusion constant decreases by increasing the SSB parameter $m/T$ indicating that the ordered phase becomes more and more stable and ``rigid". In the same figure, we show the behaviour of the sound attenuation constant $D_s$ and compare it to the hydrodynamic approximation in \eqref{att1}. Within the precision of the numerics, the agreement, in the range $m/T \ll 1$, is good.
\begin{figure}[h]
    \centering
     \includegraphics[width=5cm]{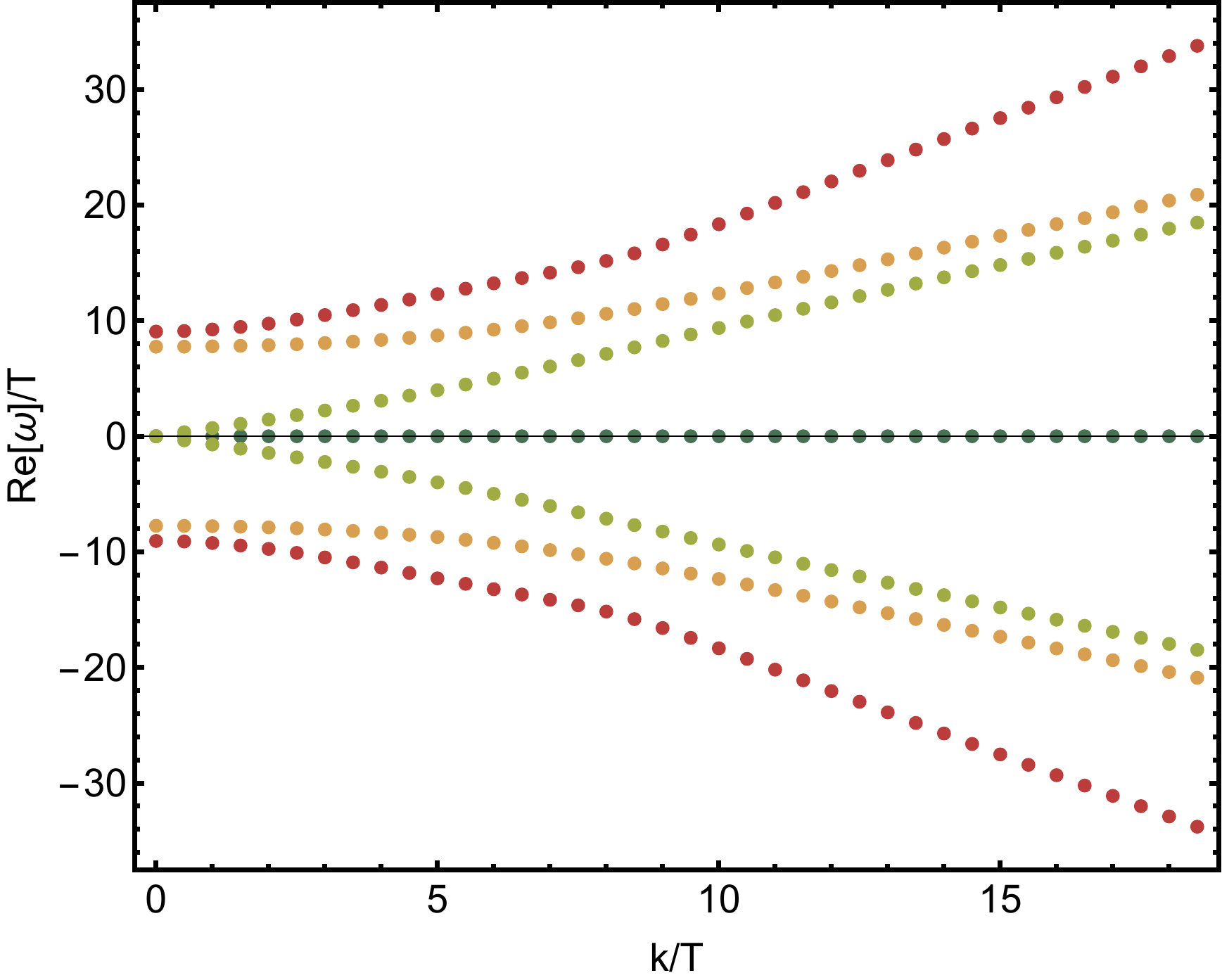}\quad  \includegraphics[width=5cm]{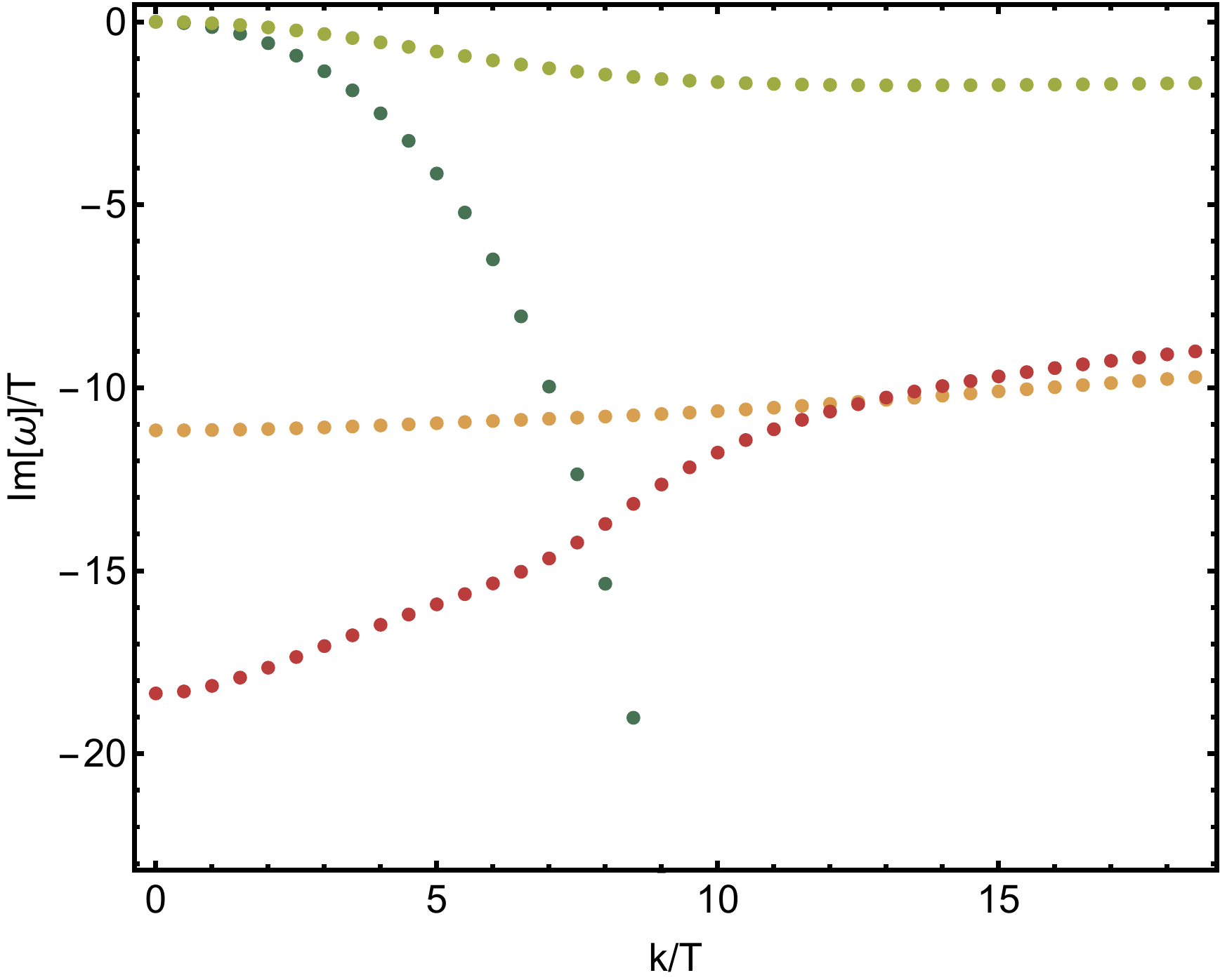}
        \caption{The modes in the longitudinal sector of model \eqref{bb} for $m/T=0.1$.}
        \label{figP}
\end{figure}
As a final task, we extend the analysis beyond the first two hydrodynamic modes until large frequencies and momenta. First, we focus on the behaviour of the hydrodynamic poles. The longitudinal sound mode \eqref{long1} interpolates smoothly between the low frequency speed $v_s^2=1/2$ to a UV relativistic dispersion relation $\omega=c\,k$, with $c=1$. This is a simple consequence of the UV behaviour of the geometry. Additionally, its imaginary part stops to increase diffusively and it asymptotes to a constant value at large momenta. This behaviour is shown in fig.\ref{figP}. At the same time, the dynamics of the higher modes in this model is very rich and complex as shown in the bottom panel of fig.\ref{figP}.\\[0.1cm]
\textbf{A digression: the Hydro$+$ formalism.}\\
The sound mode displays two interesting behaviours: (I) it interpolates between a low momentum dispersion relation $\omega=\sqrt{\frac{1}{2}}\,k$ to an high momentum relation $\omega=k$. (II) The sound attenuation does not grow arbitrarily but it saturates to a small value at high momenta. Recently, an extension of hydrodynamics (''Hydro$+$''), which considers parametrically slow modes, has been proposed in \cite{Stephanov:2017ghc} and motivated from the description of fluctuations out of equilibrium. Importantly, a new formula for the sound mode, containing those corrections, has been derived in the form of:
\begin{equation}
    \omega^2\,=\,k^2\,\left(c_s^2\,+\,\frac{\omega}{\omega\,+\,i\,\gamma}\,\Delta\right).\label{steph}
\end{equation}
The scale $\gamma$ is the relaxation of the putative additional slow mode and it defines the scale at which the corrections become important. The second parameter $\Delta$ describes the increase in the sound speed due to the stiffening of the equation of state.\\
Interestingly, the phenomenological equation \eqref{steph} reproduces perfectly the trend of our numerical data until very large momenta; see fig.\ref{hydrofig}.
\begin{figure}
    \centering
    \includegraphics[width=0.45\linewidth]{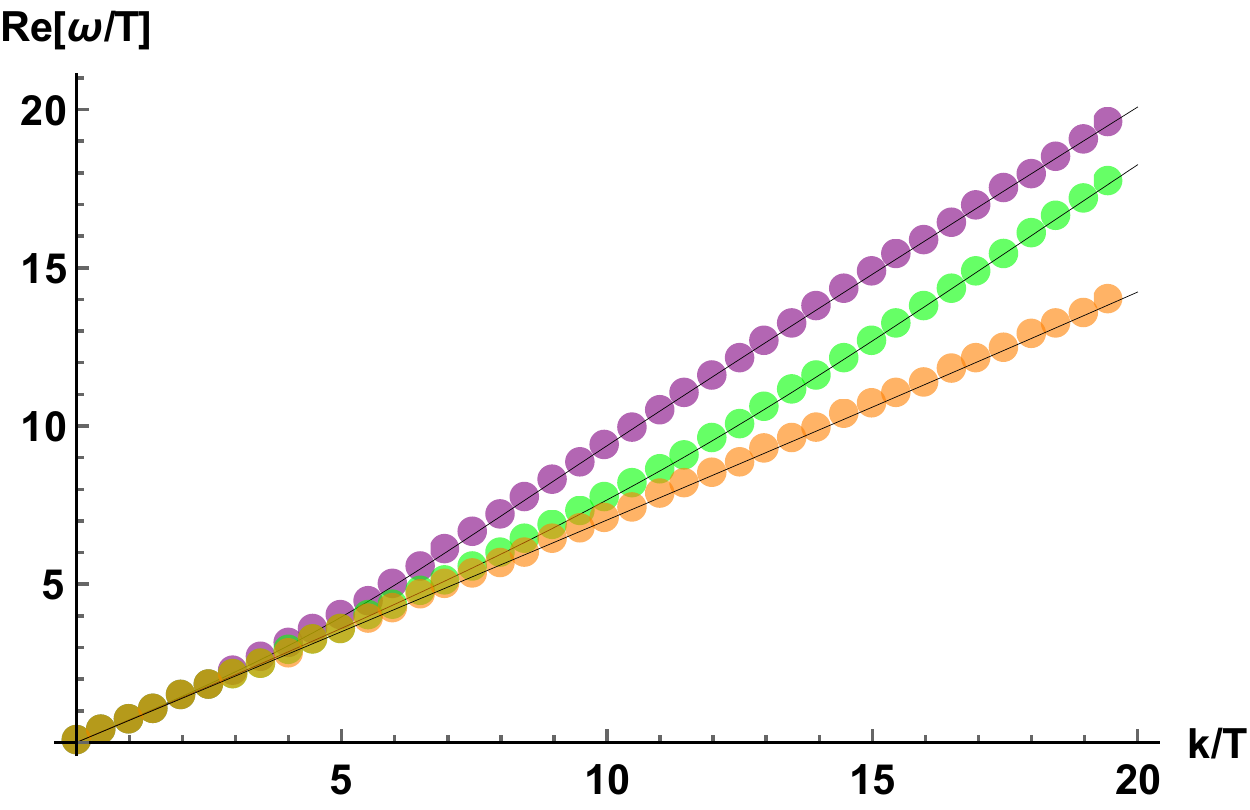}%
    \quad
    \includegraphics[width=0.45\linewidth]{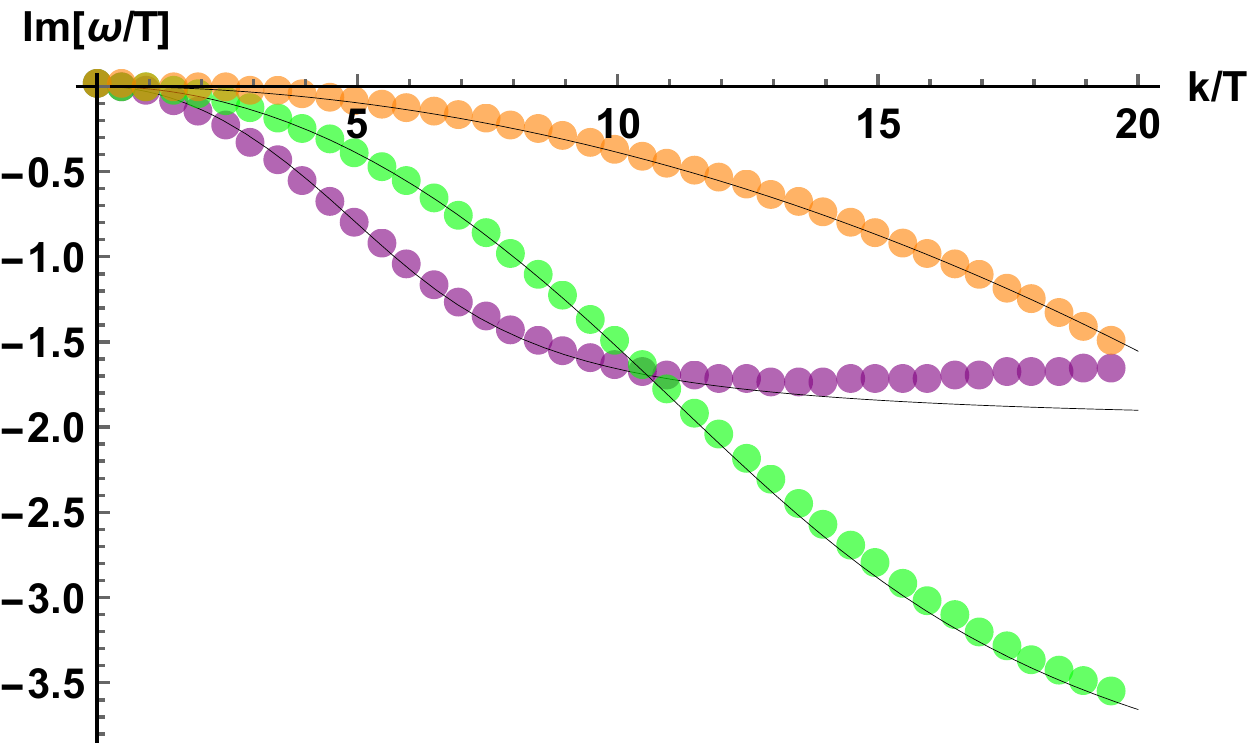}
    \caption{The numerical data for the lowest sound mode for $m/T=0,10,50$. The black line is the fit to the Hydro$+$ formula \eqref{steph}.}
    \label{hydrofig}
\end{figure}
The value of $\gamma$ grows by increasing the dimensionless parameter $m/T$. For larger $m/T$, the linear behaviour $\omega=\sqrt{\frac{1}{2}}\,k$ extends towards larger momenta. It would be interesting to investigate this point further by computing the coefficients of Hydro$+$ from first principles, and by considering the fluctuations out of equilibrium in our specific setup.
\section{Pseudo-phonons in holographic fluids}
After focusing on two holographic fluid models exhibiting respectively explicit (section \ref{fluidEXB}) and spontaneous (section \ref{fluidSSB}) breaking of translations, we are interested in analyzing their interplay. In order to do that, we consider the following potential:
\begin{equation}
    V(X,Z)\,=\,\alpha\,Z\,+\,\beta\,Z^2 \label{pp1}
\end{equation}
where $\alpha$ and $\beta$ are free tunable, and dimensionless, parameters. The model still represents a fluid system, because of the absence of any $X$ dependence in the potential \eqref{pp1}. For $\beta=0$ the breaking is purely explicit, like in section \ref{fluidEXB}, while for $\alpha=0$ the breaking is purely spontaneous, as in section \ref{fluidSSB}. Our main interest regards the regime $\alpha/\beta \ll 1$, which realizes the pseudo-spontaneous breaking of translational symmetry. For more details related to the importance and the definition of this regime we refer to \cite{Alberte:2017cch,Ammon:2019wci} and references therein.\\
The typical effect coming from the interplay of explicit and spontaneous symmetry breaking is the appearance of a mass term for the Goldstone bosons. This mass term, usually defined as \textit{pinning frequency} $\omega_0$, is just the consequence of the GMOR relation and it is the analogous of the Pion mass \cite{Burgess:1998ku}. Pinned charge density waves \cite{RevModPhys.60.1129} are a typical condensed matter example.  The value of the mass depends both on the explicit $\langle EXB \rangle$ and the spontaneous $\langle SSB \rangle$ breaking scales and it obeys the famous Gell-Mann-Oakes-Renner relation (GMOR) relation \cite{PhysRev.175.2195} :
\begin{equation}
    \omega_0^2\,=\,\langle EXB \rangle\,\langle SSB \rangle.
\end{equation}
The validity and role of this relation for phonons have been already discussed in the context of field theory and holography in \cite{Ammon:2019wci,Alberte:2017cch,Andrade:2017cnc,Li:2018vrz,Donos:2019tmo,Musso:2018wbv,Amoretti:2018tzw,Amoretti:2016bxs}.\\
Moreover, it has been shown in the previous literature \cite{Alberte:2017cch,Amoretti:2018tzw,Ammon:2019wci}, that, in the transverse sector, the pseudo-Goldstone modes appear from an interesting dynamics between two light modes, governed by the expression:
\begin{equation}
    \left(\Gamma\,-\,i\,\omega\right)\,\left(\bar{\Omega}\,-\,i\,\omega\right)\,+\,\omega_0^2\,=\,0 \, ,
\end{equation}
\begin{equation}
    \omega_{\pm}\,=\,-\,\frac{i}{2}\,\left(\bar{\Omega}\,+\,\Gamma\right)\,\pm\,\frac{1}{2}\,\sqrt{4\,\omega_0^2\,-\,\left(\Gamma\,-\,\bar{\Omega}\right)^2}\,.\label{modes}
\end{equation}
which was obtained from hydrodynamics in \cite{Delacretaz:2017zxd}. In the previous equations, $\Gamma$ is the momentum relaxation rate already discussed. The new ingredient is encoded in the new relaxation scale $\bar{\Omega}$, which has been the topic of many recent discussions (see for example \cite{Ammon:2019wci}). The parameter $\bar{\Omega}$ is small only in the pseudo-spontaneous regime and it is controlled by the ratio of the EXB and SSB scales:
\begin{equation}
    \bar{\Omega}\,\sim\,\frac{\langle EXB \rangle}{\langle SSB \rangle}\label{checkc}
\end{equation}
as suggested in \cite{Ammon:2019wci}, and recently confirmed in \cite{Donos:2019txg}. Importantly, this novel relaxation mechanism is not due to the proliferation of topological defects or the dynamics of dislocations as studied in \cite{Delacretaz:2017zxd}.\\


The presence of two relaxation scales $\Gamma$ and $\bar{\Omega}$ can be understood as follows. The first quantity is simply the momentum relaxation rate, which appears as a consequence of the explicit breaking of spacetime translations:
\begin{equation}
    \cancel{x^I\,\rightarrow x^I\,+\,a^I}\,\quad \Longrightarrow\,\quad \Gamma\,\neq\,0
\end{equation}
and it directly relates to the gravitational sector, where the momentum operator is encoded.
The second, and more interesting, relaxation scale $\bar{\Omega}$ comes from the scalar sector, as already suggested in \cite{Ammon:2019wci,Amoretti:2018tzw,Donos:2019txg}. It arises because of the explicit breaking of the internal shift symmetry of the St\"uckelberg fields:
\begin{equation}
    \cancel{\phi^I\,\rightarrow \phi^I\,+\,b^I}\,\quad \Longrightarrow\,\quad \bar{\Omega}\,\neq\,0
\end{equation}
which is a global bulk symmetry. When the coupling between the two sectors, mediated by the graviton mass, is switched on, the two relaxation scales interplay between each other and they produce the pseudo-phonon degrees of freedom.\\
The aim of this section is to test the validity of the discussion above in a concrete model which breaks translations pseudo-spontaneously but displays invariance under VPDiffs, \textit{i.e.} is a fluid.
\subsection*{The transverse sector}
We start by considering the transverse sector of the fluctuations. For the purely explicit and purely spontaneous case we refer to sections \ref{tran1} and \ref{tran2}. Let us start by briefly summarizing the results in those two cases. In the SSB case, the only hydrodynamic mode in the transverse sector is a diffusive mode. In the EXB scenario, on the contrary, the diffusive mode acquires a finite damping and collides with a second non-hydrodynamic mode producing a $k$-gap.
\begin{figure}[H]
    \centering
    \includegraphics[width=7cm]{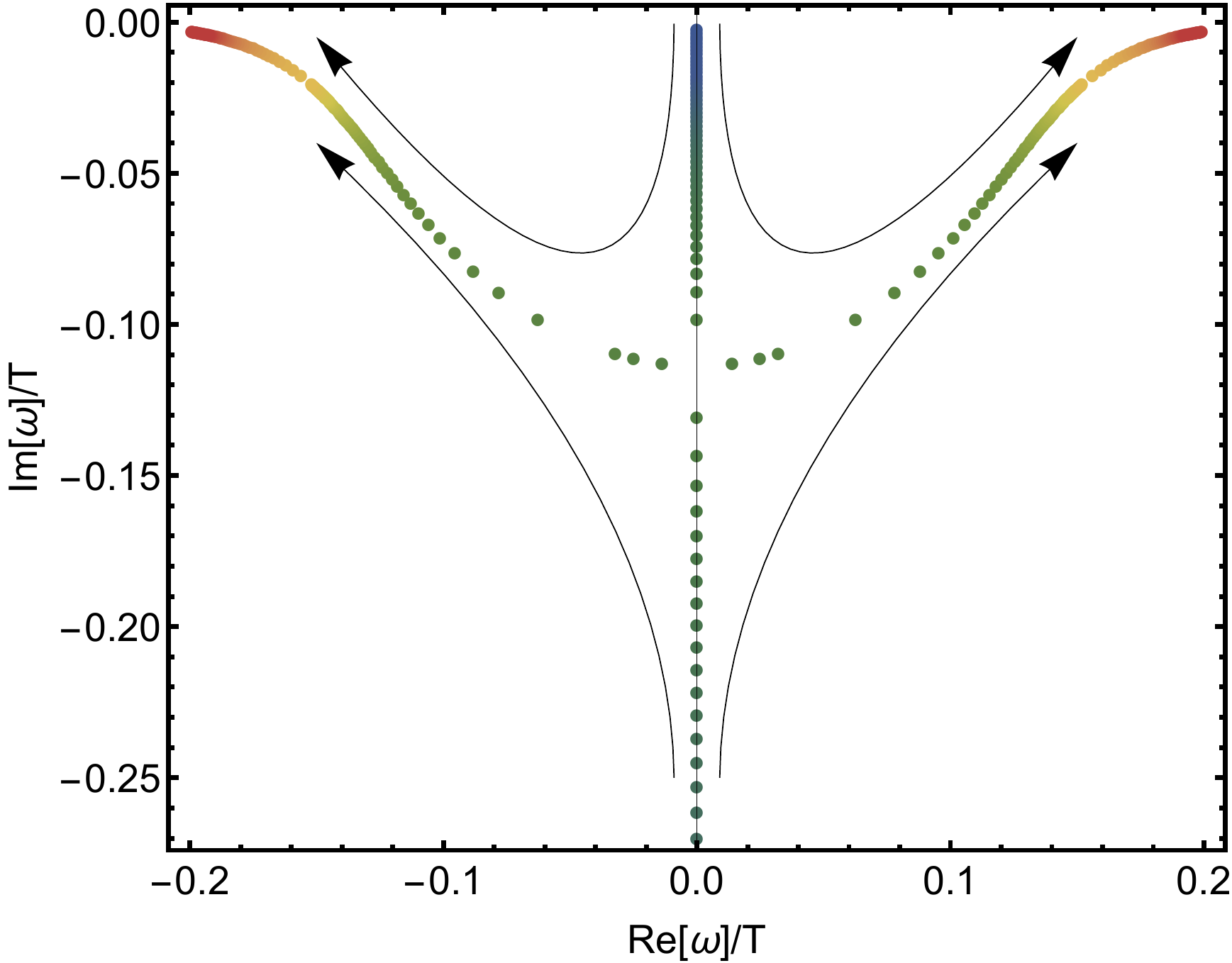}
        \caption{The dynamics of the lowest two modes in the transverse sector for $m/T=0.3,\ \alpha=0.05$ and increasing the parameter $\beta$ from zero to large values. The collision between the two poles happens at $\beta \sim 1$ and it produces the two light pseudo-phonons with a finite real part and a small damping. The arrows guide the eyes in the direction of increasing $\beta$.}
        \label{full1}
\end{figure}
Concretely, we consider a situation with a soft explicit breaking, $\langle EXB \rangle \ll 1$. This is simply achieved by demanding the UV graviton mass to be small (see \cite{Alberte:2017cch,Ammon:2019wci}), and more precisely $m \alpha/T \ll 1$. Given this limit, at $\beta=0$, the lowest hydrodynamic mode has a negligible damping $\Gamma \sim 0$. We then increase the value of the dimensioless parameter $\beta$ from zero to very large values. Doing so, we enter in the pseudo-spontaneous regime, defined by $
    \langle EXB \rangle/\langle SSB \rangle \,\ll\,1$.
The results are shown in fig.\ref{full1}. The pseudo-diffusive hydrodynamic mode $\omega \sim -\,i\,D\,k^2$ goes down along the imaginary axes upon increasing $\beta$. More importantly, a second mode, which for small $\beta$ was strongly damped, comes up. At a certain critical value $\beta^*$, the two modes collide on the imaginary axes and create a pair of off-axes poles with a finite real part, the pseudo-phonons. Increasing the SSB scale further, the damping (imaginary part) of those poles becomes smaller while their mass (real part) grows.
\begin{figure}[H]
    \centering
   \includegraphics[width=5.2cm]{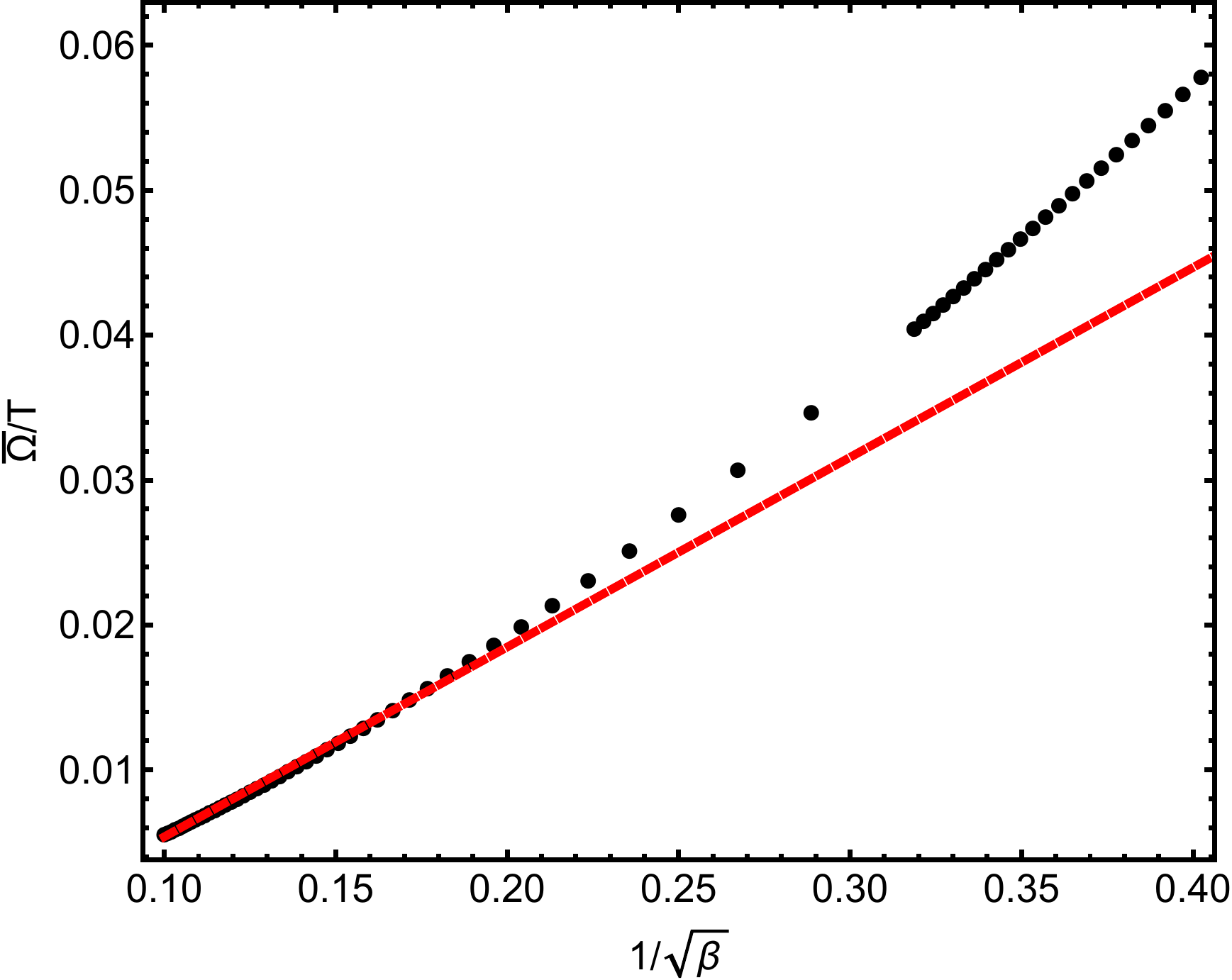}\quad  \includegraphics[width=5.2cm]{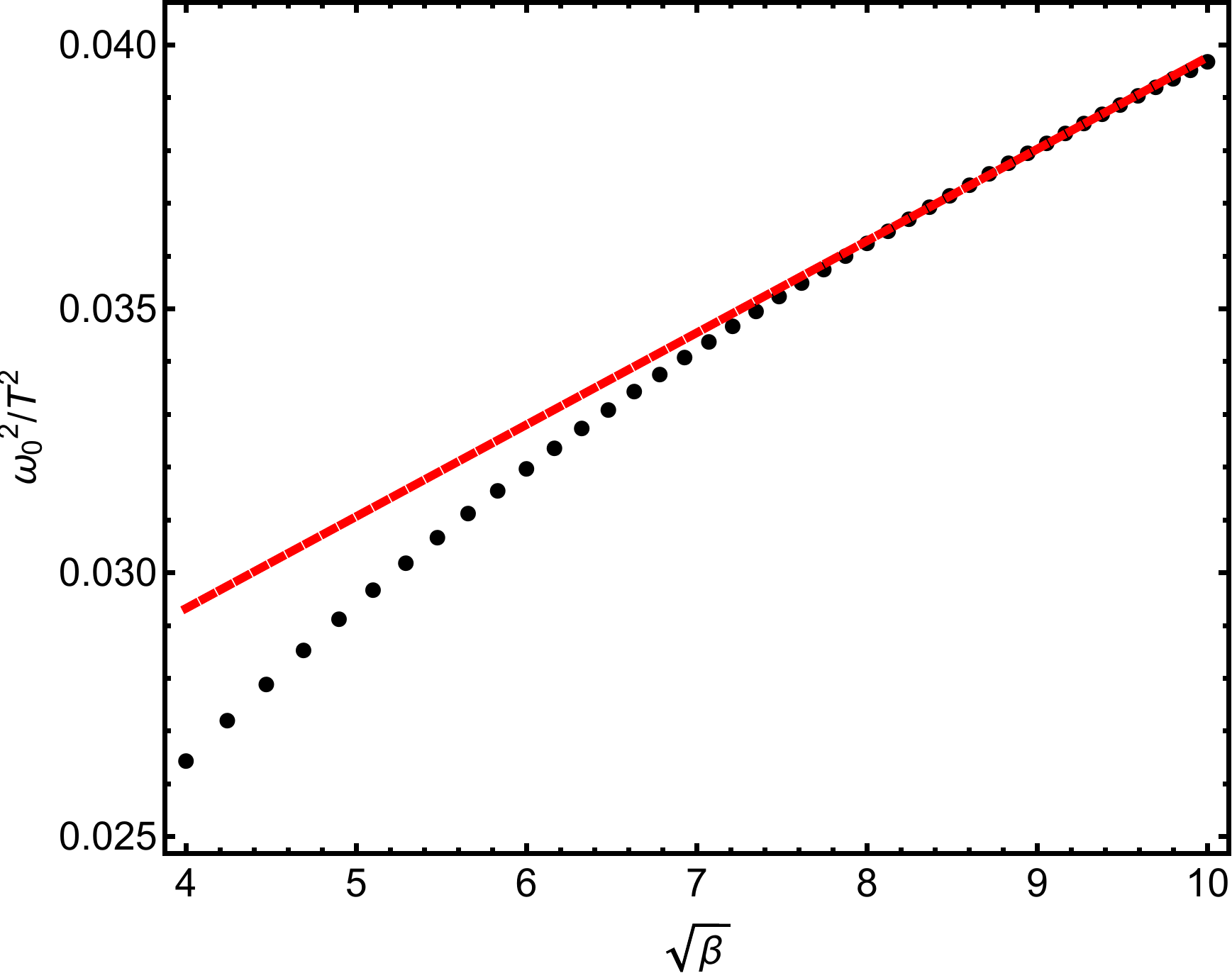}
        \caption{Dependence of the novel relaxation scale $\bar{\Omega}$ and the pinning frequency $\omega_0$ in function of the SSB parameter $\langle SSB \rangle \sim \sqrt{\beta}$. The other parameters are set to $m/T=0.3,\ \alpha=0.05$ such that $\langle EXB \rangle \ll 1$. The pseudo-spontaneous regime coincides with $\beta \gg 1$. There, the red lines guide the eyes towards the linear scalings.}
        \label{ffit}
\end{figure}
These two off-axes poles are exactly the pseudo-Goldstone modes expected in the pseudo-spontaneous regime. The situation is very similar to what is already observed in the solid models in \cite{Alberte:2017cch,Ammon:2019wci}. The second mode, appearing in the spectrum, is the one related to the novel relaxation time scale $\bar{\Omega}$ and coming from the scalar fields sector.\\
At this point, we want to fit the data in fig.\ref{full1} using the hydrodynamic formula \eqref{modes}, and taking the approximation $\Gamma \sim 0$, motivated by the small explicit breaking regime. The SSB scale is directly related to the parameter $\beta$ and more precisely $\langle SSB \rangle \sim \sqrt{\beta}$ (see \cite{Ammon:2019wci} for details). We show the dependence of the pinning frequency $\omega_0$ and the relaxation scale $\bar{\Omega}$ in terms of the spontaneous breaking scale in fig.\ref{ffit}. From there, we can immediately conclude (see red lines in fig.\ref{ffit}), that in the pseudo-spontaneous regime the following relations hold:
\begin{equation}
    \bar{\Omega}\,\sim\,\frac{\langle EXB \rangle}{\langle SSB \rangle}\,,\quad \quad \omega_0^2\,\sim\,\langle EXB \rangle\,\langle SSB \rangle
\end{equation}
This result is obtained by noticing that:
\begin{equation}
    \langle EXB \rangle\,=\,m\,\sqrt{\alpha}\,,\quad \langle SSB \rangle\,=\,m\,\sqrt{\beta},
\end{equation}
and by using the numerical results shown in fig.\ref{ffit},  $\bar{\Omega}\,\sim\,1/\sqrt{\beta}$ and $\omega_0^2\,\sim\,\sqrt{\beta}$. The dependence of $\bar{\Omega},\omega_0^2$ with respect to the other parameter $\alpha$ is not shown directly in the text but it is checked to be consistent with eq.\eqref{ffit}. The analysis is performed in the same way of \cite{Ammon:2019wci}.\\
In summary, our numerical results support that:
\begin{enumerate}
    \item In the holographic model under consideration the GMOR relation holds. The mass squared of the pseudo-Goldstone bosons relates to the spontaneous symmetry breaking scale linearly, as expected.
    \item The novel relaxation parameter $\bar{\Omega}$ is inversely proportional to the SSB scale, as already suggested in \cite{Ammon:2019wci}.
\end{enumerate}
Let us pause to discuss an interesting phenomenon. In this model, at zero explicit breaking, there are no propagating modes. As in every fluid, the transverse phonons have zero speed of propagation, and they become simply diffusive. This suggests that in dissipative systems one could have Goldstone bosons which are diffusive, as already studied in \cite{Minami:2018oxl}.\\
Nevertheless, when adding a source of EXB breaking, light and underdamped propagating modes appear, the pseudo-phonons. Given the fact that their speed is zero in absence of EXB, and that they are totally diffusive, it is an interesting question to understand what happens in the pseudo-spontaneous regime. We are not aware of any study of this sort for diffusive Goldstone bosons like that of \cite{Minami:2018oxl}. In order to analyze this feature, we take a different limit with respect to what discussed above; we start from the purely SSB regime ($\alpha=0$) and continuously increase the strength of the EXB.
The numerical outcomes are shown in fig.\ref{kdeptr}. 
\begin{figure}[H]
    \centering
   \includegraphics[width=5cm]{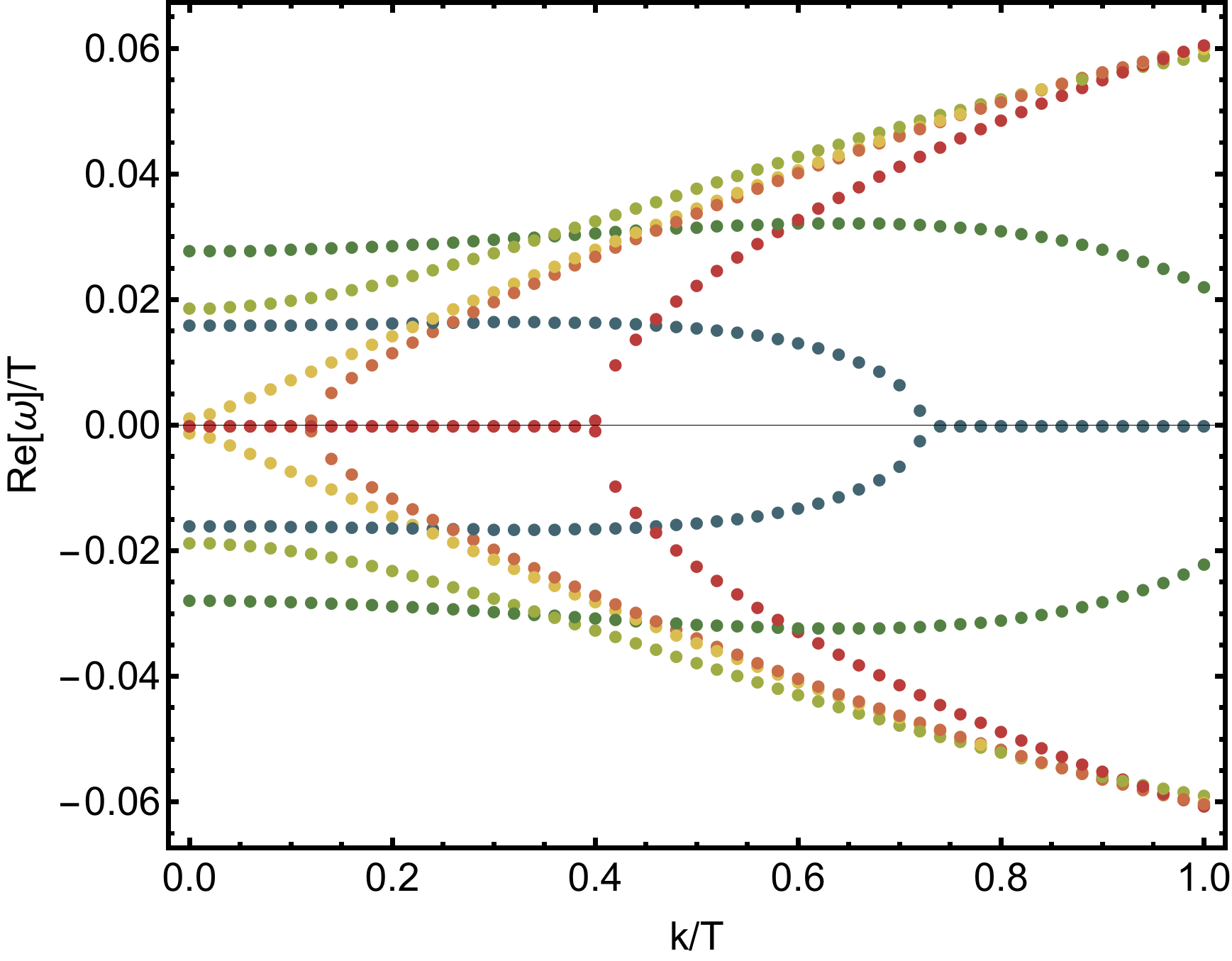}\quad  \includegraphics[width=5cm]{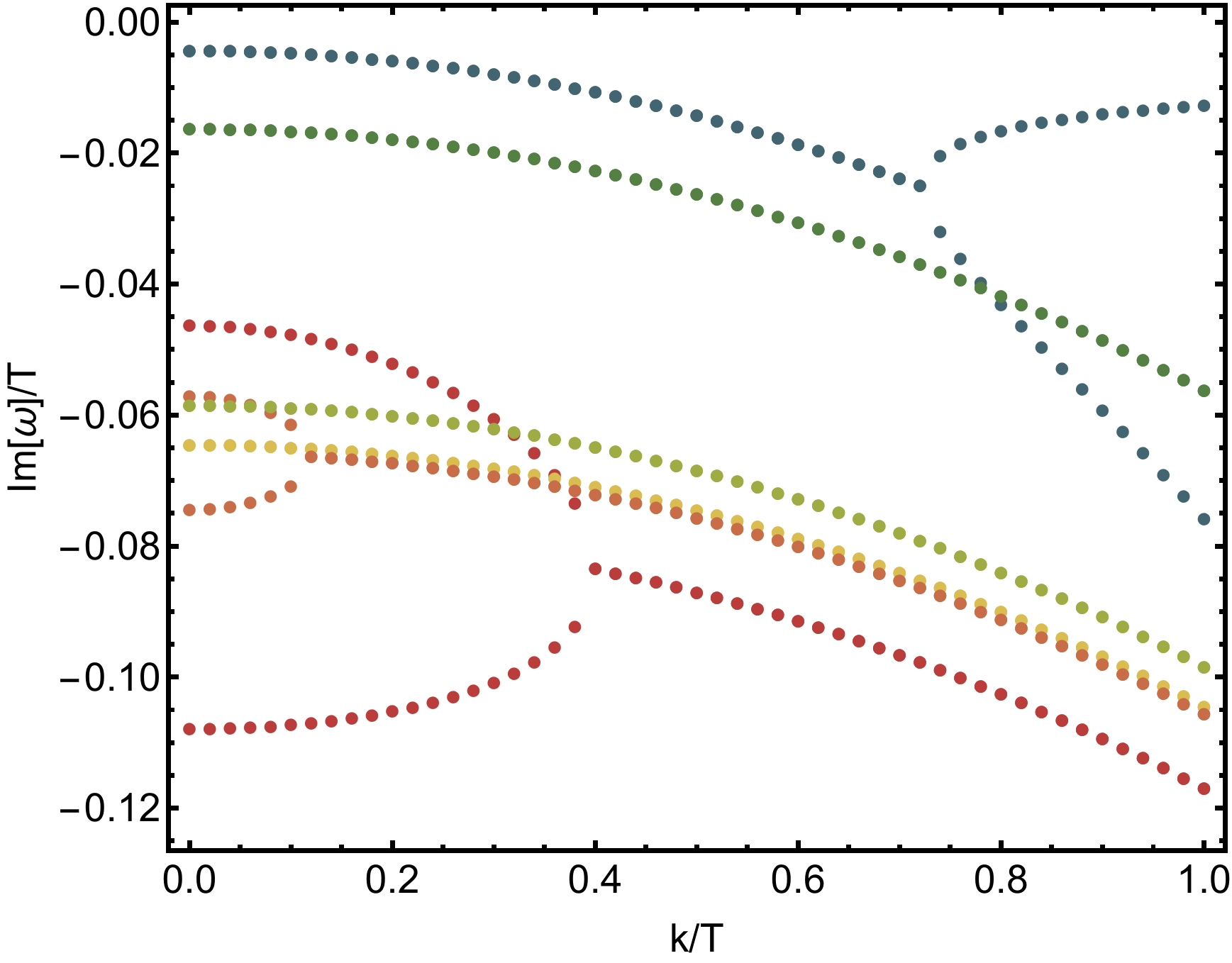}
        \caption{The lowest modes in the transverse sector for $m/T=0.1,\,\beta=1, \,\alpha\in\{0.001,0.151\}$ (blue to red). At small EXB, \textit{i.e.} small $\alpha$, a gapped phonon is present in the spectrum. Going to large EXB, such mode is destroyed and the $k-$gap, typical of the pure EXB case, appears.}
      \label{kdeptr} 
\end{figure}
At $\alpha=0$, the breaking is purely spontaneous, and the dispersion relation of the lowest modes is the one shown in section \ref{fluidSSB}, \textit{i.e.} a purely diffusive shear mode. It corresponds to the transverse phonon mode with zero propagating speed, as expected in fluids. Once a small source of explicit breaking is introduced, the massless Goldstone acquire a finite mass $\omega_0$, usually referred to as the pinning frequency\footnote{Notice that the real part of the pseudo-phonons does not corresponds exactly to the pinning frequency $\omega_0$. From eq.\eqref{modes}, we immediately obtain:
\begin{equation}
    Re(\omega)\,=\,\sqrt{\omega_0^2\,-\,\frac{1}{4}\,\left(\Gamma\,-\,\Omega\right)^2}
\end{equation}}, which satisfies the GMOR relation. Increasing the EXB scale $\sim \alpha$ further, the pseudo-phonon is destroyed and the $k-$gap phenomenon enters the dynamics. In this regime, we recover, as expected, the results for the purely EXB breaking presented in section \ref{tran1}. Let us notice that the dynamics of the modes can be understood phenomenologically by solving the simple equation:
\begin{equation}
    \omega^2\,+\,i\,\omega\,D\,k^2\,+\,i\,\omega\,\Gamma\,=\,\omega_0^2\,+\,v^2\,k^2 \label{ff}
\end{equation}
where $\Gamma$ and $\omega_0$ clearly depend on the EXB scale, see \cite{Ammon:2019wci}, as:
\begin{equation}
    \Gamma\,\sim\,\langle EXB \rangle ^2,\quad  \omega_0^2\,\sim\,\langle EXB \rangle, 
\end{equation}
\begin{figure}
    \centering
    \includegraphics[width=5cm]{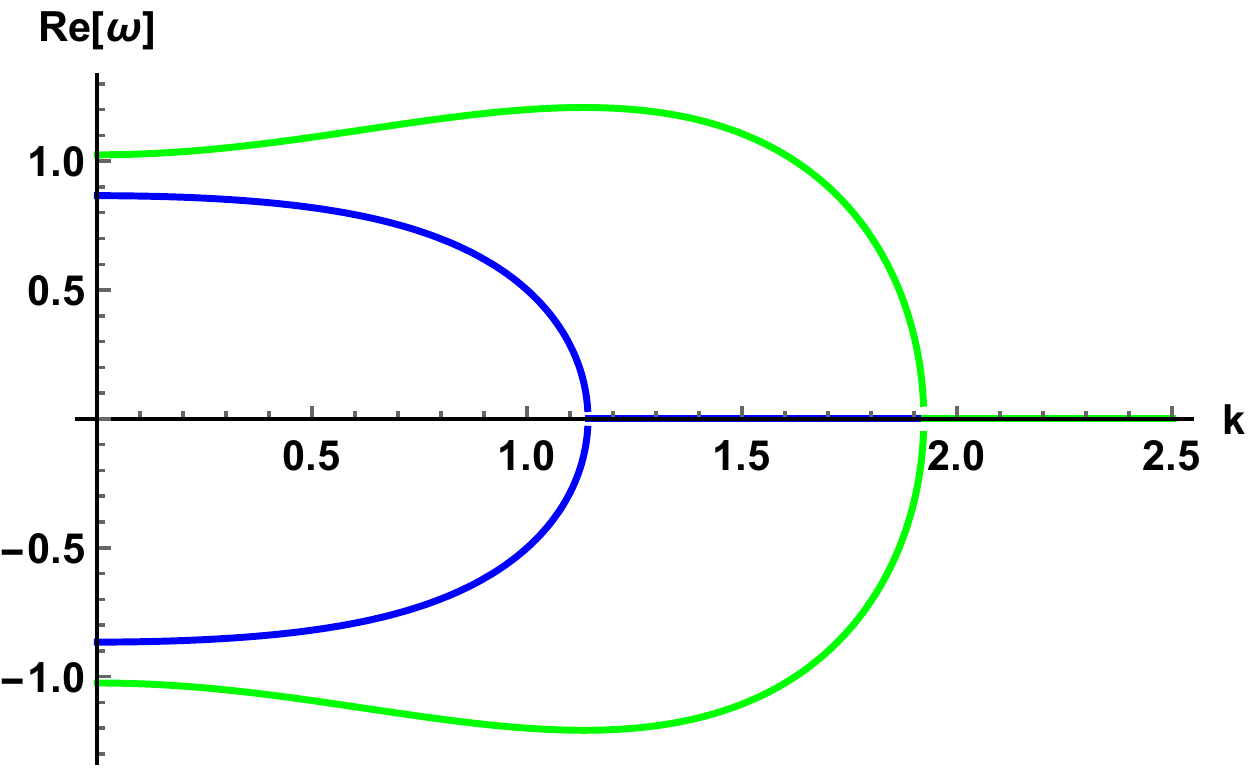}
    \quad
    \includegraphics[width=5cm]{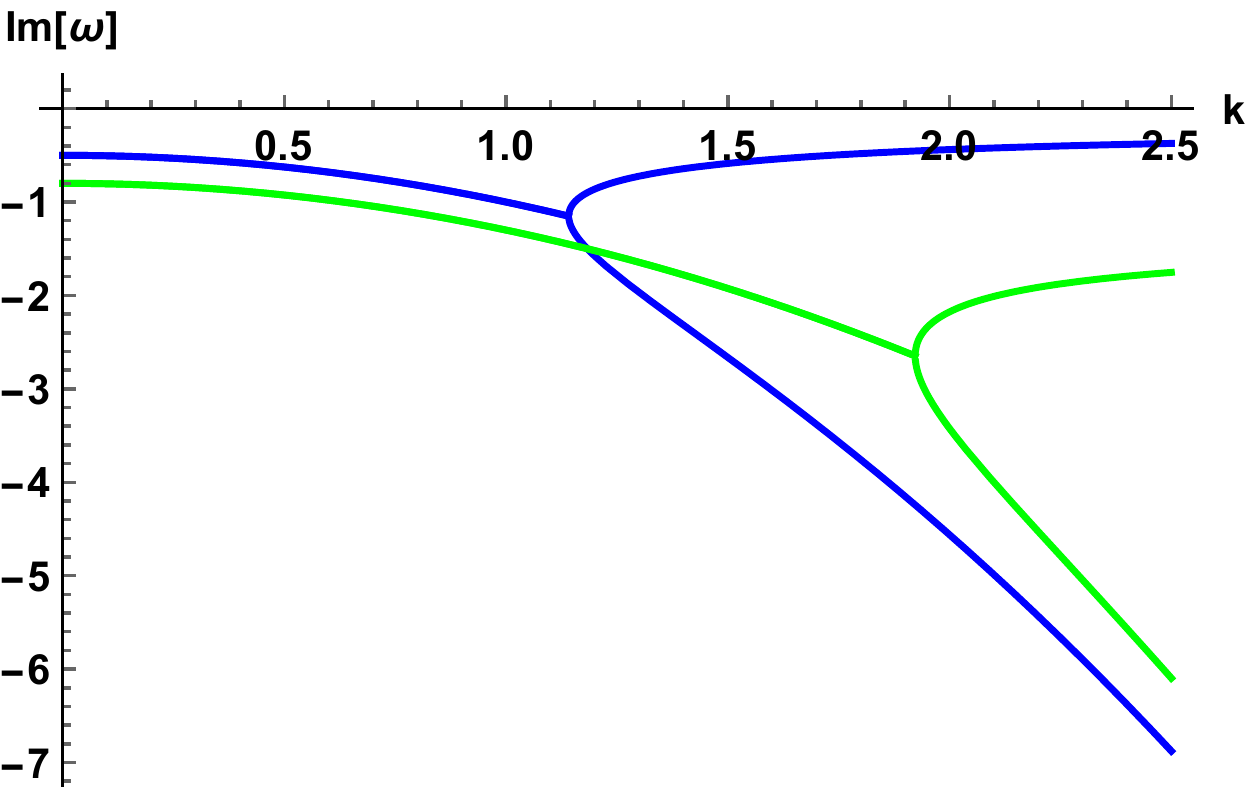}
    \caption{A cartoon of the transverse pseudo-phonons dynamics in liquids using eq.\eqref{ff}. The two lines are constructed by mimicking the increasing of the EXB parameter $\alpha$. More specifically, from blue to green line, both the damping $\Gamma$, the pinning frequency $\omega_0$ and the emergent speed $v$ are increased. The behaviour is very similar to what observed in fig.\ref{kdeptr} (see blue and green lines). This suggest that the emergent speed $v$ grows with the EXB scale.}
    \label{ffFIG}
\end{figure}
Notice that in order to reproduce the qualitative trend of the data in fig.\ref{kdeptr} a diffusive term $\sim D$ is necessary in eq.\ref{ff}. Such a term is not present in the pseudo-phonons description for solids (see for example \cite{Alberte:2017cch}). That is responsible for the presence of the flat band at small explicit breaking.
Now the important question is what is the scale controlling the emerging speed $v$? Clearly, it can not be the shear elastic modulus as in solids, since in liquids that is just zero, $G=0$. Equation \eqref{ff} seems to predict quite well the behaviour of the modes at small explicit breaking, as shown in the cartoon fig.\ref{ffFIG}. The speed of propagation, in order to fit the qualitative behaviour of the data, seems to grow as well with the explicit breaking parameter. More precisely, we observe a qualitative trend of the type:
\begin{equation}
    v\,\sim\,\frac{\langle EXB \rangle}{\langle SSB \rangle}
\end{equation}
which we are not able to explain at the moment. Notice that a contribution to the pseudo-phonons speed given by the explicit breaking was also found in holographic solids in \cite{Alberte:2017cch}. Here, this is still more surprising since at zero explicit breaking there is no propagating mode at all. Can the EXB induce a propagating shear wave in liquids? We find this observation very interesting and similar to what discussed in \cite{Baggioli:2018nnp,Baggioli:2018vfc,Grozdanov:2018fic,Baggioli:2019jcm}.
It would be valuable, to show with field theory methods that diffusive Goldstone bosons, like in \cite{Minami:2018oxl}, acquire a small propagating speed (together with a mass and a damping), when EXB is introduced.\\

A significant advance in the understanding of the new phase relaxation mechanism discussed in this paper has been proposed in \cite{Amoretti:2018tzw,Andrade:2018gqk}.
In particular, a (supposedly) universal relation between the phase relaxation rate $\bar{\Omega}$, the pinning frequency $\omega_0$ and the Goldstone diffusion $\xi$ :
 \begin{equation}
        \bar{\Omega}\,\sim\,\mathrm{M}^2\,\xi\,\sim\,\frac{\omega_0^2\,\chi_{PP}}{G}\,\xi \label{universe}
    \end{equation}
    has been suggested. In eq.\eqref{universe}, $\mathrm{M}$ is the mass of the pseudo-Goldstone mode.\\
    This relation can be directly guessed from the scaling of the various quantities. The pinning frequency scales like $\omega_0^2 \sim \langle EXB \rangle \langle SSB \rangle$ in accordance with the GMOR relation, the shear elastic modulus scales like $G \sim \langle SSB \rangle ^2$, while the momentum susceptibility and the Goldstone diffusion constant are $\mathcal{O}(0)$ in these scales. Using the observation made in \cite{Ammon:2019wci}, that $\bar{\Omega} \sim \langle EXB \rangle /\langle SSB \rangle$, the equation above is certainly verified:
    \begin{equation}
        \frac{\langle EXB \rangle}{\langle SSB \rangle}\,\sim\, \langle EXB \rangle \langle SSB \rangle \,\frac{1}{\langle SSB \rangle^2}\,\quad \text{\CheckmarkBold}
    \end{equation}
    In order to proceed, we have used the numerical values of $\bar{\Omega}$ and $\omega_0^2$ extracted from the QNMs, as described previously. Additionally, we have used the formula:
    \begin{equation}
        \frac{\xi}{G}\,=\,\frac{4\,\pi\,s\,T^2}{2\,m^2\,\chi_{PP}^2\,\left(V_X(1,1)\,+\,2\,V_Z(1,1)\right)} \label{formulaxi}
    \end{equation}
    which was derived in \cite{Amoretti:2019cef} and confirmed numerically for our models in \cite{Ammon:2019apj}. Finally, we have used the fact that $\chi_{PP}=3/2\, \epsilon$, where $\epsilon$ is the energy density of the system \cite{Alberte:2017oqx}.\\
    
\begin{figure}
    \centering
    \includegraphics[width=7.4cm]{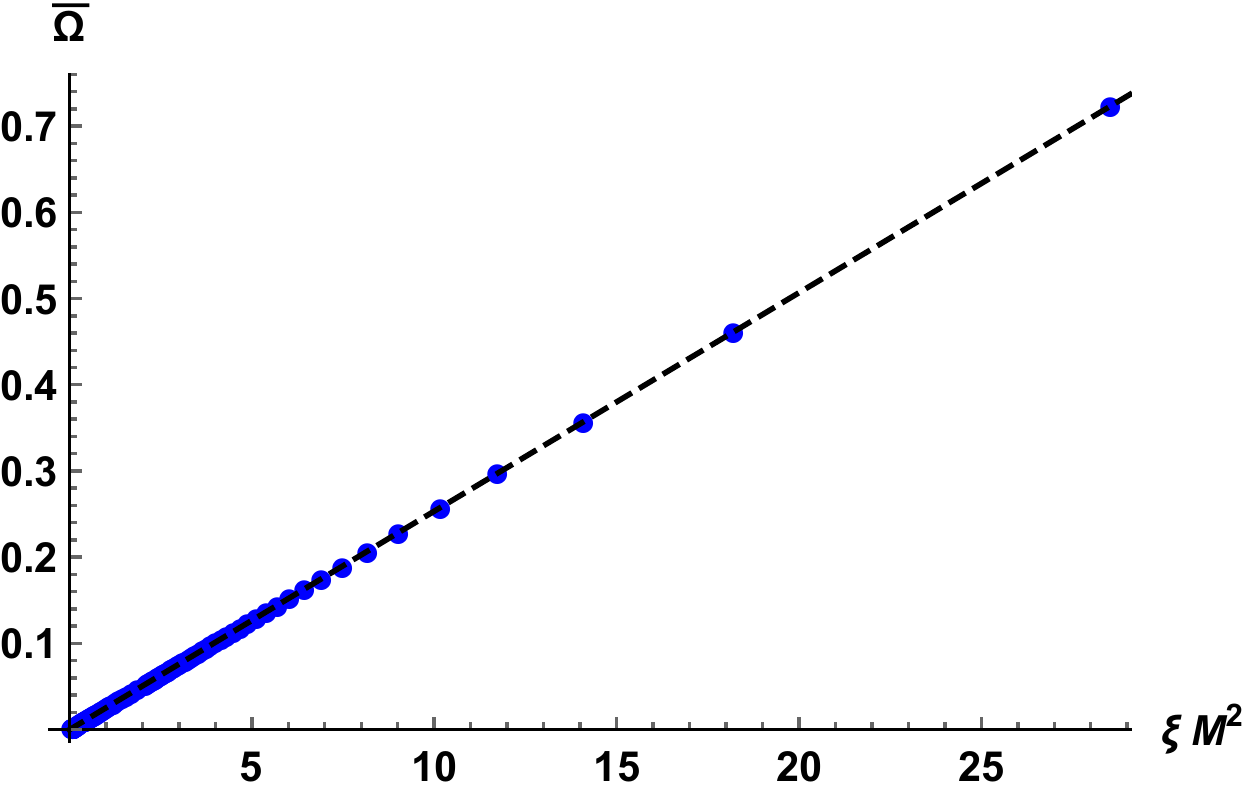}
    \caption{The novel phase relaxation scale in the fluid model $V(X,Z)=\alpha_f Z\,+\,\beta_f Z^2$. We fix $m/T=0.3,\alpha_f=0.05$ and dial $\beta_f$. This is a successful numerical check of the relation \eqref{universe}. The validity of this relation has been tested for more values of the free coefficients of the model.}
    \label{fig_fluid}
\end{figure}

    We have tested the validity of eq.\eqref{universe} numerically and the results are displayed in fig.\ref{fig_fluid}. The data confirm the validity of this relation in our holographic fluid models\footnote{For the same test in holographic solid models see \cite{Ammon:2019wci}.}.
The validity of the relation \eqref{universe} is a valuable guidance to understand the fundamental nature of the novel phase relaxation mechanism from an effective field theory point of view.
\subsection*{The longitudinal sector}
\begin{figure}[h]
    \centering
   \includegraphics[width=5cm]{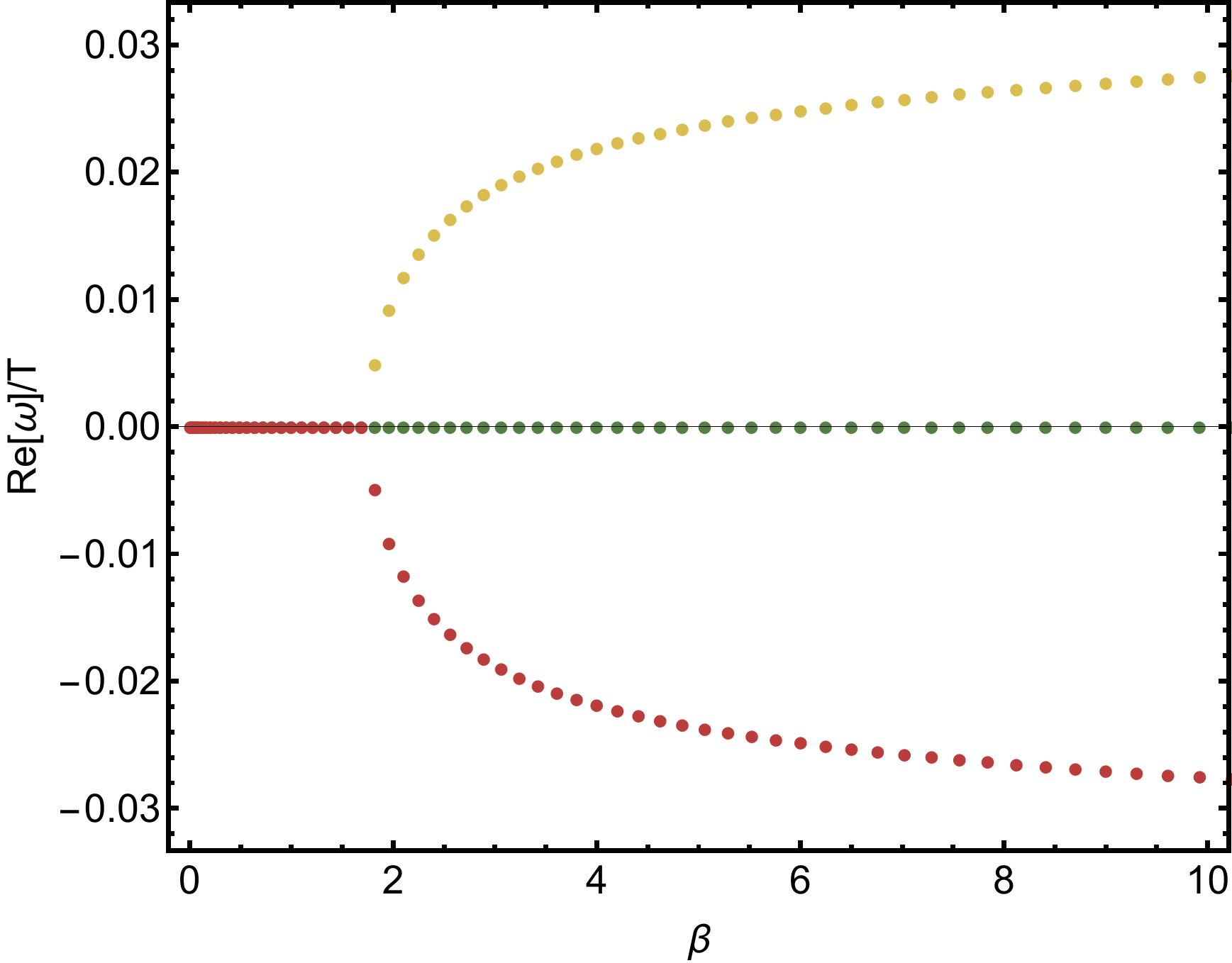}\quad  \includegraphics[width=5cm]{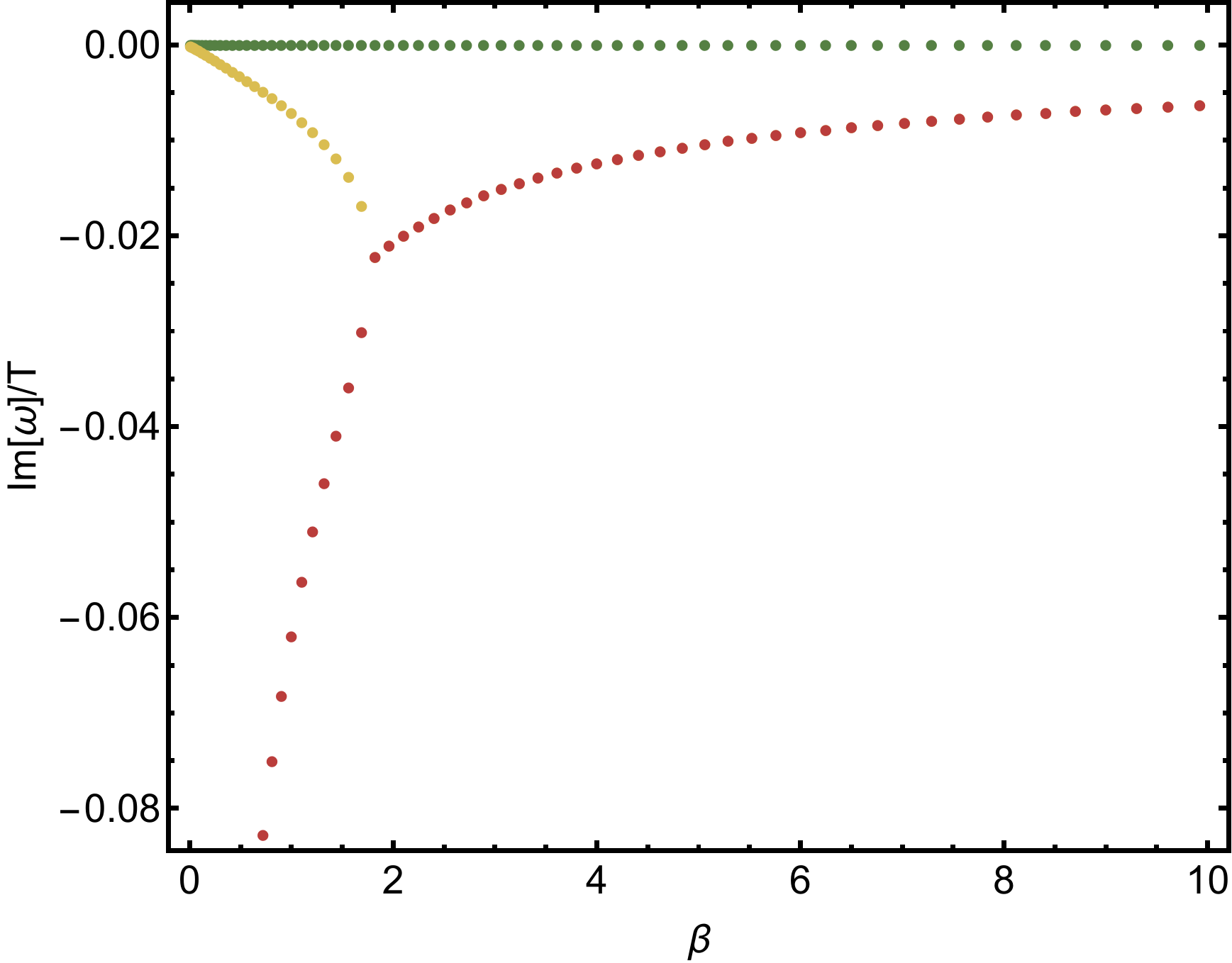}\\[0.3cm]
    \includegraphics[width=7cm]{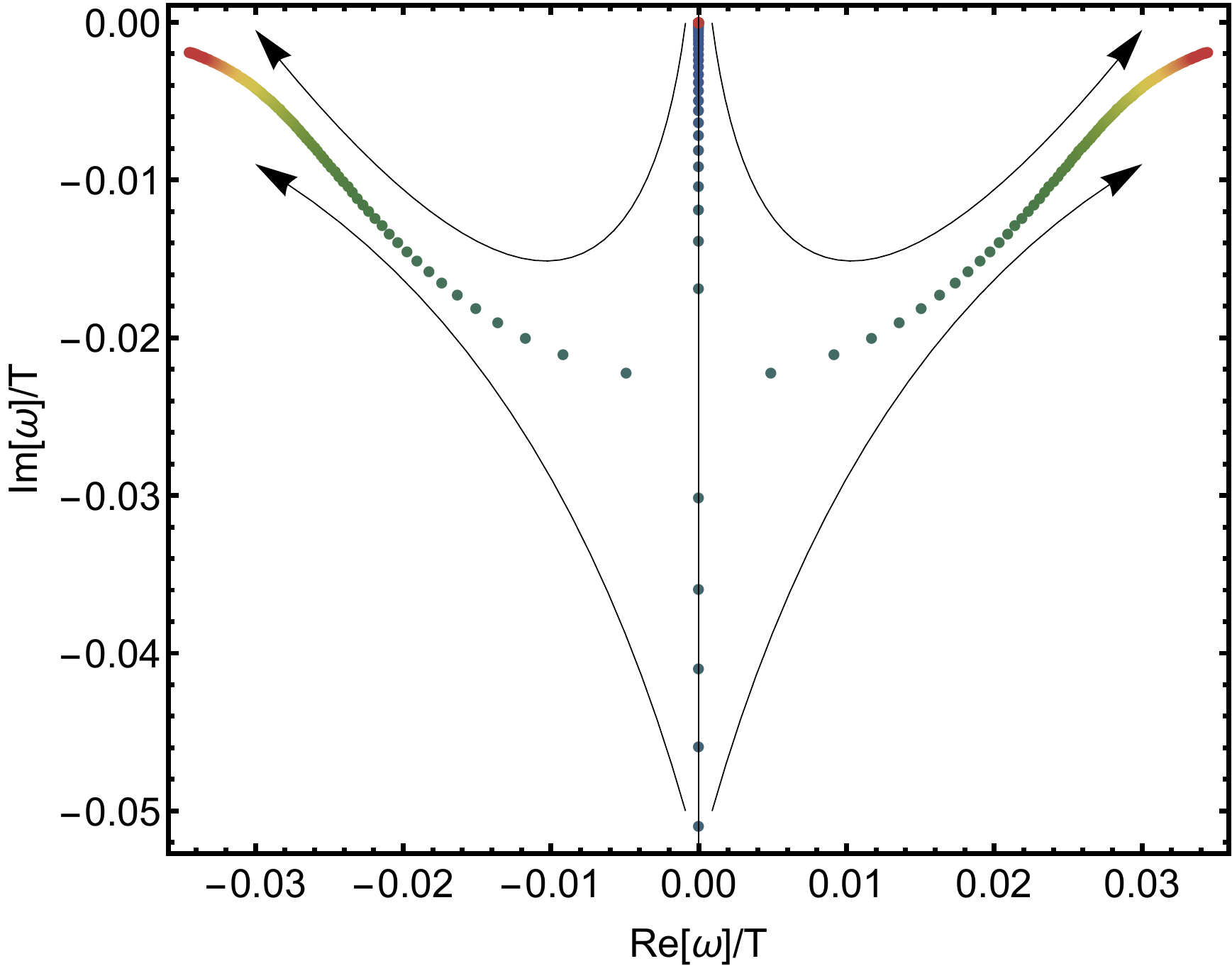}
        \caption{The two lowest mode in the longitudinal sector for $m/T=0.1,\ \alpha=0.01$ and increasing the parameter $\beta$ from zero to large values. The collision of the poles and the presence of the light pseudo-phonons are evident. Also, the damping of the crystal diffusion mode is consistent with zero. The arrows in the bottom panel guide the eyes in the direction of increasing $\beta$.}
        \label{full2}
\end{figure}

\begin{figure}[H]
    \centering
    \includegraphics[width=6.5cm]{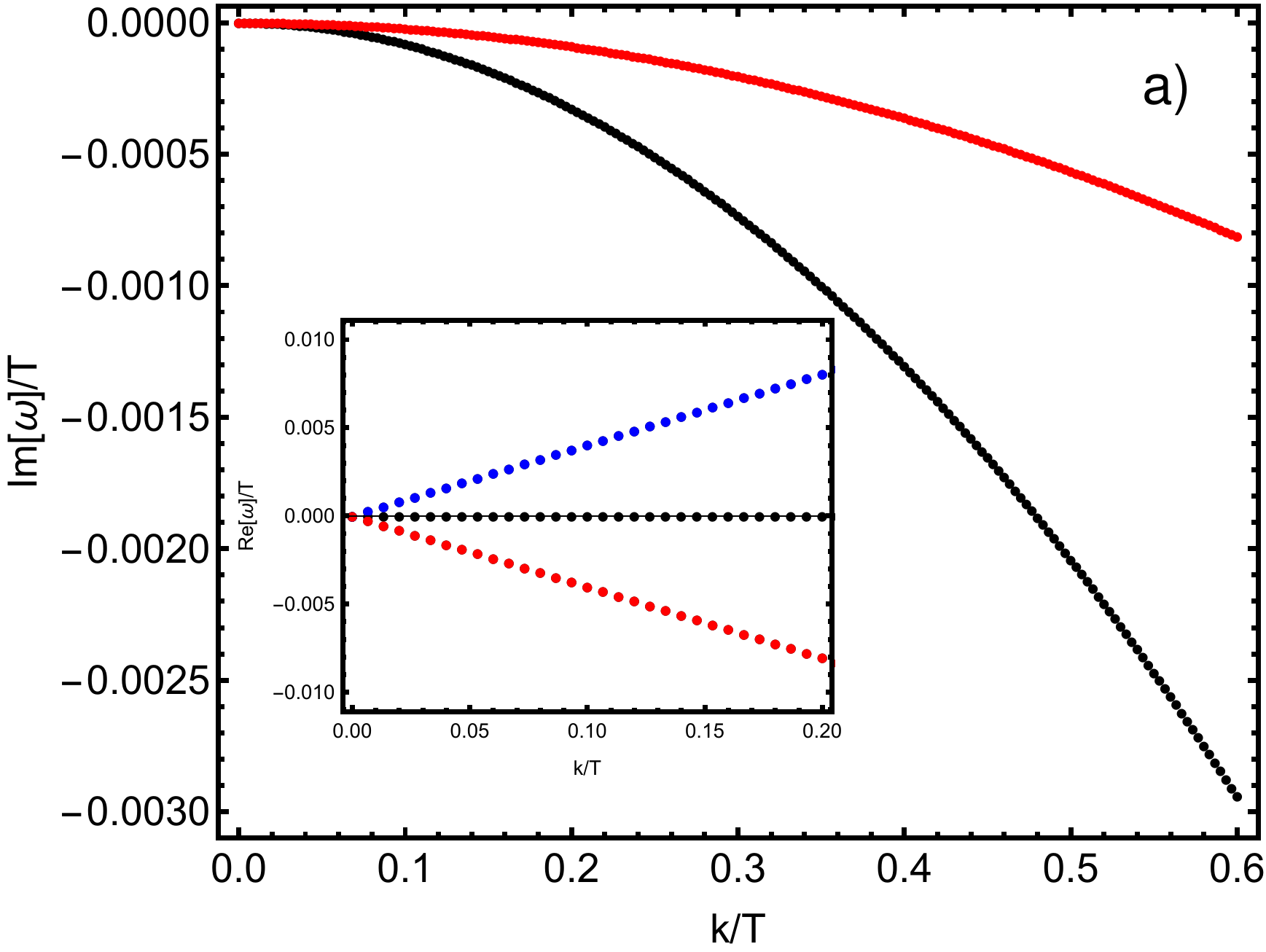}\quad \includegraphics[width=6.3cm]{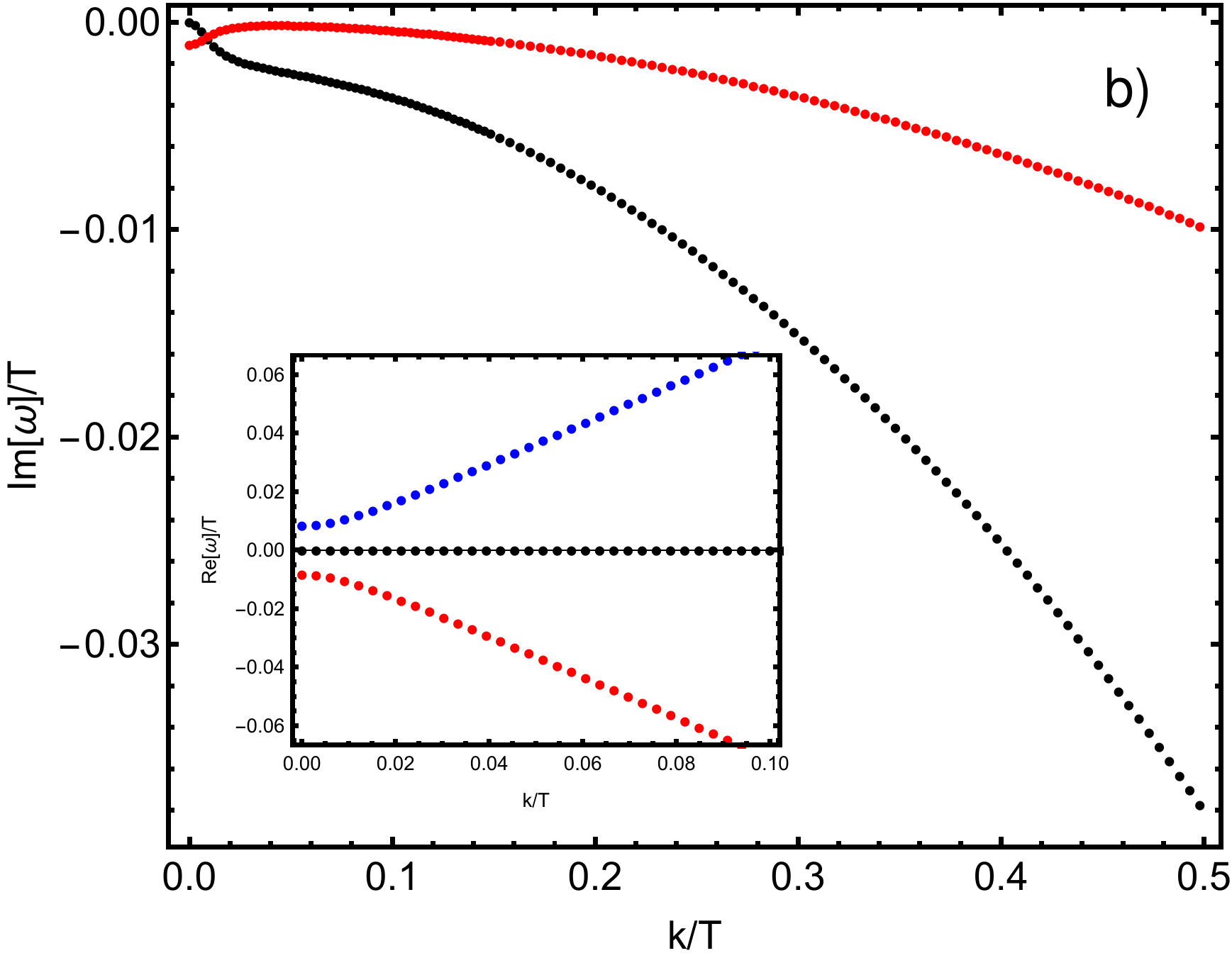}\\
      \includegraphics[width=6.5cm]{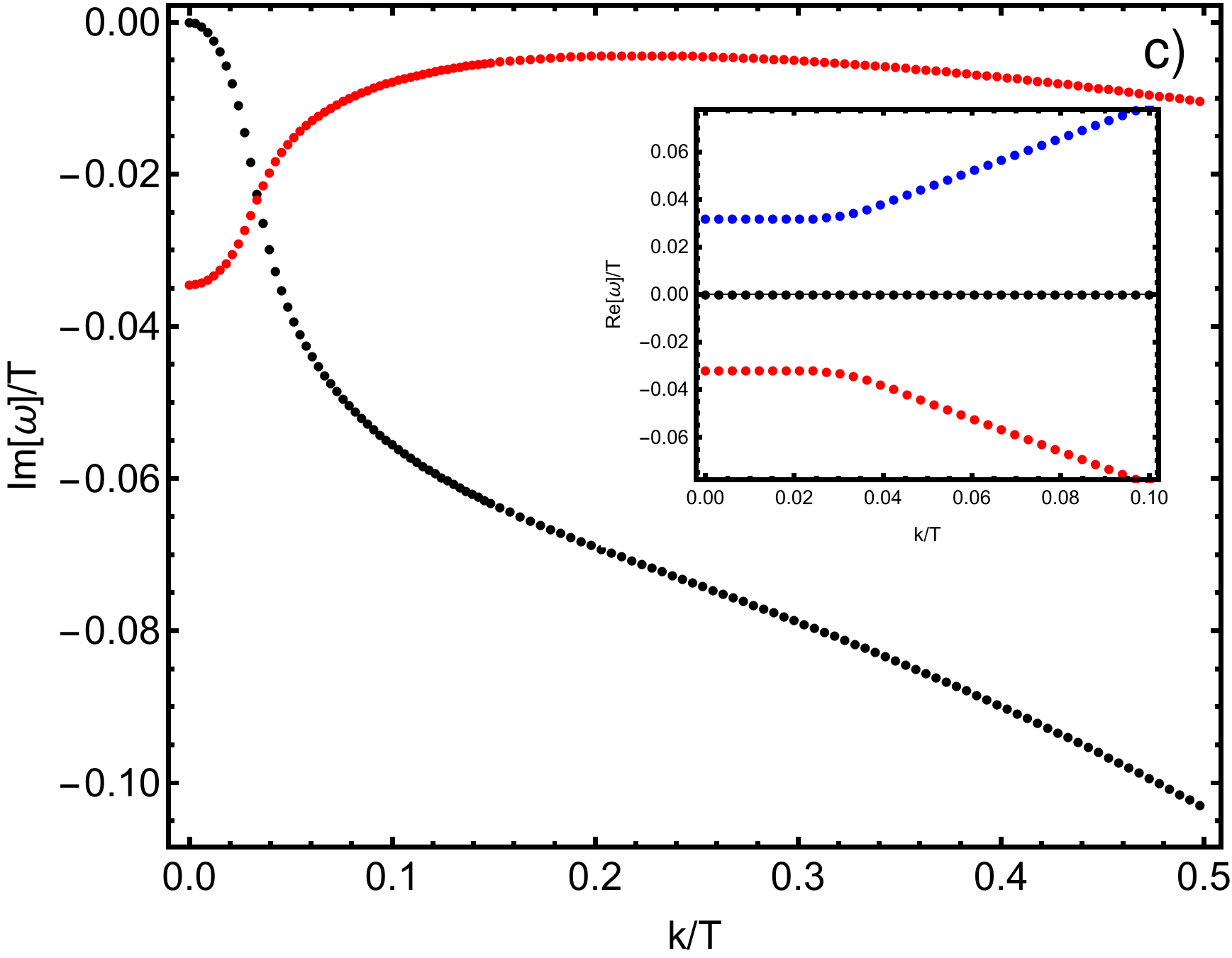}\quad \includegraphics[width=6.5cm]{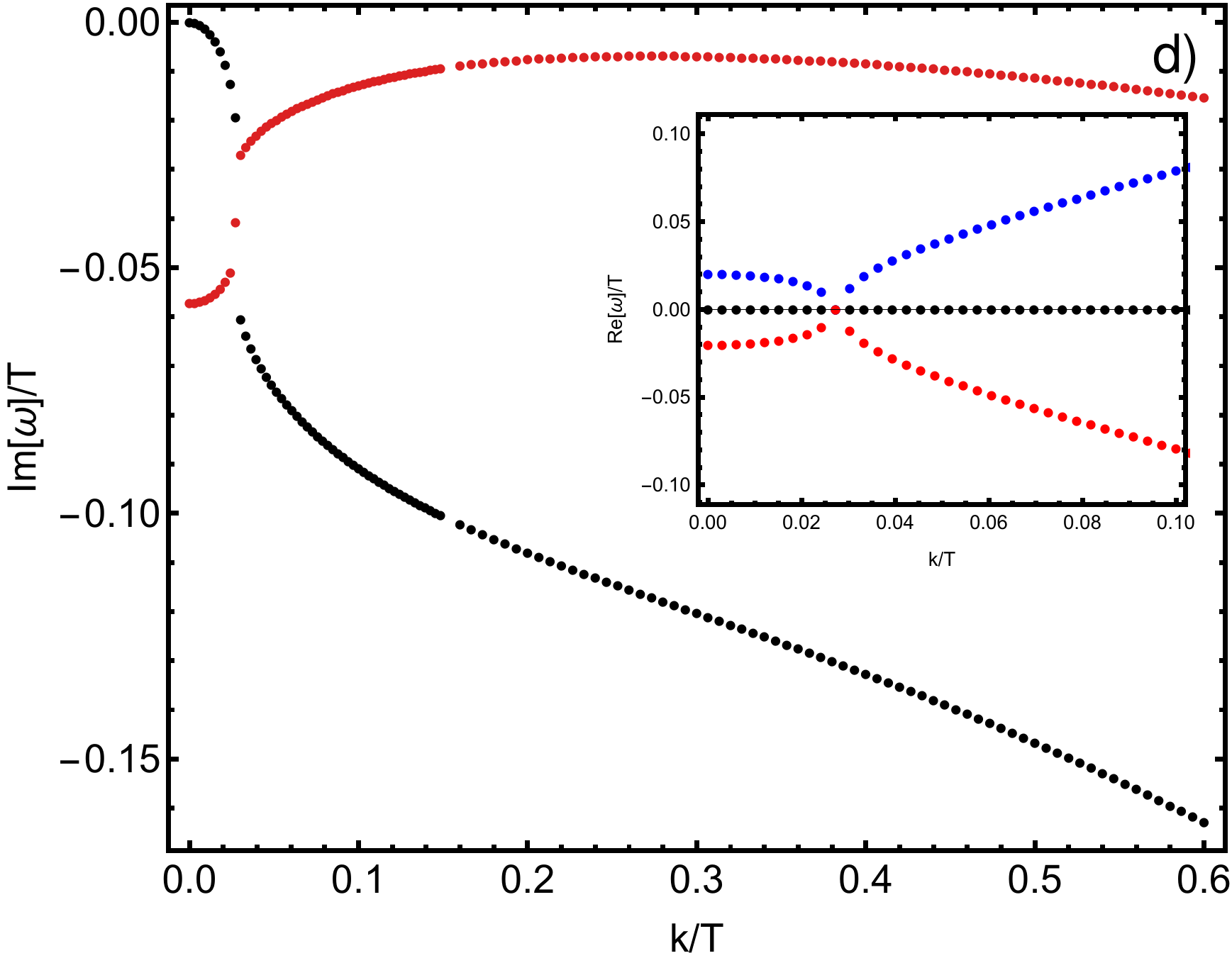}\\
      \includegraphics[width=6.5cm]{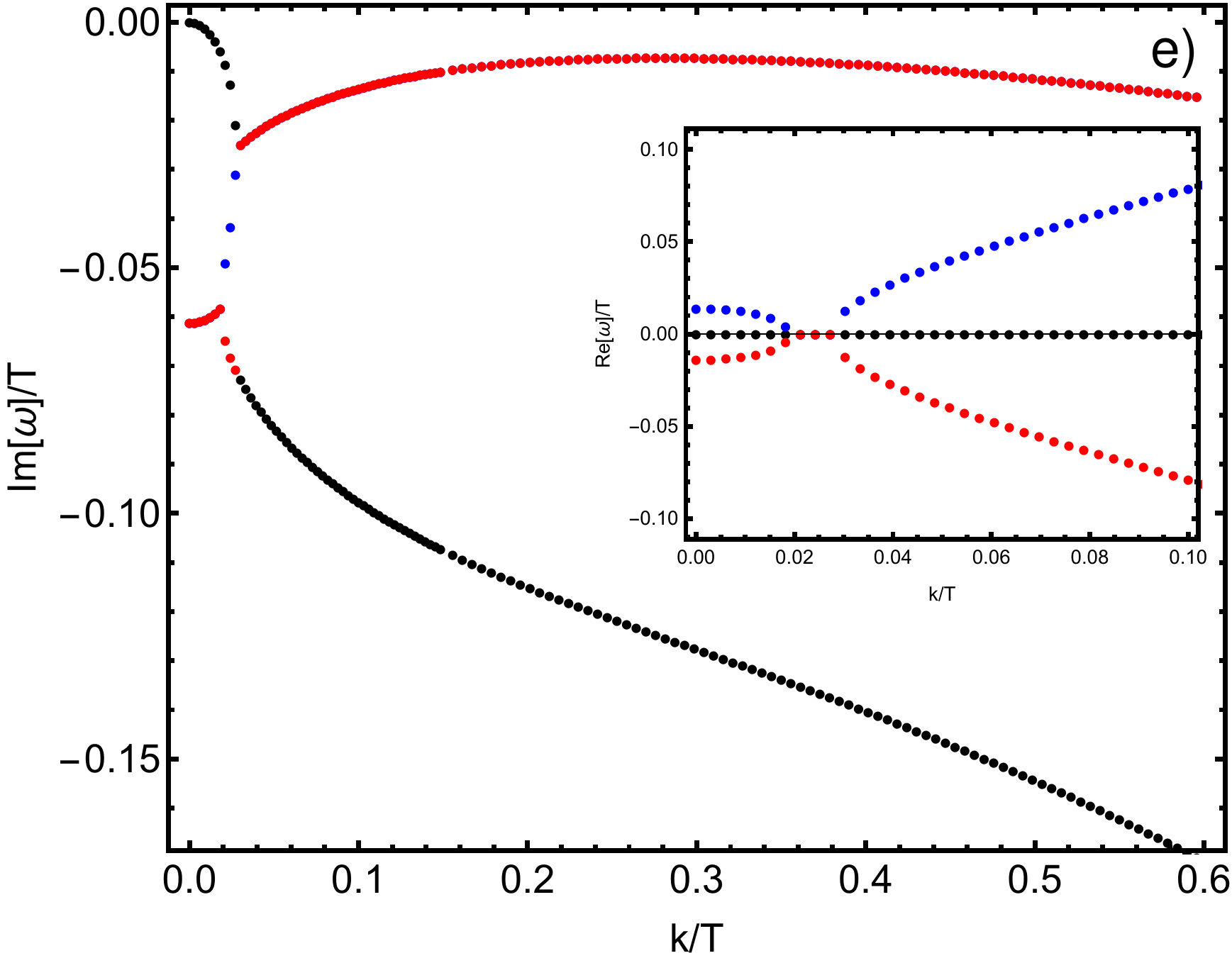}\quad \includegraphics[width=6.5cm]{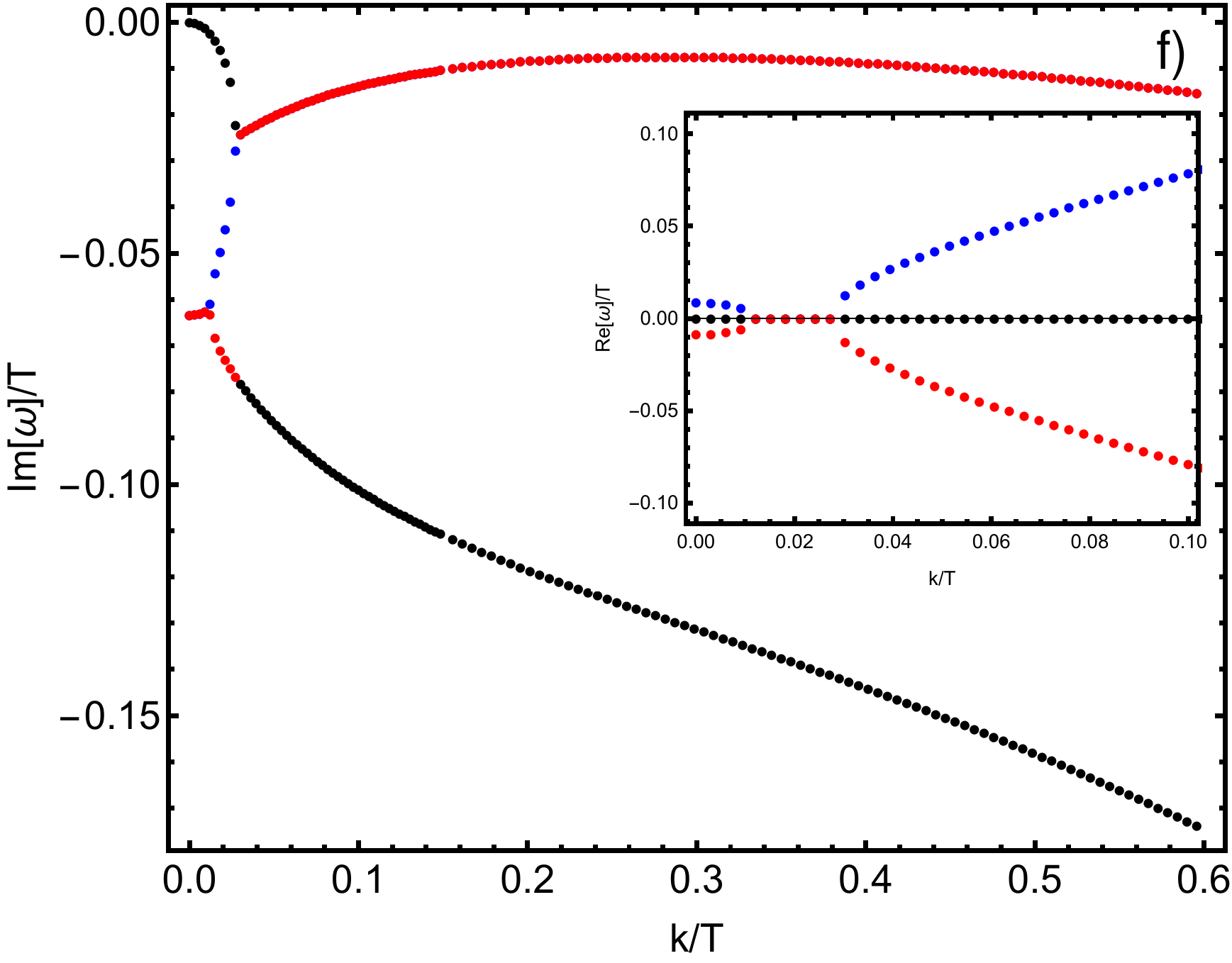}\\
      \includegraphics[width=6.5cm]{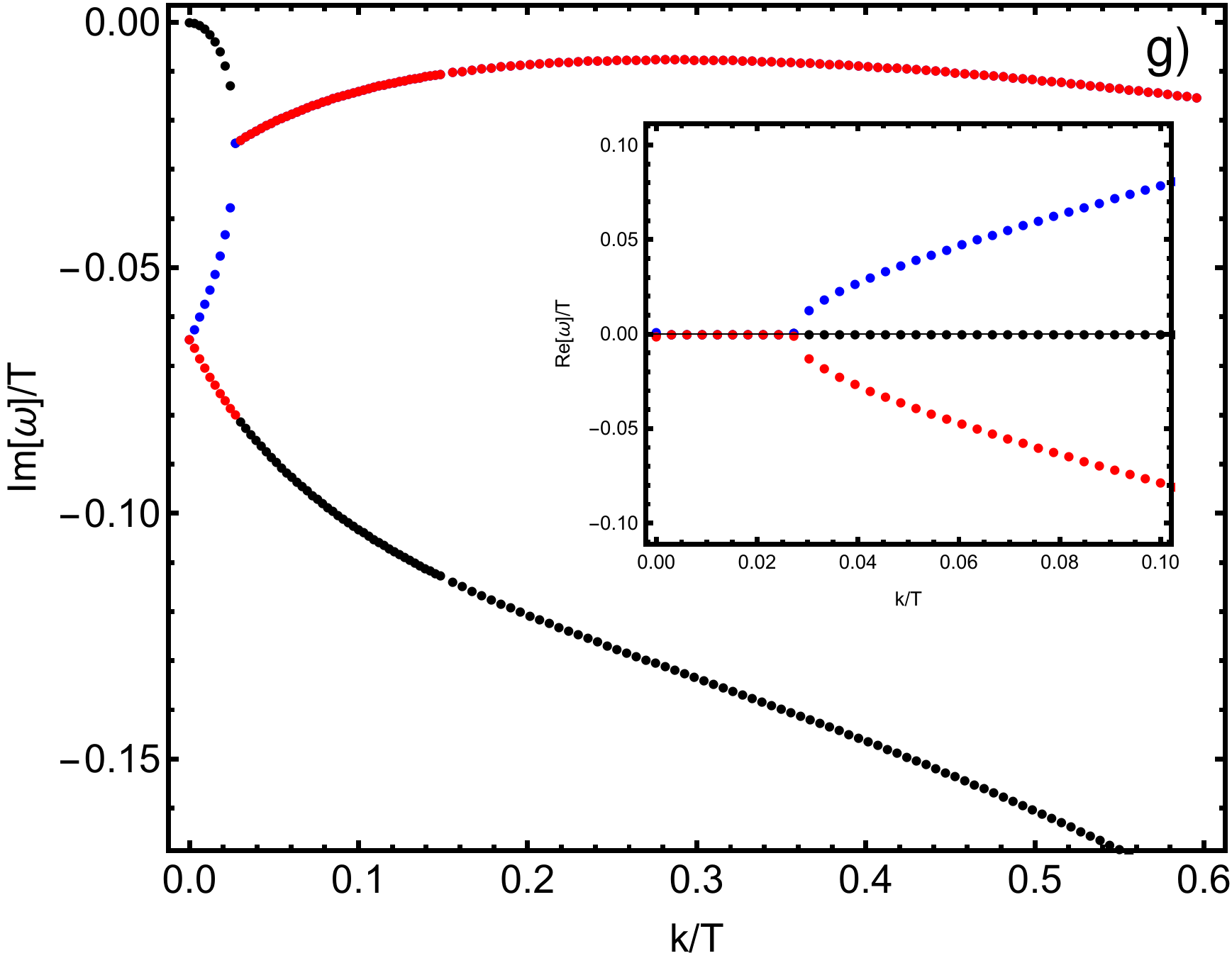}\quad \includegraphics[width=6.5cm]{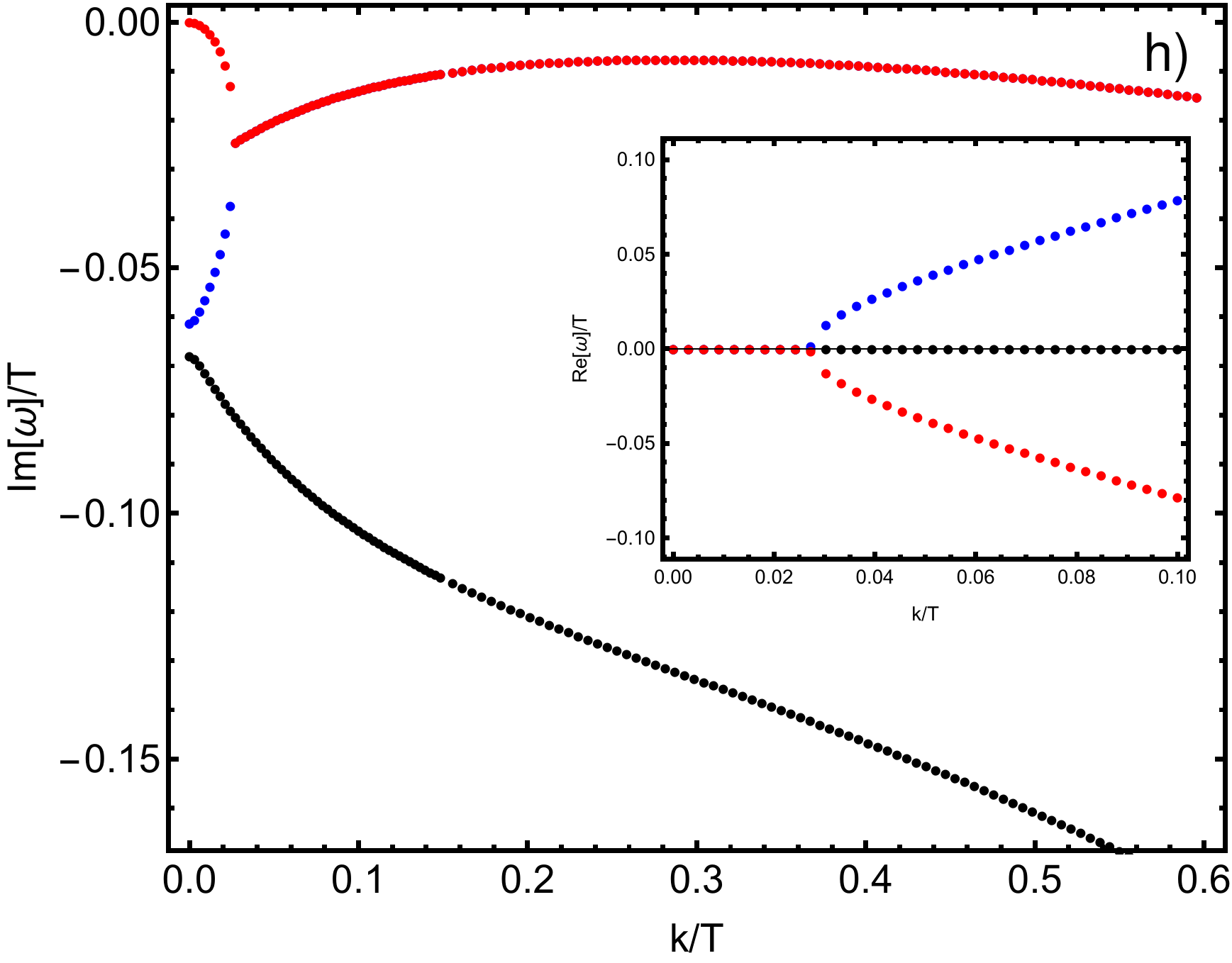}
    \caption{The lowest modes in the longitudinal sector for $\alpha=\{0,0.1185\}, m/T=0.1$ and $\beta=5$. The EXB strength increases from panel a) to panel h). In panel a) the crystal diffusion mode is shown in black color. Increasing further the EXB, the damping of the black mode in panel h) becomes larger and larger.}
    \label{test}
\end{figure}

In this section we consider the dynamics of the longitudinal collective modes in a fluid model exhibiting pseudo-spontaneous breaking of translations. To the best of our knowledge, no discussions about the longitudinal modes in the regime of pseudo-spontaneous breaking are present in the literature so far (neither in solid nor in fluid models). We follow the same logic as in the previous section. We fix the explicit breaking scale to be very small from the beginning and we gradually increase the SSB scale. The results for the lowest modes are shown in fig.\ref{full2}. We observe that: (i) the Goldstone modes acquire a finite mass and they become pseudo-phonons; (ii) the pinning frequency $\omega_0$ satisfies the GMOR relation; (iii) the relaxation scale obeys the simple scaling $\bar{\Omega}\,\sim \langle EXB \rangle /\langle SSB \rangle$, as it happens in the transverse sector. 
Despite the strong similarities with the transverse sector discussed in the previous section, one important difference appears. As we can see in fig.\ref{full2}, in this case, at zero or very small EXB, there is an additional hydrodynamic mode. Such a mode represents the crystal diffusion mode discussed and observed in \cite{Delacretaz:2017zxd,Andrade:2017cnc} and it appears to be completely decoupled from the dynamics of the other modes. In absence of explicit breaking, its dispersion relation is of the type:
\begin{equation}
    \omega\,=\,-\,i\,D_\phi\,k^2\,
\end{equation}
and it represents an additional hydrodynamic mode, corresponding to the conservation of the phase of the Goldstone modes. What happens to this mode by dialing the explicit breaking parameter $\alpha$, or in other words in the pseudo-spontaneous regime? The motivation behind this question is that in presence of an honest phase relaxation mechanism, such a mode should acquire a finite damping:
\begin{equation}
    \omega\,=\,-\,i\,\Omega_{rel}\,-\,i\,D_\phi\,k^2 \label{tt}
\end{equation}
where $\Omega_{rel}$ relates to the relaxation of the phase of the Goldstone modes \cite{Delacretaz:2017zxd}. In \cite{Ammon:2019wci}, it has been argued that in the pure SSB this is not the case. This suggests that no phase relaxation mechanism driven by topological defects or the proliferation of dislocations is present in this and similar models \cite{Amoretti:2018tzw,Andrade:2018gqk}. Nevertheless, several works \cite{Ammon:2019wci,Andrade:2018gqk,Amoretti:2018tzw,Donos:2019txg}, noticed that in the pseudo-spontaneous regime a novel relaxation parameter $\bar{\Omega}$, playing an analogous role to $\Omega_{rel}$, appears. It is therefore valuable to test these assumptions here, by varying the parameter $\alpha$ in the pseudo-spontaneous regime. In summary, we want to prove if the relaxation time for the crystal diffusion mode is given by the same relaxation scale appearing in the pseudo-phonons dynamics, $  \Omega_{rel}\,\overset{?}{=}\,\bar{\Omega}\,\sim\,\frac{\langle EXB \rangle}{\langle SSB \rangle}$,
which vanishes for zero explicit breaking. More precisely, we fix the SSB parameter $\beta$ and we increase the EXB, $\alpha$, from zero to a finite and small value. The results are presented in fig.\ref{test}. The explicit breaking is increased from panel a) to panel h). In panel a), $\alpha=0$, and the breaking is purely spontaneous. The mode displayed with the black colour is the crystal diffusion mode. The other mode is the longitudinal sound.
In panel b) we introduce a small amount of EXB and we notice that the crystal diffusion mode does not acquire any damping. On the contrary, it is the sound mode that becomes softly gapped and damped. Moreover, the crystal diffusion mode and the sound mode now cross each other at a certain momentum. At this stage, we already notice that the phase relaxation rate $\bar{\Omega}$, proportional to the EXB, does not imply a simple damping term for the crystal diffusion mode.\\ Increasing further the EXB scale $\sim \alpha$, the dynamics become more complicated and, at a certain value $\alpha^\star$, the sound mode splits into two new modes (see panel (g)). One of the two modes, produces the $k-$gap dynamics appearing in panel h). The other one becomes a pseudo-diffusive mode with zero real part, whose damping increases with the EXB. At very large of the EXB, this last mode decouples from the low energy dynamics and what we are left with is just the simple $k-$gap dynamics that we expect in the purely explicit regime.
\begin{figure}[H]
   \centering
   \includegraphics[width=5cm]{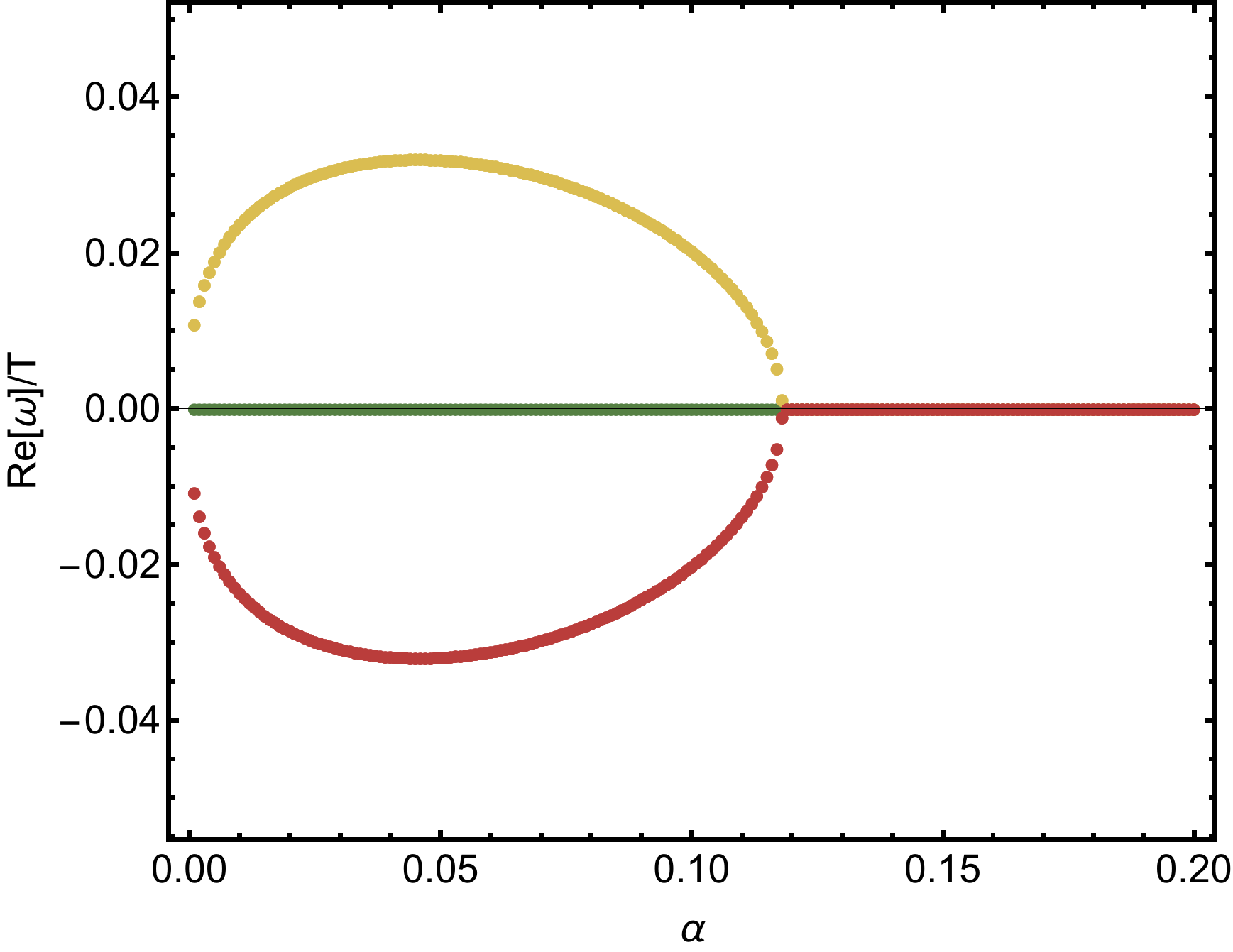}\quad  \includegraphics[width=5cm]{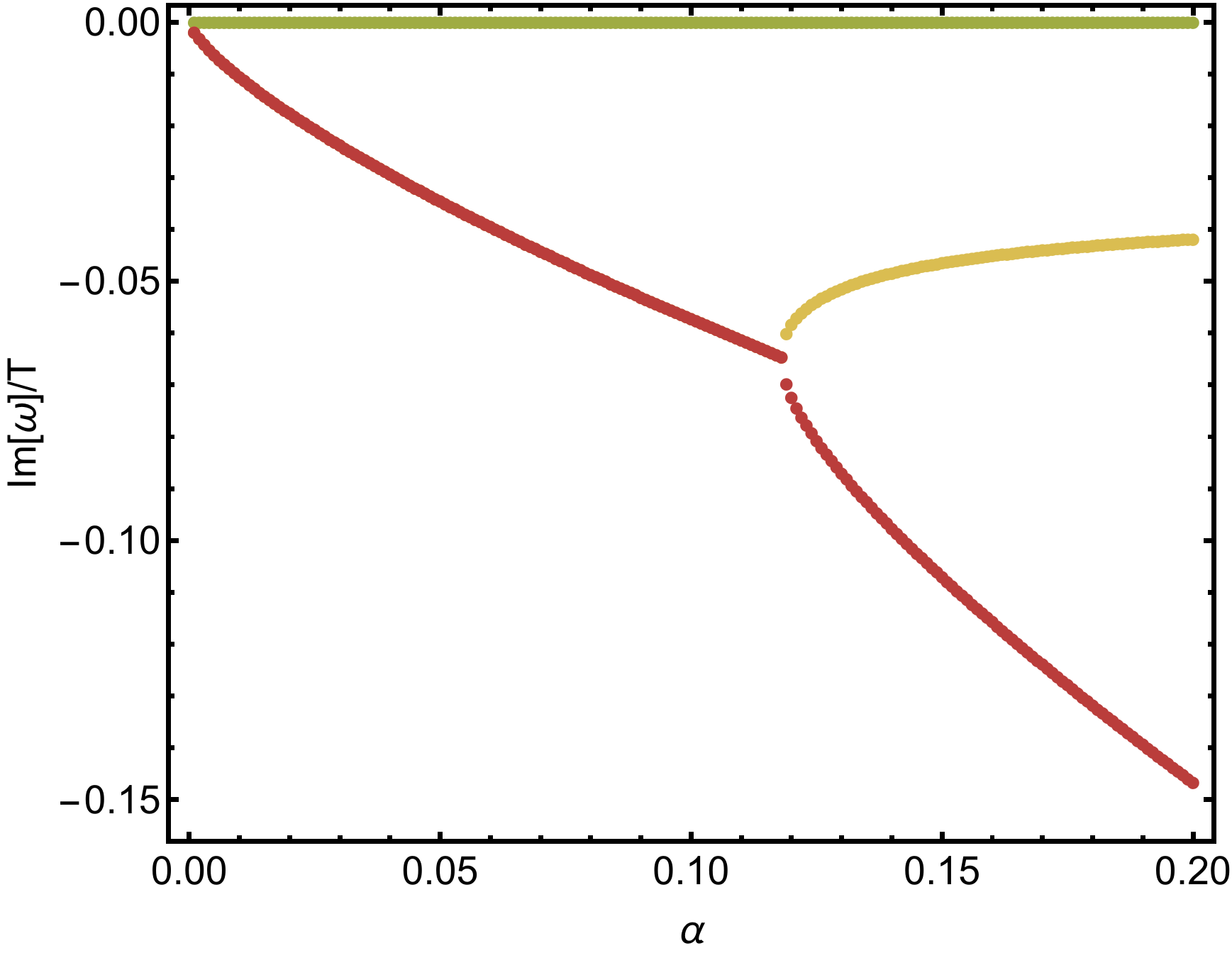}
        \caption{Lowest modes in the longitudinal sector of the fluid model $V(Z)=\alpha Z+ \beta Z^2$. The parameter are fixed to $m/T=0.1,\ \beta=5$ and the EXB scale $\sim \alpha$ varies.}
        \label{pp}
\end{figure}
The behaviour just described is more evident in fig.\ref{pp}.\\
In summary, our numerical results prove that the dynamics of the crystal diffusion mode is complicated and highly entangled with the dynamics of the longitudinal sound.
The dynamics of the crystal diffusion mode is not described by the simple expression:
\begin{equation}
    \omega\,=\,\,\underbrace{-\,\bar{\Omega}}_{\sim \langle EXB \rangle}\,-\,i\,D_\phi\,k^2\,+\,\dots
\end{equation}
as hinted in \cite{Delacretaz:2017zxd}. Additionally, our results are at odds with the idea presented in \cite{Amoretti:2018tzw,Andrade:2018gqk} that the novel relaxation scale $\bar{\Omega}$ plays the same role of the proper phase relaxation $\Omega$ given by the topological elastic defects. We do not believe that the full dynamics of the system can be understood by just adding various contributions to the total phase relaxation rate as:
\begin{equation}
    \Omega_{rel}\,=\,\underbrace{\Omega}_{\sim\,\langle SSB\rangle}\,+\,\underbrace{\bar{\Omega}}_{\sim\,\langle EXB\rangle}\,+\,\dots
\end{equation}
\section{Longitudinal pseudo-Goldstones in holographic solids}
In this last section we look back at the solid model described in the previous literature. In particular, we consider the solid potential
\begin{equation}
    V(X,Z)\,=\,\alpha\,X\,+\,\beta\,X^N\,,\quad N>5/2
\end{equation}
which was discussed in \cite{Alberte:2017cch,Ammon:2019wci}. At large $\beta \gg 1$, this setup realizes the pseudo-spontaneous breaking of translational invariance and it displays a light gapped and damped transverse pseudo-phonon mode. Here, we aim to complete its description, by discussing in detail the longitudinal spectrum of the collective modes.\\
At $\beta=0$, the breaking is purely explicit and the collective modes have been analyzed in \cite{Davison:2014lua}. At $\alpha=0$, on the contrary, the breaking is purely spontaneous and it has been recently analyzed in detail in \cite{Ammon:2019apj}. Here, we are interested in the interplay and the crossover between these two situations.\\

\begin{figure}[h]
    \centering
    \includegraphics[width=6.5cm]{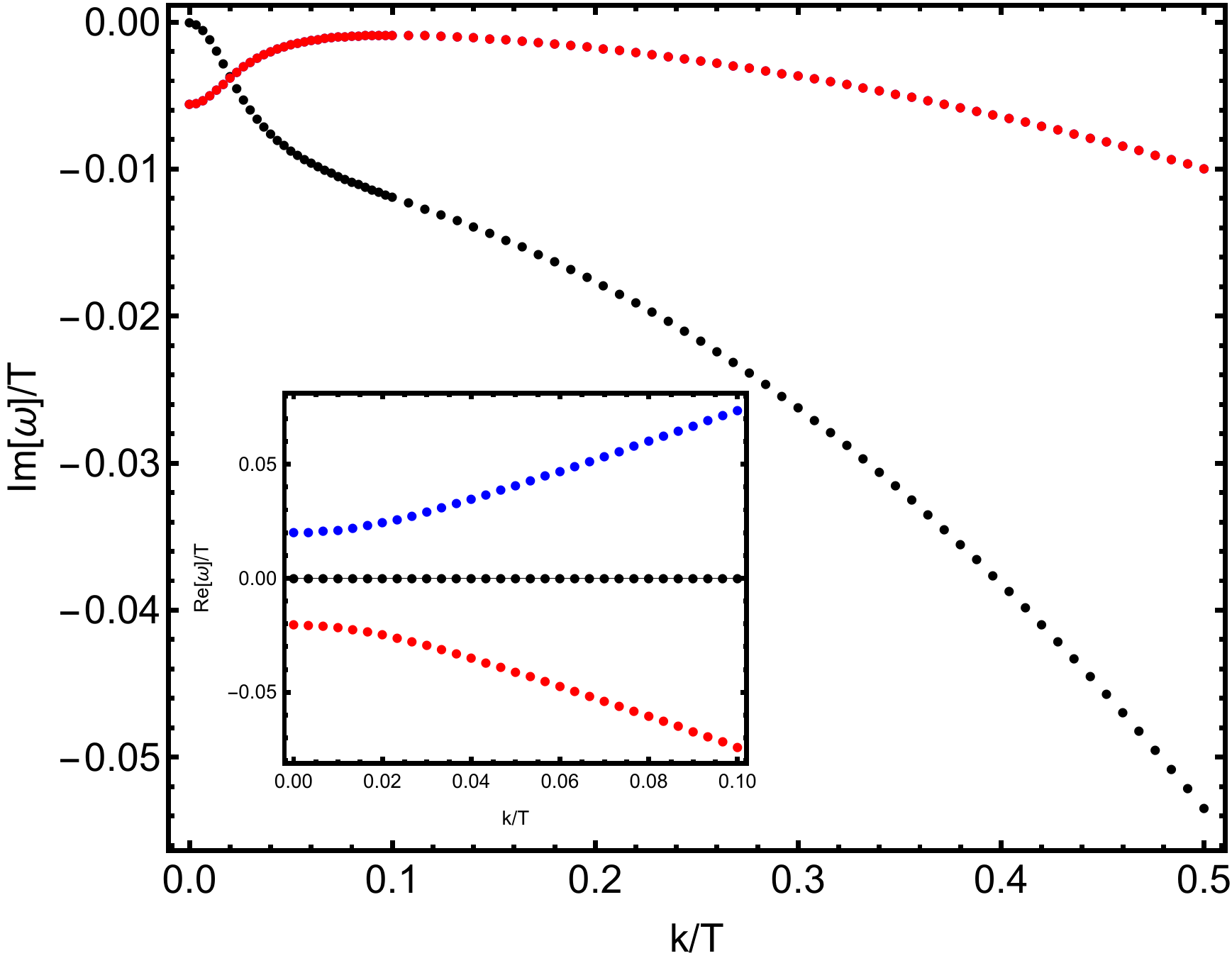}\quad \includegraphics[width=6.5cm]{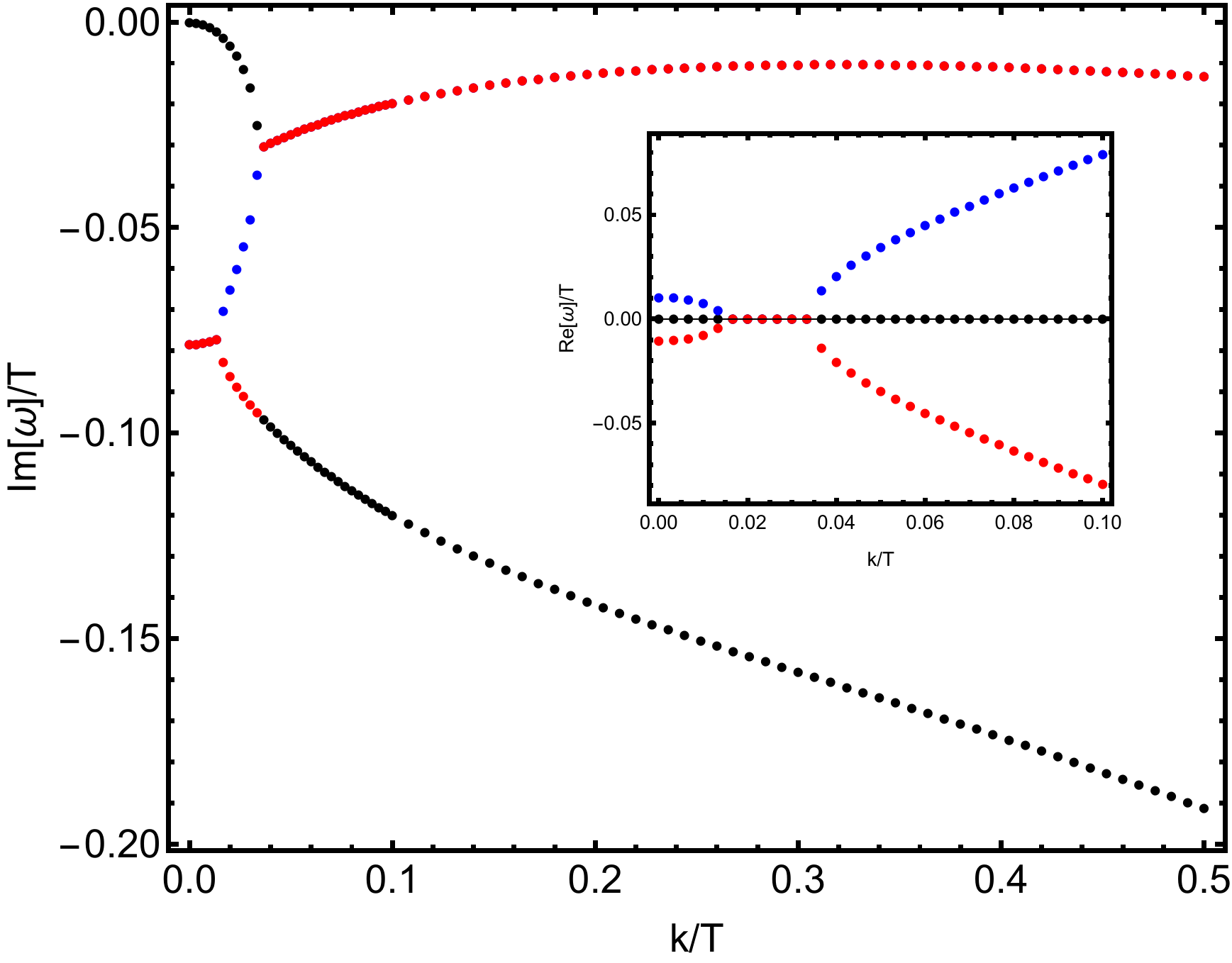}\\
      \includegraphics[width=6.5cm]{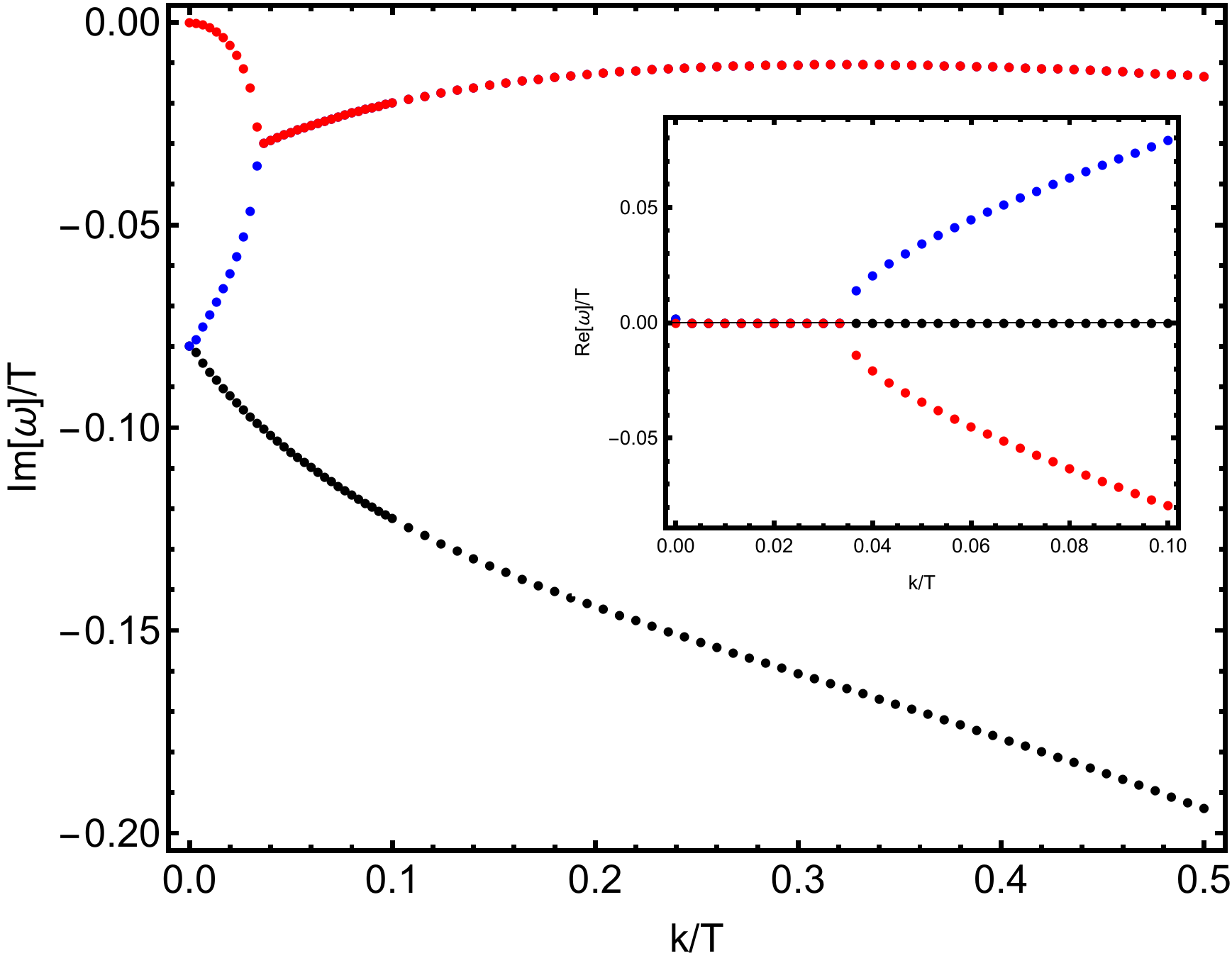}\quad \includegraphics[width=6.5cm]{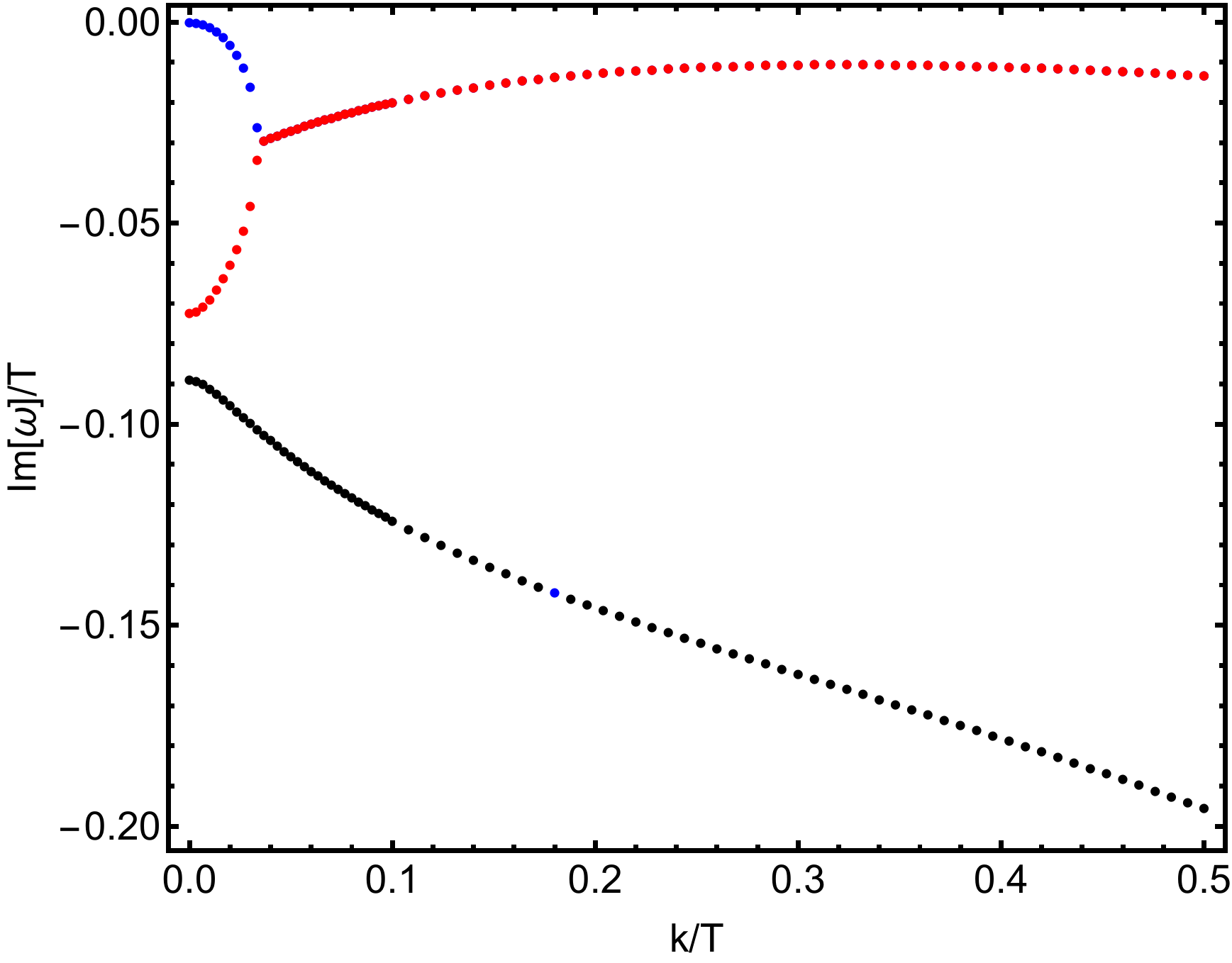}
    \caption{The lowest modes in the longitudinal sector of the solid model $V(X)=\alpha X+\beta X^5$ for $\beta=5, m/T=0.1$ and $\alpha \in [0.0005,0.0412]$ (from top to bottom).} 
    \label{finalf}
\end{figure}

With no surprise, the dynamics of the longitudinal collective modes is completely analogous to the fluid case discussed in the previous section. Given the qualitative similarities with the fluid model, we just remind here of the main features:
\begin{itemize}
    \item The dynamics of the lowest modes is perfectly described by the hydrodynamic formula \eqref{modes}. Two modes, governing by the momentum relaxation rate $\Gamma$ and the novel relaxation scale $\bar{\Omega}$, collide on the imaginary axes and produce two off-axes modes. Those are exactly the light pseudo-phonons expected by symmetry arguments.
    \item The Gell-Mann-Oakes-Renner (GMOR) relation holds true.
    \item The novel relaxation time scale obeys the relation:
    \begin{equation}
        \bar{\Omega}\,\sim\,\frac{\langle EXB \rangle}{\langle SSB \rangle }
    \end{equation}
    as claimed recently in \cite{Ammon:2019wci}.
\end{itemize}
At the same time, we analyzed the dynamics of the crystal diffusion mode in the pseudo-spontaneous regime. Also in this case, we find results which are completely analogous to those of the previous section as shown in fig.\ref{finalf}.
\\
All in all, this study of the solid model,in the pseudo-spontaneous regime, confirms the original idea that fluids and solids do not qualitative differ within the longitudinal sector. \\
Moreover, the results in the solid model give further evidence regarding the nature of the novel phase relaxation mechanism encoded in the $\bar{\Omega}$ parameter. They confirm that the involved dynamics can not be simply understood and explained by using the hydrodynamic framework of \cite{Delacretaz:2017zxd} nor the idea suggested in \cite{Amoretti:2018tzw,Andrade:2018gqk} of just replacing $\Omega \rightarrow \bar{\Omega}$.
\section{Conclusions}
\begin{figure}[h]
    \centering
  \tcbox[colframe=blue!30!black,
           colback=white!30]{ \includegraphics[width=0.6\linewidth]{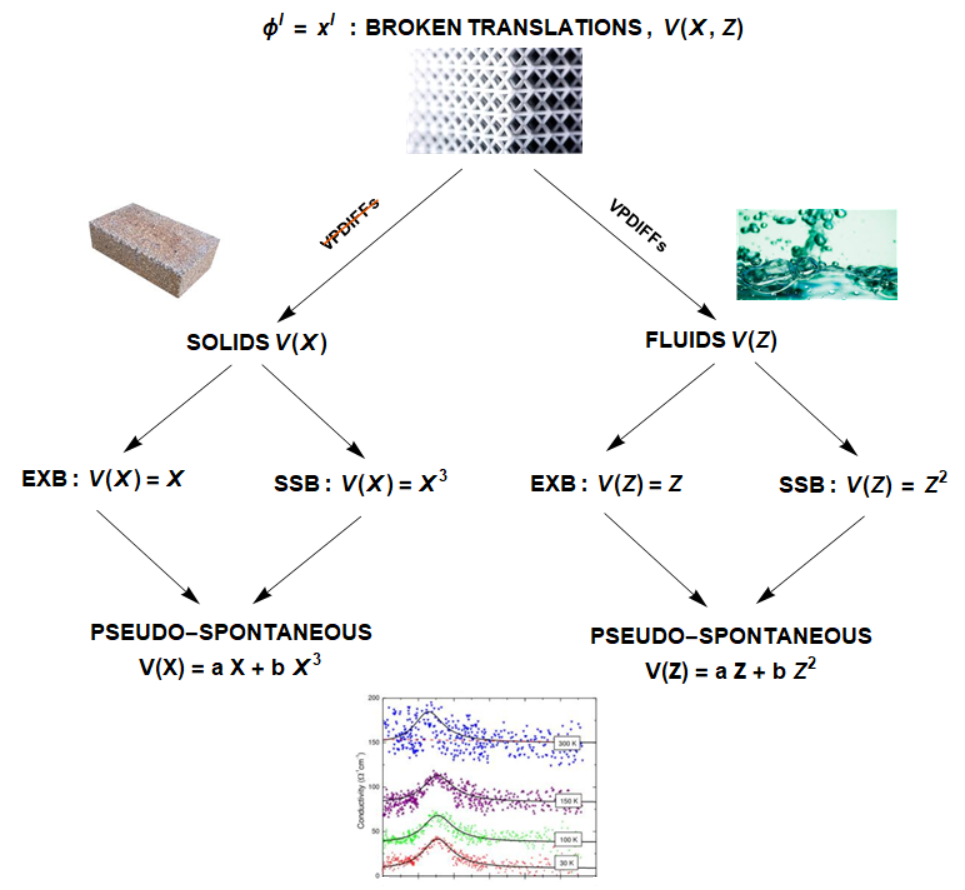}}
    \caption{A schematic diagram of all the holographic massive gravity models with broken translations. The right branch of this diagram have been completed in this manuscript.}
    \label{lastfig}
\end{figure}

In this work we provide a detailed and comprehensive description of simple holographic models with broken translations (see the schematic representation in fig.\ref{lastfig}). We complete the previous analysis by studying models with fluid symmetry, \textit{i.e.} invariant under internal volume preserving diffeomorphisms. This last class has not received lot of attention in the literature so far, but it certainly exhibits interesting properties, briefly analyzed in \cite{Alberte:2015isw,Alberte:2016xja}. We focus on the dynamics of the collective modes (quasinormal modes) in both the transverse and longitudinal sectors.\\

First, we analyze the QNMs of two fluid models exhibiting respectively explicit and spontaneous breaking of translational invariance. Our main findings are:
\begin{enumerate}
    \item In both fluid models the shear modulus is zero ($G=0)$ and the viscosity-to-entropy ratio saturates the KSS bound $\eta/s=1/4\pi$, as already observed in \cite{Alberte:2016xja}. This is the consequence of the vanishing of the mass for the helicity-2 graviton component. It also indicates that momentum dissipation does not necessarily induce the violation of the KSS bound.
    \item In case of EXB (section \ref{fluidEXB}), the dynamics of the fluid model at small frequencies and momenta ($\omega/T, k/T \ll 1$) is very similar to the solids described in \cite{Davison:2014lua}. Both a $k-$gap phenomenon \cite{Baggioli:2018vfc} and a coherent-incoherent transition \cite{Kim:2014bza} appear. Nevertheless, the fluid model displays a more complex structure beyond the hydrodynamic limit, which is due to its higher derivative nature compared to the solid counterpart.
    \item In the case of SSB (section \ref{fluidSSB}), the distinction between fluid and solid is much more evident. In the fluid model, no propagating transverse shear waves are present in the hydrodynamic regime. The speed of transverse sound is simply zero because of the absence of a finite shear modulus. The transverse sector dynamics is similar to what observed in simple relativistic hydrodynamics: a single and simple shear diffusion mode. The longitudinal spectrum contains a sound mode and a crystal diffusion mode like the solid counterpart. Nevertheless, the speed of the longitudinal sound mode is constant $v_L^2=1/2$ and surprisingly identical to the conformal field theory result \cite{Policastro:2002tn}.
\end{enumerate}
The fluid model with SSB of translations is interesting and it presents some points for discussion. What is the difference between this model and the relativistic hydrodynamics dual to the Schwarzschild background \cite{Policastro:2002se,Policastro:2002tn} ? The hydrodynamic modes are exactly the same. Nevertheless, in this case the diffusive mode in the transverse spectrum has to be thought as a Goldstone mode. More precisely, it corresponds to a double pole, a couple of transverse phonons with zero propagating speed. It would be interesting to understand these modes better in terms of diffusive Goldstone bosons. As discussed in \cite{Minami:2018oxl}, in dissipative systems, the appearance of such modes can be understood in terms of standard field theory. The same action, constructed using the determinant $Z$, has already been used as an EFT for dissipative fluids in \cite{deBoer:2015ija,Crossley:2015evo,Nicolis:2015sra,Endlich:2012vt,Dubovsky:2011sj}. A more detailed comparison to our results would be definitely valuable.\\

 We then proceed by considering a fluid model which display the interplay between the EXB and the SSB of translations. We discuss in detail both the transverse and longitudinal sectors. Our main findings can be summarized in:
 \begin{enumerate}
     \item Both the transverse and longitudinal sectors display the presence of light pseudo-phonons, as expected by symmetry arguments.
     \item The appearance of the pseudo-Goldstone modes is produced by the collision of two purely immaginary poles. One of them is controlled by the momentum relaxation rate $\Gamma$, while the other by the novel relaxation scale $\bar{\Omega}$.
     \item The pseudo-Goldstone bosons obey the Gell-Mann-Oakes-Renner (GMOR) relation.
     \item The novel relaxation scale satisfies the scaling relation:
     \begin{equation}
         \bar{\Omega}\,\sim\,\frac{\langle EXB \rangle}{\langle SSB \rangle}
     \end{equation}
     suggested in \cite{Ammon:2019wci}, where $\langle EXB \rangle,\langle SSB \rangle$ are respectively the explicit and spontaneous scales.
 \end{enumerate}
 In addition we confirm the validity of the universal relation:
 \begin{equation}
        \bar{\Omega}\,\sim\,\mathrm{M}^2\,\xi\,\sim\,\frac{\omega_0^2\,\chi_{PP}}{G}\,\xi
    \end{equation}
    proposed in \cite{Amoretti:2018tzw,Andrade:2018gqk}. which might be a promising hint towards a fundamental understanding of this novel phase relaxation dynamics.\\

Interestingly, the interplay of SSB and EXB, produces softly gapped and damped pseudo-phonons even when the original Goldstone boson is purely diffusive and not propagating. Our numerics suggest that a small propagating speed for the pseudo-phonons appears and it relates to the EXB scale. We are not aware of any field theory computation confirming this picture. It would be interesting to add a source of EXB in the analysis of \cite{Minami:2018oxl} and check the dispersion relation of the expected pseudo-Goldstone modes.\\

 Finally, we discuss the longitudinal sector of solid and fluid holographic models in the pseudo-spontaneous regime. With no big surprise, the results are qualitative identical for the two cases.
 \begin{figure}[h]
     \centering
     \includegraphics[width=0.5\linewidth]{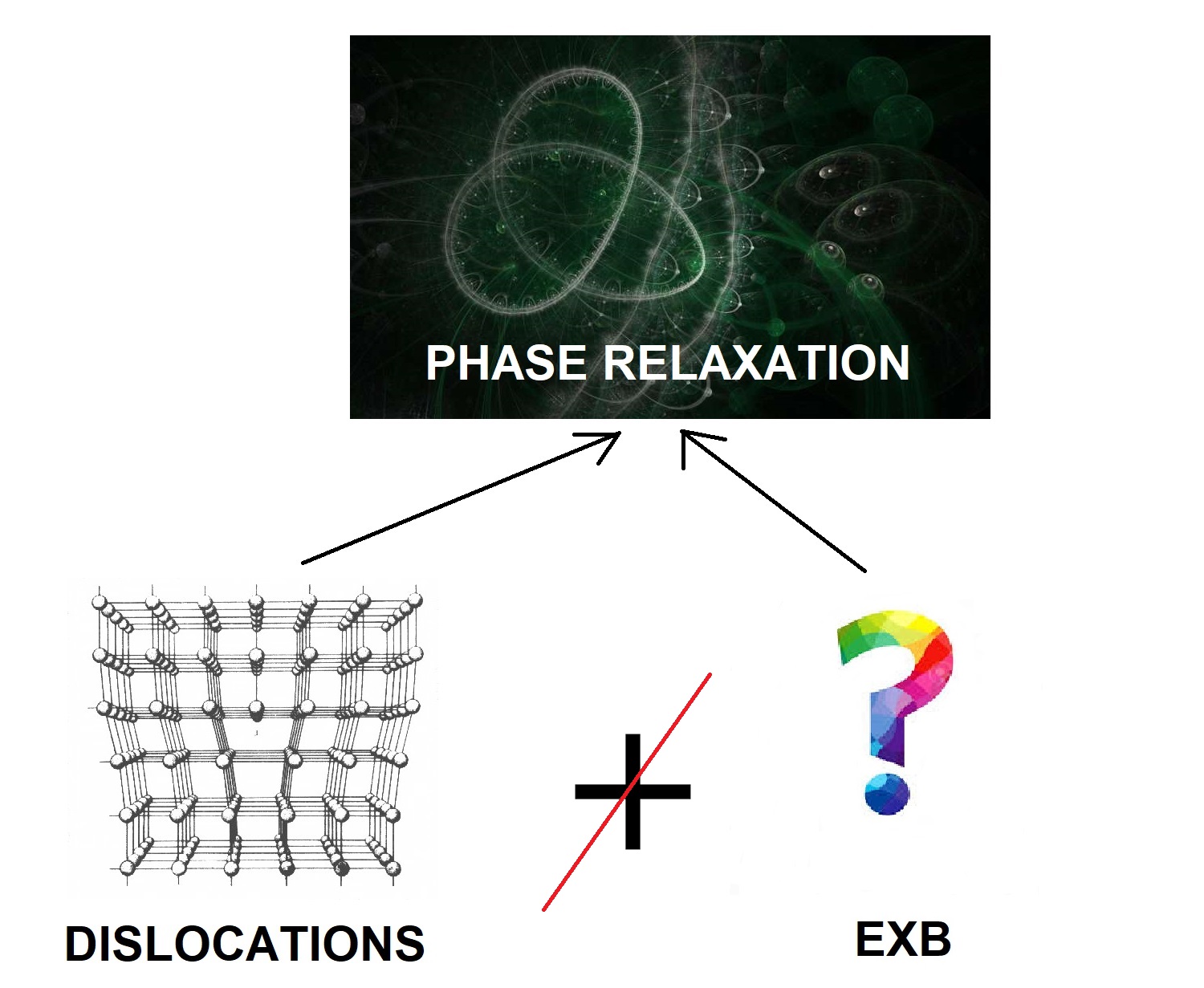}
     \caption{Two possible sources for phase relaxation. 1) the presence and proliferation of topological defects, such as dislocations, as contemplated in \cite{Delacretaz:2017zxd}. 2) the interplay between spontaneous and explicit breaking of the symmetry, observed in the holographic models \cite{Ammon:2019wci,Alberte:2017cch,Andrade:2018gqk,Amoretti:2018tzw}. The microscopic mechanism and the hydrodynamic description of the second scenario are still unknown. Our results suggest that the two mechanism can not be simply added together into the total phase relaxation rate, since their dynamics appears quite different.}
     \label{fig:ultimo}
 \end{figure}\\
 Our numerical results show that the dynamics of the crystal diffusion mode in the pseudo-spontaneous regime is quite complex and not consistent with the simple expression:
 \begin{equation}
     \omega\,=\,-\,i\,\Omega_{rel}\,-\,i\,D_\phi\,k^2\label{wow}
 \end{equation}
 where $\Omega_{rel}$ is the total relaxation rate of the Goldstone phase. In particular, the crystal diffusion mode does not immediately acquire a finite damping term as a consequence of the introduction of a small EXB source. The full phenomenology involves a non trivial interplay between the crystal diffusion mode and the longitudinal sound, which clearly deserves more investigation. The numerical results can not be explained simply by assuming a simple sum of the various phase relaxation mechanisms as suggested in \cite{Andrade:2018gqk,Amoretti:2018tzw}. At this stage, no complete hydrodynamic framework is able to describe successfully our observations is present in the literature.\\

Let us add some more remarks about the nature of the relaxation time scale $\bar{\Omega}$. This quantity originates from the St\"uckelberg sector and it is related to the explicit breaking of the global internal shift symmetry:
\begin{equation}
    \phi^I\,\,\,\rightarrow\,\,\,\phi^I\,+\,c^i
\end{equation}
where $c^I$ is just a constant shift. This represents another proof of the uncorrelation between this scale and the presence of topological elastic defects. As explained in \cite{Grozdanov:2018ewh}, the presence of dislocations or defects in the EFT will break the conservation of the higher form currents:
\begin{equation}
    J_I^{n_1\,,\dots,\,n_d}\,\equiv\,\epsilon^{n_1\,,\dots,\,n_d,\,\nu}\,\partial_\nu \phi_I \label{break}
\end{equation}
whose associated charge:
\begin{equation}
    \mathcal{N}\,=\,\int_{S^1}\,\star\,J
\end{equation}
represents indeed the density of defect lines. This density is clearly zero in our model and in the models of \cite{Amoretti:2018tzw,Andrade:2018gqk}.\\
 
 It would be interesting now to try to use the framework discussed in a more phenomenological way to test the proposal of \cite{Delacretaz:2016ivq} explaining the electric transport properties of bad metals via the interplay of EXB and SSB and to investigate in more detail the relation between these models and the physics of the Boson peak and glassy dynamics \cite{Baggioli:2018qwu,baggioli2018soft,baggioli2019hydrodynamics}. These models could possibly be relevant for the study of phonons at quantum criticality, as recently discussed in the condensed matter community \cite{2018arXiv180906495P,2019arXiv190200516S}.
 
One immediate possible direction is to break isotropy with the introduction of an external magnetic field. The interplay between the external source and the dynamical phonons should provide interesting outcomes as discussed in \cite{Delacretaz:2019wzh}. Finally, one could study time dependent configurations which might be relevant for viscoelastic materials, extending the study of \cite{Baggioli:2018bfa,Andrade:2019zey}. Moreover, one could also consider more complicated spatial configurations for the scalars $\phi^I$, like for example vortex configurations relevant for the study of elastic defects, such as dislocations \cite{Beekman:2016szb}.\\

We leave these questions for the near future.
\section*{Acknowledgements}
We thank Tomas Andrade, Daniel Arean, Ulf Gran, Marcus Torns\"{o} and Tobias Zingg for conversations linked to the topic of this manuscript.
We thank Martin Ammon, Tomas Andrade, Victor Cancer Castillo, Sean Gray, Amadeo Jimenez, Weijia Li and Oriol Pujolas for fruitful collaborations on related projects and always useful comments and discussions. We express our gratitude to Martin Ammon, Amadeo Jimenez and Oriol Pujolas for reading a preliminary version of the manuscript. M.B. acknowledges the support of the Spanish MINECO’s “Centro de Excelencia Severo Ochoa” Programme under grant SEV-2012-0249. S.G. gratefully acknowledges financial support by the DAAD (German Academic Exchange Service) for a \textit{Jahresstipendium f\"ur Doktorandinnen und Doktoranden}  (One-Year Research grant for doctoral candidates). M.B. thanks the International Institute of Physics IIP of Natal, Brazil, for the (very) warm hospitality during the completion of this work.
\appendix
\section{Equations of motions}\label{app1}
In this appendix we provide the detail of the computations performed in the main text. To keep the equations simple, we will not consider the most general case $V(X,Z)$, but only the two separate cases $V(X),V(Z)$. We consider the momentum aligned in the $y$ direction, $\vec{k}=(0,0,k)$. For a more complete analysis see \cite{Alberte:2015isw}.
\subsection*{Transverse sector}
In the transverse sector we consider the following set of bulk fluctuations:
\begin{equation}
    \{\delta \phi_x,\,h_{tx},\,h_{xy}\}
\end{equation}
where we assumed the radial gauge $h_{xu}=0$.\\[0.25cm]\newpage
\textbf{V(Z) case}\\[0.25cm]
Assuming the fluid potential $V(Z)$, we obtain the following equations of motion:
\begin{align}
   & V'\,\left( \delta\phi_x'\, \left(f'+2\, i\, \omega \right)+f \delta\phi_x''+h_{tx}'\right)+4 u^3\, V'' \left(f\, \delta\phi_x'+h_{tx}+i\, \omega\,  \delta\phi_x\right)=0\\
    & h_{tx} \left(2\, u\, f'-6 f+k^2 u^2+4 m^2 u^4\, V'-2 m^2\, V+6\right)\nonumber\\&+u \left(-u\, f\, h_{tx}''+2 f\, h_{tx}'+k\, u\, \omega\,  h_{xy}+4\, i\, m^2\, u^3\, \omega\,  \delta\phi_x'\, V'-i\, u\, \omega\, h_{tx}'\right)=0\\
 &u \left(h_{xy}' \left(-u\, f'+2\, f-2\, i\, u\, \omega \right)-u f h_{xy}''-i\, k\, u\,h_{tx}'\right)\nonumber\\&+2 h_{xy} \,\left(u\, f'-3 f-m^2\, V+i\, u\, \omega +3\right)+2\, i\, k\, u\, h_{tx}=0\\
 &i\, k\, u\, h_{xy}'-u \left(4\, m^2\, u^2\, \delta \phi_x'\, V'+h_{tx}''\right)+2 h_{tx}'=0
\end{align}
where we use the short notation $V'\equiv d V(Z)/dZ,\,V'' \equiv d^2 V(Z)/dZ^2$ and we have omitted the arguments of the various functions.\\[0.25cm]
\textbf{V(X) case}\\[0.25cm]
For the solid potential $V(X)$, we obtain the following equations of motion:
\begin{align}
&-2(1-u^2\,V''/V')h_{tx}+u\,h_{tx}'-i\,k\,u\, h_{xy}-\left( k^2\,u+2\,i\,\omega (1-u^2\,V''/V')\right)\,\delta \phi_x+u\,f\, \delta \phi_x''
\nonumber\\&+\left(-2(1-u^2\,V''/V')\,f+u\,(2 i \omega+f')\right)\delta \phi_x'
\,=0\,;\\
&2\,i\,m^2\,u^{2}\omega V'\,\delta \phi_x+u^2\, k\,\omega\, h_{xy}+(6+k^2\,u^2-2 \, m^2 (V-u^2\,V' ) \,\nonumber\\& -6 f+2uf') \, h_{tx}+\left(2\,u\,f-i\,u^2\omega\right)h_{tx}'-u^2\,f\,h_{tx}''\,=\,0 \,\,;\\
&2i\,k\,u\,h_{tx}-iku^2h_{tx}'-2\,i \, k \,m^2 \,u^{2}V'\delta \phi_x+2 h_{xy}\left(3+i\,u \, \omega-3f+uf'- \,m^2(V-u^2 V') \right)
\nonumber\\&-\left(2i\,u^2 \, \omega-2uf+u^2\,f'\right)h_{xy}'-u^2\,f\, h_{xy}''\,=\,0\,;\\
&2\,h_{tx}'-u\,h_{tx}''-2m^2\,u\,V' \, \delta\phi_x'+ik\,u\,h_{xy}'\,=\,0
\end{align}
where this time $V'\equiv d V(X)/dX,\,V'' \equiv d^2 V(X)/dX^2$.
\subsection*{Longitudinal sector}
In the longitudinal sector we consider the following set of bulk fluctuations:
\begin{equation}
    \{h_{x,\bm{s}}=1/2\,(h_{xx}+h_{yy}),\,h_{x,\bm{a}}=1/2\,(h_{xx}-h_{yy}),\,\delta\phi_y,\, h_{tt},\, h_{ty}\}
\end{equation}
where we again assumed the radial gauge:
$h_{\mu u}=0$, with $\mu\in\{t,u,x,y,z\}$. \\[0.25cm]
\textbf{V(Z) case}\\[0.25cm]
For the fluid potential $V(X)$ we obtain:
\begin{align}
 &  f'\,\delta\phi_y'\, V'+f\delta\phi_y''\, V'+4 u^3\, f\delta\phi_y'\, V''-\delta\phi_y \left(k^2\, V'+2 u^3\, \left(k^2\, u-2\, i\, \omega \right) V''\right)\nonumber\\
 &-i\, k\, h_{xx}\, \left(2 u^4\, V''+V'\right)+h_{ty}'\, V'+2\, i\, \omega\,  \delta\phi_y'\, V'+4\, u^3\, h_{ty}\, V''=0\\ &
 2 h_{ty} \left(-u f'+3 f-2 m^2 u^4 V'+m^2 V-3\right)+u (u f h_{ty}''+(-2 f+i\, u\, \omega )\, h_{ty}'+i\, k\, u\, h_{tt}'\nonumber \\ &
 -4\, i\, m^2\, u^3 \omega  \delta \phi_y\, V')+k\, u^2\, \omega\,  (h_{x,\bm{s}}+h_{x,\bm{a}})-2\, i\, k\, u\, h_{tt}=0\\
 &6\, h_{tt}+u\, \left(-u\, f'\, h_{x,\bm{s}}'+2\, f\, h_{x,\bm{s}}'+4\, i\, k\, m^2\, u^3\, \delta \phi_y \left(2\, u^4\, V''+V'\right)-i\, k\, u\, h_{ty}'+2\, i\, k\, h_{ty}\right.\nonumber\\
 &\left. -8\, m^2\, u^7\, h_{x,\bm{s}}\, V''+2\, i\, h_{x,\bm{s}} \left(\omega +2\, i\, m^2\, u^3\, V'\right)+u\, h_{tt}''-4\, h_{tt}'-2\, i\, u\, \omega  h_{x,\bm{s}}'\right)=0\\
 &h_{x,\bm{s}}\, \left(2\, u\, f'-6\, f+k^2\, u^2+4\, m^2\, u^4\, V'-2\, m^2\, V+4\, i\, u\, \omega +6\right)-u^2\, f'\, h_{x,\bm{s}}'\nonumber\\ &
 -u^2\, f'\, h_{x,\bm{a}}'+2\, u\, h_{x,\bm{a}}\, f'-u^2\, f\, h_{x,\bm{s}}''-u^2\, f\, h_{x,\bm{a}}''+4\, u\, f\, h_{x,\bm{s}}'+2\, u\, f\, h_{x,\bm{a}}'-6\ f\, h_{x,\bm{a}}\nonumber\\&
 +k^2\, u^2\, h_{x,\bm{a}}-4\, i\, k\, m^2\, u^4\, \delta \phi_y\, V'+2\, i\, k\, u\, h_{ty}-2\, m^2\, h_{x,\bm{a}}\, V-2\, i\, u^2\, \omega\,  h_{x,\bm{s}}'\nonumber \\
 &-2\, i\, u^2\, \omega\,  h_{x,\bm{a}}'-2\, u\, h_{tt}'+6\, h_{tt}+2\, i\, u\, \omega  h_{x,\bm{a}}+6\, h_{x,\bm{a}}=0\\
 &h_{x,\bm{s}} \left(2 u f'-6\, f+k^2\, u^2+4\, m^2\, u^4\, V'-2\, m^2\, V+4\, i\, u\, \omega +6\right)-u^2 f'\, h_{x,\bm{s}}'+u^2\, f'\, h_{x,\bm{a}}'\nonumber \\
 &-2\, u\, h_{x,\bm{a}}\, f'-u^2\, f\, h_{x,\bm{s}}''+u^2\, f h_{x,\bm{a}}''+4\, u\, f\, h_{x,\bm{s}}'-2\, u\, f\, h_{x,\bm{a}}'+6\, f\, h_{x,\bm{a}}+k^2\, u^2\, h_{x,\bm{a}}\nonumber\\
 &-4\,i\, k\, m^2 \,u^4\, \delta \phi_y\, V'-2\, i\, k\, u^2\, h_{ty}'+6\, i\, k\, u\, h_{ty}+2\, m^2\, h_{x,\bm{a}}\, V-2\, i\, u^2\, \omega\,  h_{x,\bm{s}}'\nonumber\\&+2\, i\, u^2\, \omega\,  h_{x,\bm{a}}'-2\, u\, h_{tt}'+6\, h_{tt}-2\, i\, u \,\omega \, h_{x,\bm{a}}-6\, h_{x,\bm{a}}=0\\
 &2\, h_{ty}'-u\, \left(i\, k\, \left(h_{x,\bm{s}}'+h_{x,\bm{a}}'\right)+4\, m^2\, u^2\, \delta \phi_y'\, V'+h_{ty}''\right)=0\\
 &h_{x,\bm{s}}''=0
\end{align}
where $V'\equiv d V(Z)/dZ,\,V'' \equiv d^2 V(Z)/dZ^2$.\\[0.25cm]
\textbf{V(X) case}\\[0.25cm]
For the solid potential $V(X)$ we get:
\begin{align}
&u f'\, \delta \phi_y'\, V'\,+2\, u^2\, f\, \delta \phi_y'\, V''+u\, f\, \delta \phi_y''\, V'-2\, f\, \delta \phi_y'\, V'-k^2\, u\, \delta \phi_y\, V'-k^2\, u^3\, \delta \phi_y\, V''\nonumber\\
&+i\, k\, u\, h_{x,\bm{a}}\, V'-i\, k\, u^3\, h_{x,\bm{s}}\, V''+2\, i\, u^2\, \omega\,  \delta \phi_y\, V''+u\, h_{ty}'\, V'+2\, i\, u\, \omega\,  \delta \phi_y'\, V'\nonumber\\
&-2\, i\, \omega\,  \delta \phi_y V'-2\, h_{ty} \left(V'-u^2\, V''\right)=0\\
&u (f \left(u f'\, h_{x,\bm{s}}'-2\, f\, h_{x,\bm{s}}'-2\, i\, k\, m^2\, u^3\, \delta \phi_y\, V''-u\, h_{tt}''+4\, h_{tt}'\right)\nonumber\\
&+k\, h_{ty} \left(i\, u\, f'-2\, i\, f+2\, u\, \omega \right)+h_{x,\bm{s}} \left(2\, m^2\, u^3\, f\, V''+\omega\,  \left(i\, u\, f'-2\, i\, f+2\, u\, \omega \right)\right))\nonumber\\ 
&+h_{tt} \left(u \left(-u f''+4\, f'+2\, m^2\, u\, V'\right)-12\, f+k^2\, u^2-2\, m^2\, V-2\, i\, u\, \omega +6\right)=0\\
&2 h_{ty} \left(u \left(f'+m^2\, u\, V'\right)-3\, f-m^2\, V+3\right)\nonumber\\
&-u \left(u\, f\, h_{ty}''-2\, f\, h_{ty}'+i\, k\, u\, h_{tt}'+k\, u\, \omega\,  h_{x,\bm{s}}+k\, u\, \omega\,  h_{x,\bm{a}}-2\, i\, m^2\, u\, \omega\,  \delta \phi_y\, V'+i\, u\, \omega\,  h_{ty}'\right)\nonumber\\
&+2\, i\, k\, u\, h_{tt}=0\\
&h_{x,\bm{s}} \left(2\, u \left(f'+m^2\, u\, V'\right)-6\, f+k^2\, u^2-2\, m^2\, V+4\, i\, u\, \omega +6\right)-u^2\, f'\, h_{x,\bm{s}}'\nonumber\\
&-u^2\, f'\, h_{x,\bm{a}}'+2\, u\, h_{x,\bm{a}}\, f'-u^2\, f\, h_{x,\bm{s}}''-u^2\, f\, h_{x,\bm{a}}''+4\, u\, f\, h_{x,\bm{s}}'+2\, u\, f\, h_{x,\bm{a}}'-6\, f\, h_{x,\bm{a}}\nonumber\\
&+k^2\, u^2\, h_{x,\bm{a}}+2\, i\, k\, u\, h_{ty}+2\, m^2\, u^2\, h_{x,\bm{a}}\, V'-2\,m^2\, h_{x,\bm{a}}\, V-2\, i\, u^2\, \omega\,  h_{x,\bm{s}}'\nonumber\\
&-2\, i\, u^2\, \omega\,  h_{x,\bm{a}}'
-2\, u\, h_{tt}'+6 h_{tt}(u)+2\, i\, u\, \omega\,  h_{x,\bm{a}}+6 h_{x,\bm{a}}=0\\
&h_{x,\bm{s}} \left(2\, u \left(f'+m^2\, u\, V'\right)-6\, f+k^2\, u^2-2\,m^2\, V+4\, i\, u\, \omega +6\right)-u^2\, f'\, h_{x,\bm{s}}'+u^2\, f'\, h_{x,\bm{a}}'\nonumber\\
&-2\, u\, h_{x,\bm{a}}\, f'-u^2\, f\, h_{x,\bm{s}}''+u^2\, f\, h_{x,\bm{a}}''+4\, u\, f\, h_{x,\bm{s}}'-2\, u\, f h_{x,\bm{a}}'
+6\, f\, h_{x,\bm{a}}+k^2\, u^2\, h_{x,\bm{a}}\nonumber\\
&-4\, i\, k\, m^2\, u^2\, \delta \phi_y\, V'-2\, i\, k\, u^2\, h_{ty}'+6\, i\, k\, u\, h_{ty}-2\, m^2\, u^2\, h_{x,\bm{a}}\, V'+2\, m^2\, h_{x,\bm{a}}\, V\nonumber\\
&-2\, i\, u^2\, \omega\,  h_{x,\bm{s}}'+2\, i\, u^2\, \omega\,  h_{x,\bm{a}}'-2\, u\, h_{tt}'+6\, h_{tt}-2\, i\, u\, \omega\,  h_{x,\bm{a}}-6\, h_{x,\bm{a}}=0\\
&-6\, h_{tt}+u\, (u\, f'\, h_{x,\bm{s}}'-2\, f\, h_{x,\bm{s}}'-2\, i\, k\, m^2\, u^3\, \delta \phi_y\, V''+i\, k\, u\, h_{ty}'-2\, i\, k\, h_{ty}\nonumber\\
&+2\, m^2\, u^3\, h_{x,\bm{s}}\, V''-u\, h_{tt}''+4\, h_{tt}'+2\, i\, u\, \omega\,  h_{x,\bm{s}}'-2\, i\, \omega\,  h_{x,\bm{s}})=0\\
&k\, u \left(h_{x,\bm{s}}'+h_{x,\bm{a}}'\right)-i\, u\, \left(2\,m^2\, \delta \phi_y'\, V'+h_{ty}''\right)+2\, i\, h_{ty}'=0\\
&h_{x,\bm{s}}''=0,
\end{align}
where again $V'\equiv d V(X)/dX,\,V'' \equiv d^2 V(X)/dX^2$.
\section{Numerical techniques}\label{app2}

In this section we briefly review the numerical methods we applied to solve the equations of motions. For a detailed introduction see for example \cite{boyd2001chebyshev,Grandclement:2007sb}; we follow \cite{boyd2001chebyshev,Grandclement:2007sb,Grieninger:2017jxz,Ammon:2016fru,Kovtun:2005ev} Throughout this work we restrict ourselves to solving the linearized response of our system to small pertubations from the equilibrium state, the so called \textbf{Q}uasi \textbf{N}ormal \textbf{M}odes (QNMs). QNMs are the solutions to the linearized equations of motion of the scalar field and metric fluctuations subject to specific boundary conditions. On the one hand we do not allow for sources to the fluctuations which thus have to fulfill a Dirichlet boundary condition at the conformal boundary of our asymptotic AdS spacetime. On the other hand we subject the fluctuations to a ingoing wave condition at the horizon since classical black holes do not emit radiation. As a consequence of this choice the corresponding eigenvalue problem will be non-Hermitian resulting in complex eigenfrequencies. In order to solve this eigenvalue problem we apply so called pseudo-spectral methods which proved to be a very efficient and highly accurate approach.

\subsection*{(Pseudo-)Spectral methods}
 In order to solve the equations of motion we discretize them in the radial direction, using a Chebychev Lobatto grid with $N$ gridpoints. The assumption of spectral methods is that we may write the solutions to a given differential equation as a linear combination of basis functions; in our case we choose the so called Chebychev polynomials $T_k=\cos(k\,\arccos(x))$, where $x\in[-1,1]$ as basis functions. 
 In order to solve the equations by means of a pseudospectral method, we expand the unknown functions on the gridpoints in the Chebychev basis
 \begin{equation}
     X(z_i)=\sum_{k=0}^{N-1}c_i\,T_k(z_i).
 \end{equation}
 Note, that derivatives of the unknown functions simply translate to derivatives of the basis functions, which we know analytically $X'(z_i)=\sum_{k=0}^{N-1}c_i\,\left(\text{d}/\text{d}z\, T_k(z)\right)_{z=z_i}$. Plugging this in the discretized EOMs and collecting in powers of $\omega$ leads to the matrix valued equation of the form
 \begin{equation}
(\bm{A}\,\omega-\bm{B})\,\bm{x}=0,\label{mae}
 \end{equation}
 where $\bm{x}=\{\delta\phi_x,h_{tx},h_{xy}\}$ in the transversal cases and $\bm{x}=\{\delta\phi_y,h_{x,\bm{a}},h_{tt},h_{ty}\}$ in the longitudinal cases, respectively. We may solve this equation by solving the generalized eigenvalue problem, where the unknown functions are the eigenfunctions and the quasinormal modes $\omega_n$ are the eigenvalues of the matrix equation \eqref{mae}.
 
\subsection*{Convergence of the numerical methods}
In order to demonstrate that the numerical solution converges to the exact solution of the equation, we performed several consistency and accuracy checks. We present all the tests in the ``hardest"" numerical regime. The asymptotic expansion of the scalar field in the case of $V(Z)=Z^N$ is given by
\begin{align}
    \delta\phi_i=z^0\,(\phi_1z+\ldots)+z^{5-4N}(\phi_2+\ldots+\phi_\text{ll}\,\log(z)).
\end{align}
For $N=1$, we identify the $\phi_1$ with the source term $\phi_1\equiv\phi_{\bm{s}}$ and the second with the vacuum expectation value $\phi_2\equiv\phi_{\bm{v}}$, respectively; for $N=1$ however, we identify $\phi_2\equiv\phi_{\bm{s}}$ and $\phi_1\equiv\phi_{\bm{v}}$, respectively. Note, that we always have to set $\phi_{\bm{s}}$=0. If we choose the potential to be of the form $V(Z)=\alpha Z+\beta Z^2$, the asymptotic behavior is the same as for $N=1$; in the limit $\alpha\to 0$, we should recover the asymptotic behavior of $V(Z)=\beta\,Z^2$ which behaves as $\delta\phi_i\sim u^0$. In order to correctly resolve this behavior for a function $\tilde f=u\,f$, we need a lot of gridpoints around $u=0$.

\begin{figure}[h]
    \centering
    \includegraphics[width=4.5cm]{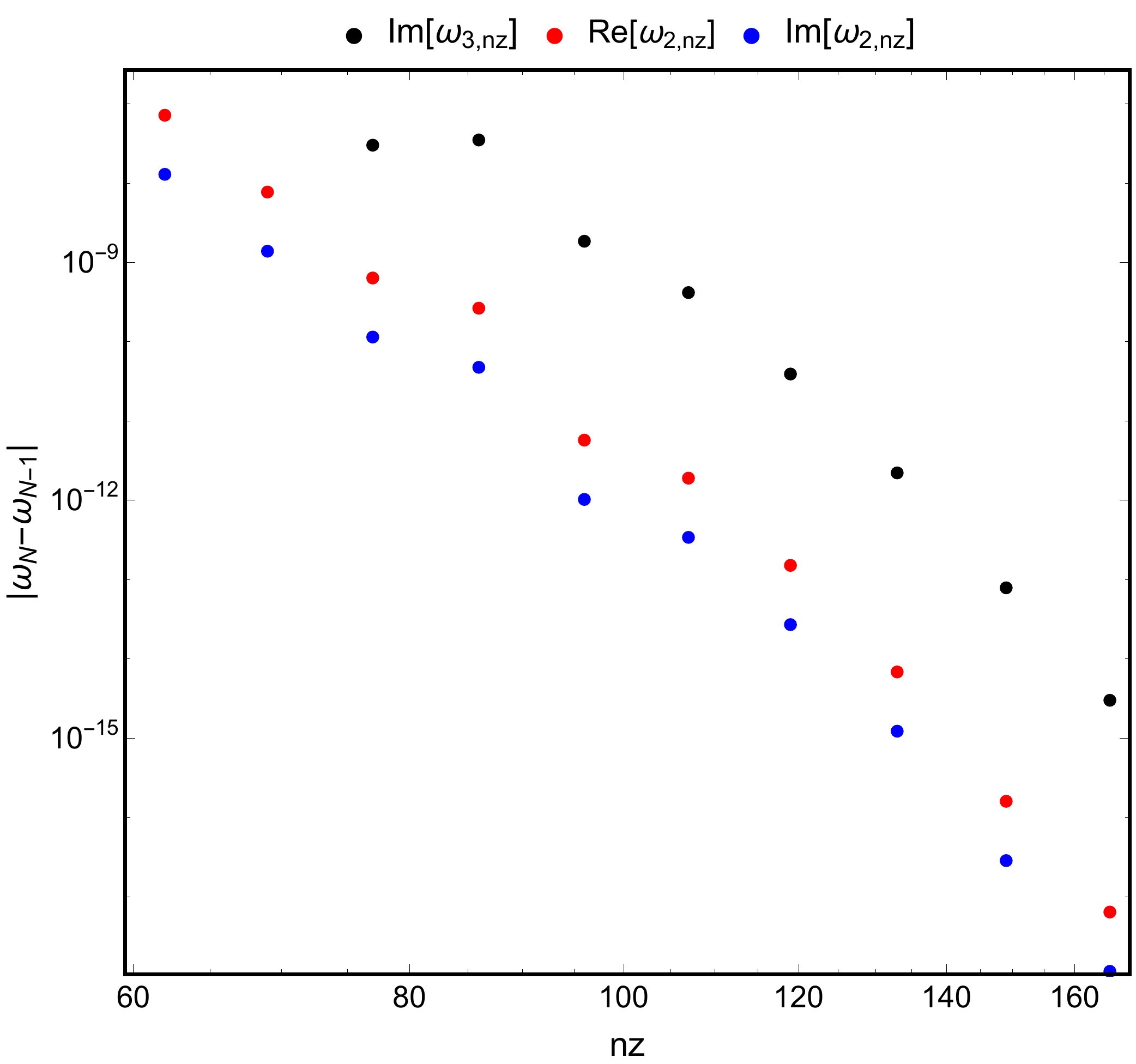}\quad \includegraphics[width=4.5cm]{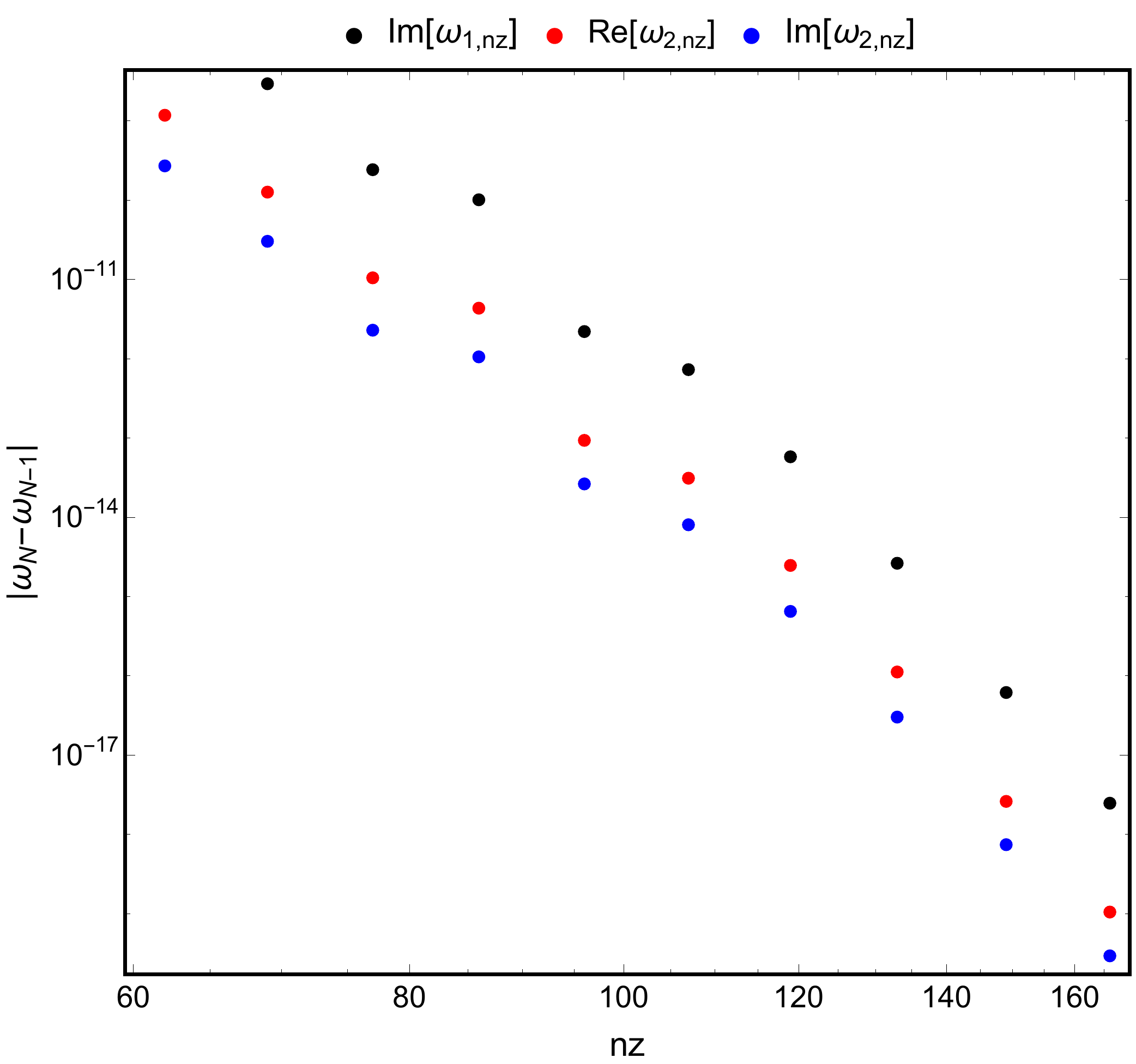}\quad
     \includegraphics[width=4.5cm]{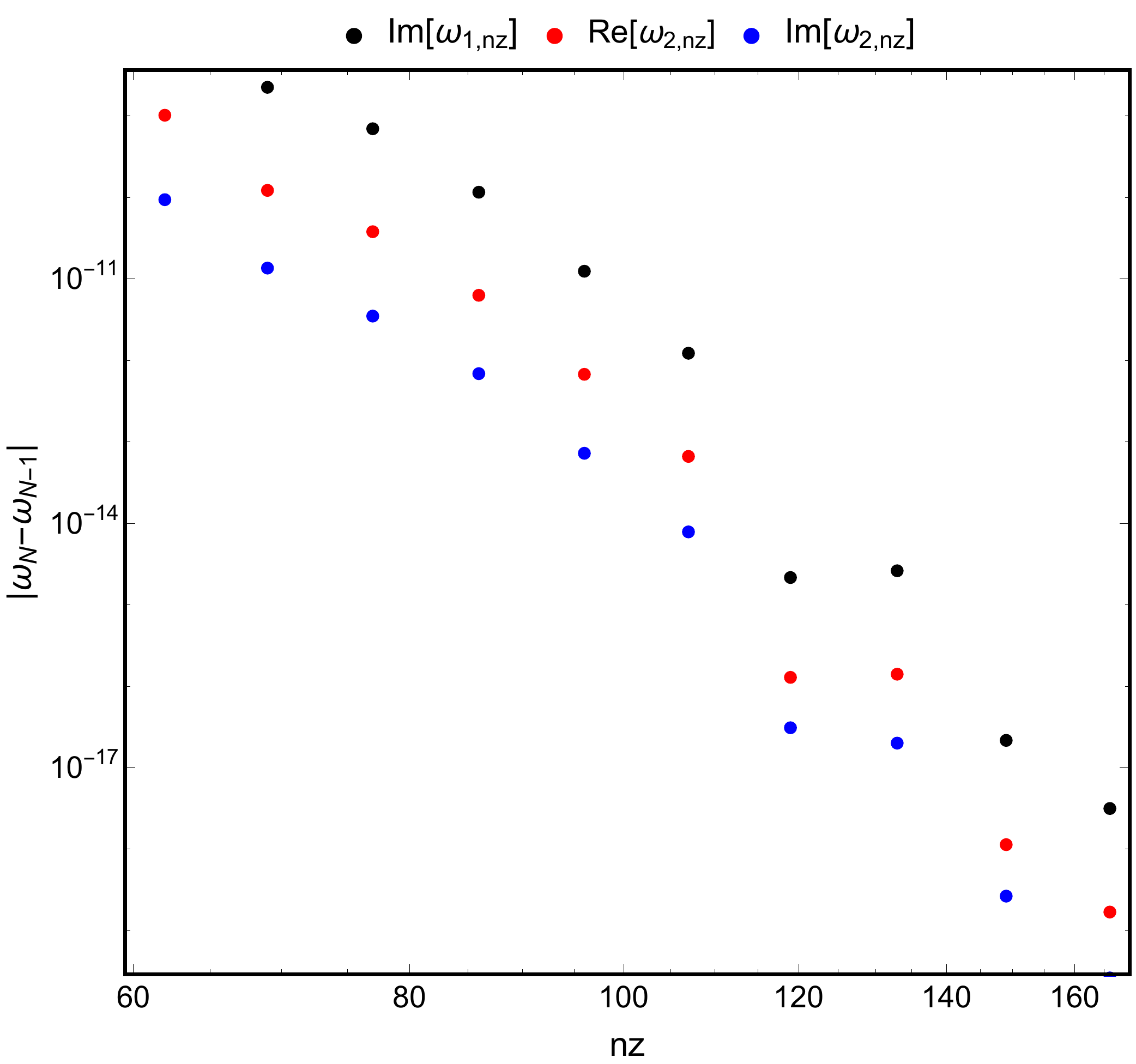}\vspace{0.2cm}
     \includegraphics[width=4.5cm]{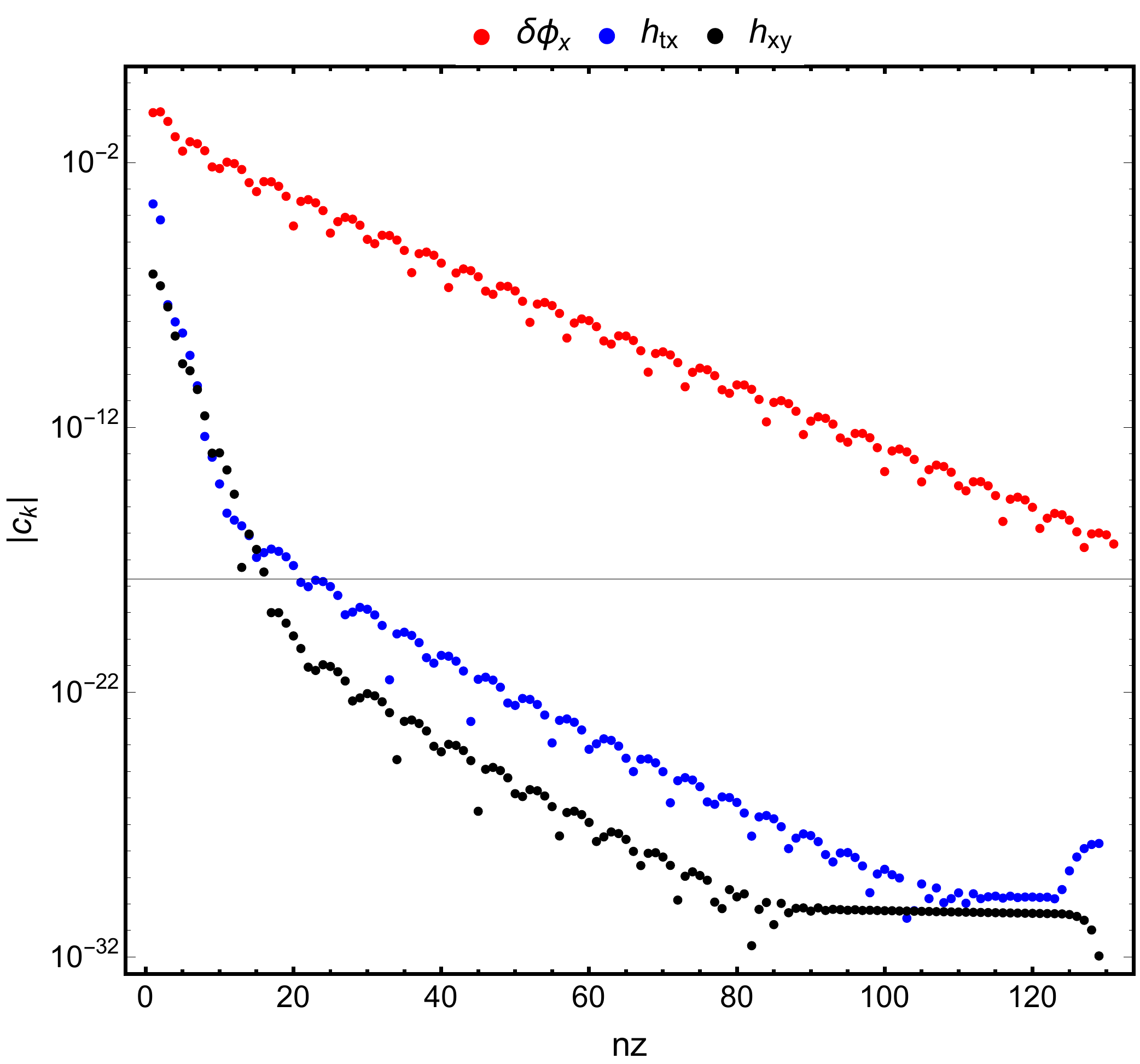}
     \includegraphics[width=4.5cm]{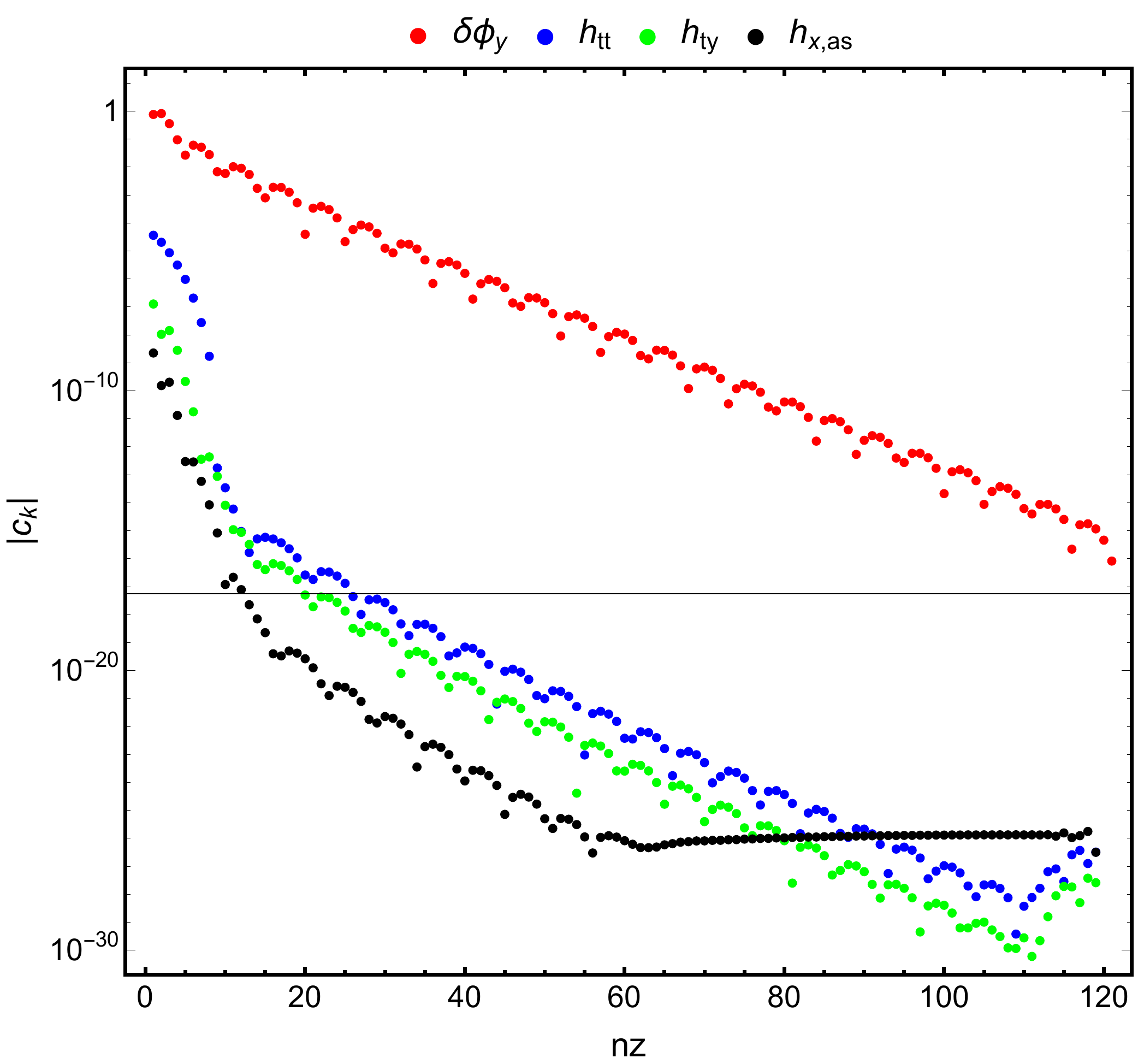}
      \includegraphics[width=4.5cm]{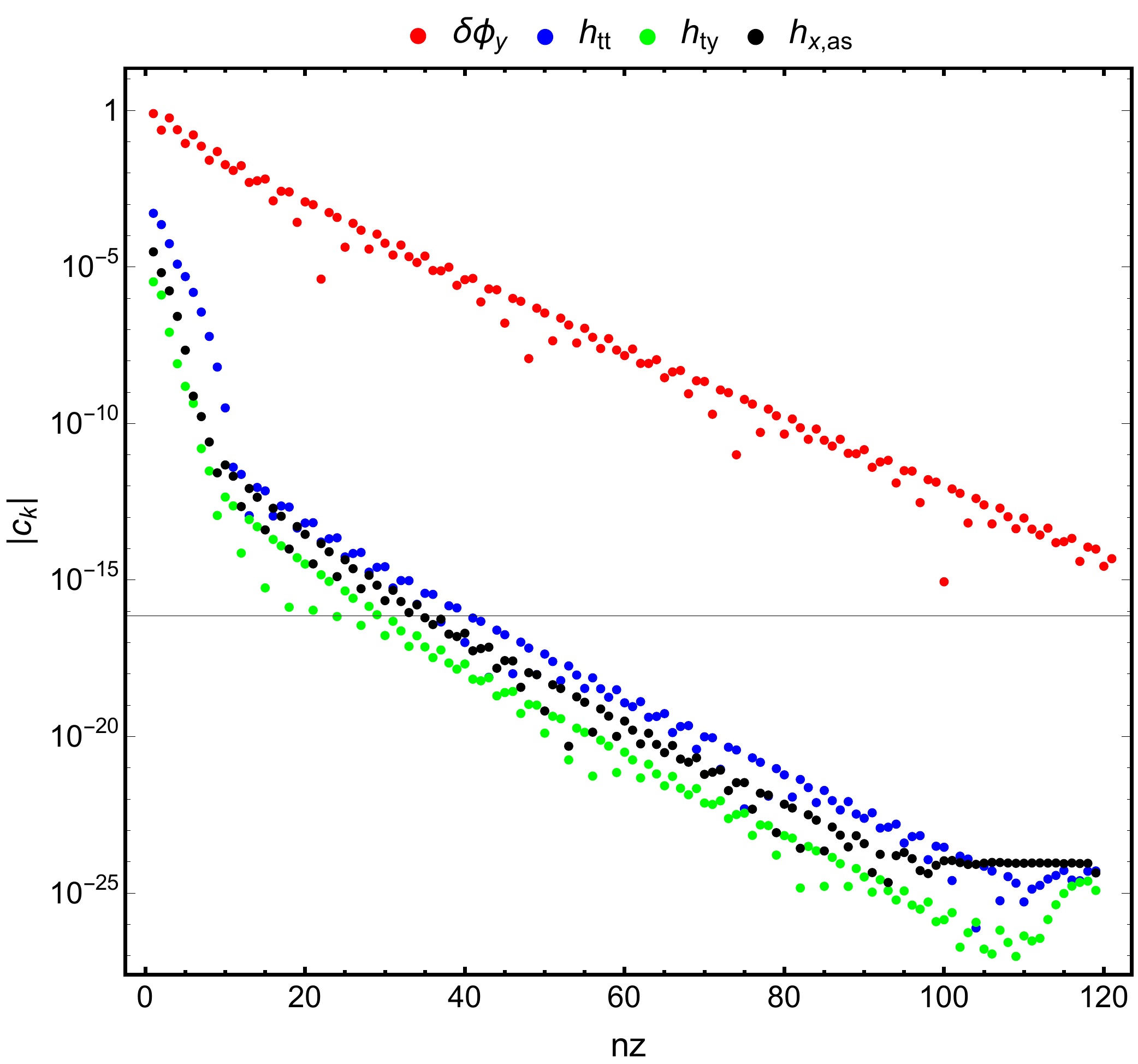}
    \caption{\textbf{Top panel:} Moving of the eigenvalues; monitored is the difference of the absolute value of the eigenvalues with increasing gridsize of the lowest QNMs. The difference decreases with higher grid resolutions. The plots correspond to the transverse sector of the $V(Z)$ model, the longitudinal sector of the $V(Z)$ model and the longitudinal sector of the $V(X)$ model (from left to right).
    \textbf{Bottom panel:} Chebychev coefficients of the eigenfunctions of the lowest QNM.}\label{convv}
\end{figure}

Keeping in mind that this is the most extreme case for our code, we present all plots for this case. First of all for an increasing number of gridpoints, the solution should converge to the exact solution. This means, that the difference between a solution with $N$ gridpoints and a solution with $N-1$ gridpoints should go to zero which we depict for the numerically hardest case in the first row of fig \ref{convv}. Second of all, we can test our solutions by plugging the numerical solution (namely the eigenvectors and eigenfunctions) back in the EOMs and constraint and check that these are fulfilled with the wanted precision. Lastly, we checked that the Chebychev coefficients drop down to the wanted precision which is depicted in the second line of fig \ref{convv}. 
\vspace{-0.7cm}
\bibliographystyle{JHEP}
\bibliography{Z}
\end{document}